\documentclass[
acmsmall,
authorversion=true
]{acmart}

\acmJournal{TOSEM}

\usepackage{pifont}

\usepackage{algorithmic}
\usepackage{graphicx}
\usepackage{textcomp}
\usepackage{xcolor}
\usepackage{array,booktabs}
\usepackage{multirow}
\usepackage[linesnumbered,ruled]{algorithm2e}
\usepackage{xspace}
\usepackage{pgfplotstable}
\usepackage{paralist}
\usepackage[binary-units=true,detect-weight=true, detect-family=true]{siunitx}
\DeclareSIUnit{\pp}{\textup{pp}}
\usepackage{datatool}

\usepackage{subfigure}
\usepackage{tikz}
\usetikzlibrary{shapes.geometric,arrows}
\usetikzlibrary{positioning,fit,calc}
\usetikzlibrary{arrows.meta}
\tikzstyle{arrow}=[draw] 
\usetikzlibrary{arrows}
\usepackage{subfigure}
\usepackage{pgfplotstable}
\pgfplotsset{compat=1.12}
\usepgfplotslibrary{statistics}
\usepackage{makecell}
\usepackage[normalem]{ulem}
\newcommand{\approach}{LAFF\xspace}
\newcommand{\myblock}[1]{\tikz[baseline=(char.base)]{
    \node[shape=circle, draw, fill=gray, opacity=.2, text opacity=1,inner sep=1pt] (char) {\vphantom{1g}#1};}}

\usepackage{hyperref}
\usepackage[hyphenbreaks]{breakurl}

\begin{document}

\title{A Machine Learning Approach for Automated Filling of Categorical Fields in Data Entry Forms}

\author{Hichem Belgacem}
\orcid{0000-0002-0521-2905}
\affiliation{\institution{University of Luxembourg}
  \country{Luxembourg}
}
\email{hichem.belgacem@uni.lu}

\author{Xiaochen Li}
\orcid{0000-0002-5068-1938}
\affiliation{\institution{Dalian University of Technology}
  \country{China}
}  
\email{xiaochen.li@dlut.edu.cn}
\authornote{This work was done while the author was affiliated with
  the University of Luxembourg, Luxembourg.}

\author{Domenico Bianculli}
\orcid{0000-0002-4854-685X}
\affiliation{\institution{University of Luxembourg}
  \country{Luxembourg}
}
\email{domenico.bianculli@uni.lu}

\author{Lionel Briand}
\orcid{0000-0002-1393-1010}
\affiliation{\institution{University of Luxembourg}
  \country{Luxembourg}
}
\affiliation{\institution{University of Ottawa}
  \country{Canada}
}  
\email{lionel.briand@uni.lu}

\setcopyright{acmcopyright}
\acmJournal{TOSEM}
\acmYear{2022} \acmVolume{1} \acmNumber{1} \acmArticle{1} \acmMonth{1} \acmPrice{}\acmDOI{10.1145/3533021}

\begin{filecontents*}{data/rq-applicability.csv}
Dataset, Algorithm,MRR-S1,Covrg-S1,MRR-R1,Covrg-R1,Train,P-Avg, P-Range
    , MFM,0.42,1.00,0.42,1.00,0.03,0, 0
	,ARM,0.47,1.00,0.53,0.99,1.26,120,8--409
NCBI,Na{\"i}veDT, 0.52, 1.00, 0.53, 1.00, 23, 1, 1--1
    ,FLS,0.55,1.00,0.54,1.00,N/A,0, 0
    ,LAFF,0.74,0.70,0.73,0.70,546,15,5--44
    ,MFM,0.64,1.00,0.64,1.00,0.15,0, 0
	,ARM,0.67,0.95,0.65,0.96,15,878,6--3074
PROP,Na{\"i}veDT, t/o, t/o, t/o, t/o, t/o, t/o, t/o
    ,FLS,0.53, 1.00, 0.58, 1.00, N/A, 0, 0
    ,LAFF,0.78,0.86,0.81,0.87,3652,138,9--317
\end{filecontents*}
\DTLloaddb{effectivenss}{data/rq-applicability.csv}
\DTLgetvalue{\MFTrain}{effectivenss}{1}{\dtlcolumnindex{effectivenss}{Train}}
\DTLgetvalue{\MFPredict}{effectivenss}{1}{\dtlcolumnindex{effectivenss}{P-Avg}}
\DTLgetvalue{\MFSeqMrr}{effectivenss}{1}{\dtlcolumnindex{effectivenss}{MRR-S1}}
\DTLgetvalue{\MFSeqMrrBGL}{effectivenss}{6}{\dtlcolumnindex{effectivenss}{MRR-S1}}

\DTLgetvalue{\ARMSeqMrr}{effectivenss}{2}{\dtlcolumnindex{effectivenss}{MRR-S1}}
\DTLgetvalue{\ARMRandMrr}{effectivenss}{2}{\dtlcolumnindex{effectivenss}{MRR-R1}}
\DTLgetvalue{\ARMSeqCov}{effectivenss}{2}{\dtlcolumnindex{effectivenss}{Covrg-S1}}
\DTLgetvalue{\ARMRandCov}{effectivenss}{2}{\dtlcolumnindex{effectivenss}{Covrg-R1}}
\DTLgetvalue{\ARMTrain}{effectivenss}{2}{\dtlcolumnindex{effectivenss}{Train}}
\DTLgetvalue{\ARMPredict}{effectivenss}{2}{\dtlcolumnindex{effectivenss}{P-Avg}}
\DTLgetvalue{\ARMSeqMrrBGL}{effectivenss}{7}{\dtlcolumnindex{effectivenss}{MRR-S1}}
\DTLgetvalue{\ARMPredictBGL}{effectivenss}{7}{\dtlcolumnindex{effectivenss}{P-Avg}}

\DTLgetvalue{\DTTrain}{effectivenss}{3}{\dtlcolumnindex{effectivenss}{Train}}

\DTLgetvalue{\LAFFTrain}{effectivenss}{5}{\dtlcolumnindex{effectivenss}{Train}}
\DTLgetvalue{\LAFFPredict}{effectivenss}{5}{\dtlcolumnindex{effectivenss}{P-Avg}}
\DTLgetvalue{\LAFFRandCov}{effectivenss}{5}{\dtlcolumnindex{effectivenss}{Covrg-R1}}
\DTLgetvalue{\LAFFRandCovBGL}{effectivenss}{10}{\dtlcolumnindex{effectivenss}{Covrg-R1}}
\DTLgetvalue{\LAFFRandMrr}{effectivenss}{5}{\dtlcolumnindex{effectivenss}{MRR-R1}}
\DTLgetvalue{\LAFFRandMrrBGL}{effectivenss}{10}{\dtlcolumnindex{effectivenss}{MRR-R1}}
\DTLgetvalue{\LAFFTrainBGL}{effectivenss}{10}{\dtlcolumnindex{effectivenss}{Train}}
\DTLgetvalue{\LAFFPredictBGL}{effectivenss}{10}{\dtlcolumnindex{effectivenss}{P-Avg}}
\newcommand{\LAFFPredictMax}{317}

\begin{filecontents*}{data/rq-component.csv}
ID, Local, Filter, MRR-S1,Covrg-S1,MRR-R1,Covrg-R1, MRR-S2,Covrg-S2,MRR-R2,Covrg-R2
\approach-LE, \ding{55}, \ding{55}, 0.54, 1.00, 0.56, 1.00, 0.74, 1.00, 0.77, 1.00
\approach-E,  \ding{51}, \ding{55}, 0.60, 1.00, 0.60, 1.00, 0.76, 1.00, 0.79, 1.00
\approach-L,  \ding{55}, \ding{51}, 0.55, 0.83, 0.61, 0.77, 0.78, 0.86, 0.79, 0.87
\approach,    \ding{51}, \ding{51}, 0.74, 0.70, 0.73, 0.70, 0.78, 0.86, 0.81, 0.87
\end{filecontents*}
\DTLloaddb{module}{data/rq-component.csv}
\DTLgetvalue{\LAFFSeqMrr}{module}{4}{\dtlcolumnindex{module}{MRR-S1}}
\DTLgetvalue{\BNSeqMrr}{module}{1}{\dtlcolumnindex{module}{MRR-S1}}
\DTLgetvalue{\LAFFRandMrr}{module}{4}{\dtlcolumnindex{module}{MRR-R1}}
\DTLgetvalue{\LAFFRandCov}{module}{4}{\dtlcolumnindex{module}{Covrg-R1}}
\DTLgetvalue{\BNRandMrr}{module}{1}{\dtlcolumnindex{module}{MRR-R1}}
\DTLgetvalue{\BNSeqMrrBGL}{module}{1}{\dtlcolumnindex{module}{MRR-S2}}
\DTLgetvalue{\BNRandMrrBGL}{module}{1}{\dtlcolumnindex{module}{MRR-R2}}

\begin{filecontents*}{data/rq-fields-ncbi-mean-mrr.csv}
NFF,mean, label
1,  0.417,mfm
1,  0.485,arm
1,  0.599,laff
2,  0.410,mfm
2,  0.564,arm
2,  0.758,laff
3,  0.270,mfm
3,  0.576,arm
3,  0.730,laff
\end{filecontents*}
\DTLloaddb{num-fields-laff-ncbi-mrr}{data/rq-fields-ncbi-mean-mrr.csv}
\DTLgetvalue{\LAFFOneFFNcbiMRR}{num-fields-laff-ncbi-mrr}{3}{\dtlcolumnindex{num-fields-laff-ncbi-mrr}{mean}}
\DTLgetvalue{\LAFFAllFFNcbiMRR}{num-fields-laff-ncbi-mrr}{9}{\dtlcolumnindex{num-fields-laff-ncbi-mrr}{mean}}
\begin{filecontents*}{data/rq-fields-ncbi-mean-pcr.csv}
NFF,mean, label
1,  1.000,mfm
1,  0.963,arm
1,  0.482,laff
2,  1.000,mfm
2,  1.000,arm
2,  0.638,laff
3,  1.000,mfm
3,  1.000,arm
3,  0.629,laff


\end{filecontents*}
\DTLloaddb{num-fields-laff-ncbi-pcr}{data/rq-fields-ncbi-mean-pcr.csv}
\DTLgetvalue{\LAFFOneFFNcbiPCR}{num-fields-laff-ncbi-pcr}{3}{\dtlcolumnindex{num-fields-laff-ncbi-pcr}{mean}}
\DTLgetvalue{\LAFFAllFFNcbiPCR}{num-fields-laff-ncbi-pcr}{9}{\dtlcolumnindex{num-fields-laff-ncbi-pcr}{mean}}

\begin{filecontents*}{data/rq-fields-prop-mean-mrr.csv}
NFF,mean,  label
1,  0.627, mfm
1,  0.631, arm
1,  0.774, laff
2,  0.629, mfm
2,  0.631, arm
2,  0.794, laff
3,  0.650, mfm
3,  0.644, arm
3,  0.793, laff
4,  0.628, mfm
4,  0.641, arm
4,  0.794, laff
5,  0.633, mfm
5,  0.648, arm
5,  0.800, laff
6,  0.635, mfm
6,  0.651, arm
6,  0.801, laff
7,  0.656, mfm 
7,  0.647, arm
7,  0.808, laff
8,  0.668, mfm
8,  0.651, arm
8,  0.808, laff
9,  0.659, mfm
9,  0.680, arm
9,  0.805, laff
10, 0.658, mfm
10, 0.660, arm
10, 0.808, laff
11, 0.635, mfm
11, 0.664, arm
11, 0.807, laff
12, 0.605, mfm
12, 0.666, arm
12, 0.813, laff
13, 0.632, mfm
13, 0.661, arm
13, 0.807, laff
14, 0.615, mfm
14, 0.670, arm
14, 0.815, laff
15, 0.667, mfm
15, 0.657, arm
15, 0.821, laff
16, 0.640, mfm
16, 0.637, arm
16, 0.804, laff
\end{filecontents*}
\DTLloaddb{num-fields-laff-prop-mrr}{data/rq-fields-prop-mean-mrr.csv}
\DTLgetvalue{\LAFFOneFFPropMRR}{num-fields-laff-prop-mrr}{3}{\dtlcolumnindex{num-fields-laff-prop-mrr}{mean}}

\DTLgetvalue{\LAFFAllFFPropMRR}{num-fields-laff-prop-mrr}{48}{\dtlcolumnindex{num-fields-laff-prop-mrr}{mean}}

\begin{filecontents*}{data/rq-fields-prop-mean-pcr.csv}
NFF, mean,  label
1,   1.000, mfm
1,   0.615, arm
1,   0.630, laff
2,   1.000, mfm
2,   0.870, arm
2,   0.703, laff
3,   1.000, mfm
3,   0.945, arm
3,   0.748, laff
4,   1.000, mfm
4,   0.971, arm
4,   0.781, laff
5,   1.000, mfm
5,   0.988, arm
5,   0.818, laff
6,   1.000, mfm
6,   0.994, arm
6,   0.853, laff
7,   1.000, mfm
7,   0.998, arm
7,   0.873, laff
8,   1.000, mfm
8,   0.999, arm
8,   0.892, laff
9,   1.000, mfm
9,   1.000, arm
9,   0.915, laff
10,  1.000, mfm
10,  1.000, arm
10,  0.933, laff
11,  1.000, mfm
11,  1.000, arm
11,  0.947, laff
12,  1.000, mfm
12,  1.000, arm
12,  0.957, laff
13,  1.000, mfm
13,  1.000, arm
13,  0.963, laff
14,  1.000, mfm
14,  1.000, arm
14,  0.974, laff
15,  1.000, mfm
15,  1.000, arm
15,  0.978, laff
16,  1.000, mfm
16,  1.000, arm
16,  0.975,laff
\end{filecontents*}
\DTLloaddb{num-fields-laff-prop-pcr}{data/rq-fields-prop-mean-pcr.csv}
\DTLgetvalue{\LAFFOneFFPropPCR}{num-fields-laff-prop-pcr}{3}{\dtlcolumnindex{num-fields-laff-prop-pcr}{mean}}

\DTLgetvalue{\LAFFAllFFPropPCR}{num-fields-laff-prop-pcr}{48}{\dtlcolumnindex{num-fields-laff-prop-pcr}{mean}}

\begin{filecontents*}{data/rq-sample-rand-ncbi.csv}
Size, MRR,      PCR
5.928,    0.400,	0.610
11.857,   0.320,	0.620
17.785,   0.630,	0.640
23.714,   0.290,	0.620
29.642,   0.730,	0.690
35.570,   0.560,	0.710
41.499,   0.730,	0.690
47.427,   0.530,	0.750
53.356,   0.730,	0.700
59.284,   0.740,	0.700
\end{filecontents*}
\DTLloaddb{sample-rand-ncbi}{data/rq-sample-rand-ncbi.csv}
\begin{filecontents*}{data/rq-sample-rand-prop.csv}
Size, MRR,      PCR
13.9557,   0.769,	0.827
27.9114,   0.797,	0.834
41.8671,   0.804,	0.824
55.8228,   0.804,	0.860	
69.7785,   0.805,	0.860	
83.7342,   0.807,	0.849
97.6899,   0.802,	0.859
111.6456,  0.806,	0.861
125.6013,  0.805,	0.864
139.557,  0.804,	0.869
\end{filecontents*}
\DTLloaddb{sample-rand-prop}{data/rq-sample-rand-prop.csv}
\begin{filecontents*}{data/rq-sample-seq-ncbi.csv}
Size, MRR,      PCR
5.928,    0.350,	0.800
11.857,   0.320,	0.760
17.785,   0.670,	0.720
23.714,   0.240,	0.820
29.642,   0.740,	0.700
35.570,   0.650,	0.690
41.499,   0.740,	0.700
47.427,   0.650,	0.700
53.356,   0.740,	0.700
59.284,   0.730,	0.690
\end{filecontents*}
\DTLloaddb{sample-seq-ncbi}{data/rq-sample-seq-ncbi.csv}
\begin{filecontents*}{data/rq-sample-seq-prop.csv}
Size, MRR,      PCR
13.955,   0.649,	0.990
27.911,   0.733,	0.990
41.867,   0.755,	0.909
55.822,   0.798,	0.870
69.778,   0.798,	0.870
83.734,   0.764,	0.918
97.689,   0.798,	0.870
111.645,  0.800,	0.866
125.601,  0.798,	0.870
139.557,  0.799,	0.870
\end{filecontents*}
\DTLloaddb{sample-seq-prop}{data/rq-sample-seq-prop.csv}
\DTLgetvalue{\NCBISeqMRRLow}{sample-seq-ncbi}{6}{\dtlcolumnindex{sample-seq-ncbi}{MRR}}
\DTLgetvalue{\NCBISeqMRRHigh}{sample-seq-ncbi}{9}{\dtlcolumnindex{sample-seq-ncbi}{MRR}}
\DTLgetvalue{\NCBIRandMRRLow}{sample-rand-ncbi}{8}{\dtlcolumnindex{sample-rand-ncbi}{MRR}}
\DTLgetvalue{\NCBIRandMRRHigh}{sample-rand-ncbi}{10}{\dtlcolumnindex{sample-rand-ncbi}{MRR}}
\DTLgetvalue{\PROPSeqMRRLow}{sample-seq-prop}{6}{\dtlcolumnindex{sample-seq-prop}{MRR}}
\DTLgetvalue{\PROPRandMRRHigh}{sample-rand-prop}{6}{\dtlcolumnindex{sample-rand-prop}{MRR}}
\DTLgetvalue{\NCBISeqPCRLow}{sample-seq-ncbi}{6}{\dtlcolumnindex{sample-seq-ncbi}{PCR}}
\DTLgetvalue{\NCBIRandPCRHigh}{sample-rand-ncbi}{8}{\dtlcolumnindex{sample-rand-ncbi}{PCR}}
\DTLgetvalue{\PROPRandPCRLow}{sample-rand-prop}{6}{\dtlcolumnindex{sample-rand-prop}{PCR}}
\DTLgetvalue{\PROPSeqPCRHigh}{sample-seq-prop}{6}{\dtlcolumnindex{sample-seq-prop}{PCR}}

\def\UrlBreaks{\do\A\do\B\do\C\do\D\do\E\do\F\do\G\do\H\do\I\do\J
\do\K\do\L\do\M\do\N\do\O\do\P\do\Q\do\R\do\S\do\T\do\U\do\V
\do\W\do\X\do\Y\do\Z\do\[\do\\\do\]\do\^\do\_\do\`\do\a\do\b
\do\c\do\d\do\e\do\f\do\g\do\h\do\i\do\j\do\k\do\l\do\m\do\n
\do\o\do\p\do\q\do\r\do\s\do\t\do\u\do\v\do\w\do\x\do\y\do\z
\do\.\do\@\do\\\do\/\do\!\do\_\do\|\do\;\do\>\do\]\do\)\do\,
\do\?\do\'\do+\do\=\do\#} 


\begin{abstract}
Users frequently interact with software systems through data
entry forms. However, form filling is time-consuming and error-prone.
Although several techniques have been proposed to auto-complete or
pre-fill fields in the forms, they provide limited support to help
users fill categorical fields, i.e., fields that require users to
choose the right value among a large set of options.

In this paper, we propose \approach, a learning-based automated
approach for filling categorical fields in data entry forms.
\approach first builds Bayesian Network models by learning field dependencies from a
set of historical input instances, representing the values of the
fields that have been filled in the past. To improve its learning
ability, \approach uses local modeling to effectively mine the local
dependencies of fields in a cluster of input instances.  During the
form filling phase, \approach uses such models to predict possible
values of a target field, based on the values in the already-filled
fields of the form and their dependencies; the predicted values
(endorsed based on field dependencies and prediction confidence) are then provided to
the end-user as a list of suggestions.

We evaluated \approach by assessing its effectiveness and efficiency
in form filling on two datasets, one of them proprietary from the banking domain. 
Experimental results show that \approach is able to provide accurate suggestions
with a Mean Reciprocal Rank value above {\LAFFRandMrr}. Furthermore,
\approach is efficient, requiring at most
{\SI{\LAFFPredictMax}{\milli\s}} per suggestion.

 \end{abstract}

\maketitle

\section{Introduction}
\label{sec:introduction}

Data entry forms are essential software user interface (UI)
elements~\cite{jarrett2009forms, sears2003data} to collect users'
inputs. 
According to statistics, approximately 70 million professionals or 59\% of all professionals in the United States need to complete on-line forms for their daily jobs~\cite{wang2017context}.
However, form filling is time-consuming and error-prone~\cite{rukzio2008automatic}.
Some domain-specific software, such as enterprise resource planning
(ERP) systems~\cite{akiki2016engineering}, 
may include data entry forms with hundreds of fields for users to fill in during the execution of specific
business processes.

The form filling process lays an extra burden on users, who have to
spend tremendous energy on inputting the right value into each
field, resulting in a task that is both slow and frustrating~\cite{gutwin2006improving}. 
This situation inevitably leads to
data quality issues (e.g., when the wrong value is provided as input
for a field), which may seriously affect data-reliant software systems and undermine business opportunities and even cause loss of human life~\cite{american2005data}. 
Existing statistics reveal that data entry typically has an error rate of  over 1\%~\cite{fan2008revival}. 
Such erroneous data is then transferred into the software system that
uses it and affects all the other connected information. 
For example, erroneous input data in retailing systems alone leads to a waste of \$2.5 billion each year for consumers~\cite{fan2008revival}. 
Further, data entry errors in spreadsheets mislead business decisions, 
causing additional costs in correcting the errors~\cite{sakal2012errors}.
Even worse, data entry has been a top cause of medication errors~\cite{american2005data}, 
which resulted in at least 24 deaths in the US in 2003~\cite{mucslu2015preventing}.

From a software engineering point of view, it is important to develop
approaches that reduce data entry errors.  Since data entry forms may
contain various types of fields (e.g., textual, numerical, and
categorical fields), various studies have proposed different  strategies to improve data quality.  One way is to
detect data quality issues by running test queries (which check the
semantic validity of the data) on the application database right after
data updates; this is the approach proposed in the context of
continuous data testing~\cite{mucslu2015preventing}, which focuses on
identifying numerical errors.  Another possible solution for data
error prevention is to relieve users from the burden of form filling,
i.e., automating the form filling process, with mechanisms that
pre-fill or auto-complete data entry forms during data entry
sessions~\cite{van2008efficient, salama2018text}.  Existing work in
this area mainly focuses on textual fields; they build language models
(e.g., sequence-to-sequence models) to learn the character or word
relation from historical textual inputs, and then provide word
auto-completion based on the letters typed in a
field~\cite{zhang2019text}.

However,  the aforementioned techniques cannot be applied to
\emph{categorical} fields, which are fields that provide
a list of candidate values (also called ``options'') from which the user has to choose (e.g.,
country).
Such fields are likely to generate data quality issues. 
For instance, empirical studies have shown that more than half
(54.5\%) of the data errors in a medical record system 
were caused by the candidate value selection error~\cite{qian2020trend}. 
Users could wrongly select the job role, the modality of care, or the drugs unintentionally~\cite{qian2020trend,khajouei2010impact},
causing potential medical negligence.
As another example, in financial systems, the correctness of selected values 
is required by regulations (e.g., for the ``purpose of business'' or
``source of wealth'' fields, when opening a bank account),
to prevent money laundering~\cite{bis2003general}.

The data entry errors in categorical fields are mainly caused by two reasons.
First, categorical fields require focused attention for users
to choose among a large set of options in a limited time.
For example, a real-world computer-assisted clinical ordering system contains a lengthy alphabetized list of 33 items, 
13 of which are category headers with subsumed (concealed) levels~\cite{horsky2003framework}. Also, a species management system may require users to select species from dozens of genera, 
with nearly a hundred valid species in a single genus~\cite{gafur2020updated}.
It is well known that users tend to make more mistakes
as more options are presented to select from~\cite{khajouei2010impact,chen2011usher}. 
Moreover, juxtaposition selection errors can happen 
when adjacent values are textually similar (e.g., in an alphabetically-sorted list)~\cite{tolley2018factors}.
Although some systems enable searchable list boxes
to help users accelerate this selection~\cite{jensen2020country},
filling categorical fields is still difficult, 
as users require significant cognitive effort to match each option with the actual value they intend to fill in.  
Many users lack a detailed conceptual model of the software system ~\cite{horsky2003framework} defined by requirements analysts and domain experts.
For example, junior healthcare providers may not remember and understand 
all the options predefined in the field of ``modality of care'' in an electronic medical record system~\cite{qian2020trend}, 
which necessitates a potential lengthy search process or may lead to the selection of an inappropriate value~\cite{horsky2003framework}.
The cognitive load is even higher when there are more options for users to compare with~\cite{chen2011usher}. 
In such cases, the time to select a value is linear with the list length~\cite{cockburn2009predictive}.

Some form filling tools~\cite{chrome, wang2017context} support filling categorical fields, by suggesting
frequently selected values in a field or values selected by a user in
some similar fields from 3rd-party software systems. Nevertheless,
they provide limited support due to the low accuracy of their 
suggestions; moreover, their usage may violate enterprise
security policies, since they rely on information from 3rd-party
software systems~\cite{winckler2011approach}.

Furthermore, existing automated form filling approaches exhibit some
limitations when dealing
with
\begin{inparaenum}[(1)]
\item forms filled following an arbitrary order, and 
\item partially filled forms.
\end{inparaenum}
The first case occurs because, while filling in a data entry form, it
is very frequent to have little or no restriction on the order of
users' inputs: a user may select any field as the next target or even
go back and modify already filled fields. Some automated form filling
approaches~\cite{hermens1994machine} require a fixed form filling
order before building a recommendation model; however, this assumption
is unrealistic from a practical standpoint.  As for the second case,
at a certain time during the data entry session, a form is usually
partially filled,  meaning that an approach for automated form
filling can only use information in currently filled fields.
However, when this is not sufficient to
predict the target, existing approaches~\cite{martinez2019using,
  hermens1994machine} yield inaccurate suggestions.

To provide effective form filling suggestions, in this paper, we
propose \emph{\approach}, a \underline{L}earning-based
\underline{A}utomated \underline{F}orm \underline{F}illing approach,
for filling categorical fields.  The basic idea of \approach is to
build machine learning models based on input instances (i.e., fields
and the corresponding values provided in input) obtained from data
entry forms that have been filled in the past (hereafter called
\emph{historical input instances}); such models represent
dependencies among fields in historical input instances. Using these
models, the already-filled fields in a data entry form can then be
used as features to predict the possible values of a given target
field.
\approach aims to
be used by developers, who can integrate it into their data entry form
implementations.

To deal with forms filled in an arbitrary order (which would result in
a huge number of combinations of filled fields (features) and target
fields to handle), \approach utilizes Bayesian Networks (BN) to mine the
dependencies between any field combinations, without
assuming, a priori, an order for form filling.  Moreover, to improve its
learning ability, \approach uses a local modeling strategy to cluster
historical input instances; further, it builds additional local BNs,
which learn fine-grained field dependencies from the clusters of
historical input instances. These local models capture additional
dependencies that might not have been captured by the model trained
on the entire historical dataset.
Once the trained models are available, \approach uses them in the form
filling suggestion phase, which occurs during the data entry session:
given a target field, \approach selects one of the available BNs and
predicts the possible values of the field based on the values in the
already-filled fields.  To deal with partially filled forms (which
might lead to inaccurate suggestions), \approach includes a
heuristic-based endorser, which determines whether the values predicted
in the previous step are accurate enough to be returned to the user,
based on the analysis of the field dependencies and of the predicted
probability distribution of the values for the target field.

We evaluated \approach using form filling records from both a public dataset and a proprietary dataset extracted from a production-grade enterprise information system in the banking domain.
The experimental results show \approach can yield a large number of
accurate suggestions with a \textit{Mean Reciprocal Rank} (\textit{MRR}) value above {\LAFFRandMrr} and a prediction coverage rate ranging from \numrange{\LAFFRandCov}{\LAFFRandCovBGL},
significantly outperforming a state-of-the-art approach based on association
rule mining by \SIrange{11}{27}{\pp} (with \si{\pp} = percentage
points) in terms of \textit{MRR} on both datasets.
Furthermore, \approach is efficient; it takes at most
{\SI{\LAFFPredictMax}{\milli\s}} to provide suggestions for input
instances of the proprietary dataset.

To summarize, the main contributions of this paper are:
\begin{itemize}
	\item The \approach approach, which addresses the problem of
          automated form filling for categorical fields, an important
          user interface challenge in many software systems. To the
          best of our knowledge, 
          LAFF is the first work to combine
          BNs with local modeling and a heuristic-based endorser to provide accurate form filling suggestions, even for arbitrary filling orders and partially filled forms.
\item An extensive evaluation assessing the effectiveness and efficiency of \approach and comparing it with the state of the art.
\end{itemize}

The rest of the paper is organized as
follows. Section~\ref{sec:background} provides a motivating example
and explains the basic definitions of automated form filling and its
challenges. Section~\ref{sec:preliminaries} introduces the basic
machine learning algorithms used in this
paper. Section~\ref{sec:approach} describes the different steps and
the core algorithms of \approach. Section~\ref{sec:evaluation} reports
on the evaluation of \approach. 
Section~\ref{sec:discussion} discusses the usefulness, practical implications, and limitations
of \approach.
Section~\ref{sec:related_work}
surveys related work. Section~\ref{sec:conclusion} concludes the
paper.

\section{Data Entry Form Filling}
\label{sec:background}

In this section, we introduce the concepts related to data entry
forms, provide a motivating example, define the problem of
automated form filling, and discuss its challenges.

\subsection{Data Entry Forms}

A data entry form is typically composed of many fields (also called input
parameters~\cite{wang2017context} or
elements~\cite{akiki2016engineering}), which can be of different
types: textual, categorical, numerical, and file.  Textual and
numerical fields collect free text and numerical values respectively; users can
freely input any value that is compliant with the field input
validation rules.  Categorical fields provide a list of candidate values
from which the user has to choose (e.g., country); the source
of candidate
values is defined statically~\cite{aggarwal2007database}.
File fields are used to upload files, such as images and videos.
During software design, these fields are associated with specific UI
widgets, based on the corresponding type.  For example, developers can
use a list box or combo box to collect categorical
data~\cite{w3cschools2017html}.

\subsection{Motivating Examples}
\label{sec:motivating-examples}
Although software users frequently fill forms, such activity
is time-consuming and error-prone~\cite{rukzio2008automatic}.
In the following, we describe two examples illustrating the main
  challenges faced while filling data entry forms with categorical fields.

\textbf{Example 1}: \textit{Users require focused attention to choose options from categorical fields}.

Alison is a student majoring in biology. 
She uses information management platforms 
(such as NCBI~\cite{barrett2012bioproject} and SAIKS~\cite{gafur2020updated}) 
to record the basic information of biological samples. 
Given a biological sample (e.g., the genus \textit{Pratylenchus}), 
she needs to fill fields ``sex'', ``tail shape'', and ``species name'' of the sample in a data entry form from such a system. 
All three fields are categorical with predefined values.
After filling the first two fields, Alison starts to select the species name.
However,  the genus \textit{Pratylenchus} currently includes over 80 valid species~\cite{gafur2020updated}. 
She has to scroll down the list to check for the relevant species.
Although she can search a species by inputting the first letter of the name,
many similar species names are presented (e.g., \textit{pratensisobrinus}, \textit{pseudocoffeae}, \textit{pseudofallax}, and \textit{pseudopratensis}). 
She still requires focused attention to choose among these textually similar options in a limited time.
According to existing studies, about half of the data entry errors are caused by selection error~\cite{qian2020trend,westbrook2013safety}

\begin{figure}[tb]
	\centering{
\tikzstyle{field} = [rectangle, minimum width=1.2cm, minimum height=0.7cm, text centered, text width=2.3cm, draw=black, fill= white!30]

\tikzstyle{label} = [rectangle, minimum width=1.7cm, minimum height=0.35cm, text centered, text width=2.5cm, align=right, draw=white, fill= white!30]

\tikzstyle{title_node} = [rectangle, minimum width=0.5cm, minimum height=0.35cm, text centered, text width=4.5cm, draw=white, fill= white!30]

\tikzstyle{text_node} = [rectangle, minimum width=2.5cm, minimum height=0.7cm, text centered, text width=2.5cm, fill=white!30]

\tikzstyle{round_rect} = [rectangle, rounded corners, minimum width=1.5cm, minimum height=0.5 cm,text centered, draw=black, fill=white!30]

\newcommand{\DrawTriangle}[1][]{\begin{tikzpicture}[overlay,remember picture]
	\filldraw[fill=black] (90:1.2ex) -- (170:0.7ex) -- (10:0.7ex) --cycle;
	\end{tikzpicture}
}

\resizebox{1\textwidth}{!}{
\begin{tikzpicture}[node distance=3mm, >=latex]
	
	\node (tb1) [field]{Gibson};
	\node (f1) [label, left= of tb1] {Name of contact person};
	\node (tb2) [field, below=of tb1, text width=1.1cm, minimum width=1.1cm, xshift=-0.6cm] {20};
	\node (f2) [label, left= of tb2] {Monthly income};
	\node (f2_unit) [label, right= of tb2, xshift=-2mm, align=left, text width=1.3cm] {\emph{k euro}};
	\node (tb3) [field, below=of tb2, xshift=0.6cm] {Private};
	\node (f3) [label, left= of tb3] {Legal entity};
	\node (tb4) [field, below=of tb3] {Leasing};
	\node (f4) [label, left= of tb4] {Company type};
	\node (tb5) [field, below=of tb4, text=red] {\textbf{?}};
	\node (f5) [label, left= of tb5] {Primary field of activity};
	
	\node (icon_tb3) [isosceles triangle,
	isosceles triangle apex angle=60, draw, rotate=270, fill=gray!120, minimum size =0.1cm, right=of tb3, xshift=-0.25cm, yshift=-0.6cm]{};
	
	\node (icon_tb5) [isosceles triangle,
	isosceles triangle apex angle=60, draw, rotate=270, fill=gray!120, minimum size =0.1cm, right=of tb5, xshift=-0.25cm, yshift=-0.6cm]{};
		
	\node (submit) [round_rect, below= of tb5, xshift=0.35cm, minimum height=0.7cm, text centered, text width=1.6cm, fill=gray!30] {Submit};
	
	\node (cancel) [round_rect, left= of submit, xshift=-0.35cm, minimum height=0.7cm, text centered, text width=1.6cm, fill=gray!30] {Cancel};
	
	\node (ui) [draw=none, fit= (tb1) (tb2)(tb3)(tb4)(tb5)(f1)(f2)(f3)(f4)(f5)(submit) ] {};
	
	\node (title)[title_node, font=\fontsize{12}{0}\selectfont, above=of ui] {\textbf{Data entry form \emph{F}}};
	
	\node (form) [draw=black, fit= (ui)(title)] {};
		
	\draw [decorate, decoration={brace, aspect=0.37, amplitude=10pt, raise=4pt}, yshift=0pt] (tb1.north east) -- (tb4.south east)  node [black, xshift=0cm] { };
	
	\node (model) [round_rect, right= of f2_unit, xshift=2cm, minimum height=1cm, text centered, text width=2.3cm] {\textbf{Model $\mathbf{M}$}};
	
	\draw [arrow,->] (tb2) ++(2.3,0) -- (model) node[midway, text width=2.5cm, text centered] {$F^f$\\ filled fields} ;
	
	\draw [arrow,->] (model) |- (tb5) node[anchor=north east, above, pos=0.80] {predict};

	\node (tab1) [shape=rectangle, draw=white, right= of submit, xshift=3.9cm, yshift=1cm] {
		\scriptsize
		\begin{tabular}{ccccccc}
		\toprule
		
		\multirow{2}{*}{\textbf{$f_1$: name}}&\textbf{$f_2$: in-}&\multirow{2}{*}{\textbf{$f_3$: entity}}&\textbf{$f_4$: company}&\textbf{$f_5$: primary }&\textbf{submission}\\
		&\textbf{come}&&\textbf{type}&\textbf{activity}&\textbf{time}\\
		\midrule
		Alice &20 &Public &Investment&Financial Service&20180101194321\\
		\midrule
		Bob&21 & Public &Investment&Leasing Service&20180101194723 \\ 
		\midrule
		...&... & ... &...&...&  \\ 
		\midrule
		Frank&40&Public &Leasing &Financial Service&20180102132016\\ 
		\bottomrule
		\end{tabular}
	};
	
	\draw [arrow,->] (submit)  -- ++(5.1, 0)(tab1) node[above, pos=0.4, text width=2.5cm, text centered] {submission} ;
	
	\node(history)[text_node, above= of tab1, text width=5cm, yshift=-0.4cm]{Historical input instances $I^F(t)$};
	
	\draw [arrow,->] (history)  |- (model) node[below, pos=0.75, text width=2.5cm, text centered] {train} ;

\end{tikzpicture}
}
 }
	\caption{The Automated Form Filling Problem}
	\label{fig:def}
\end{figure}
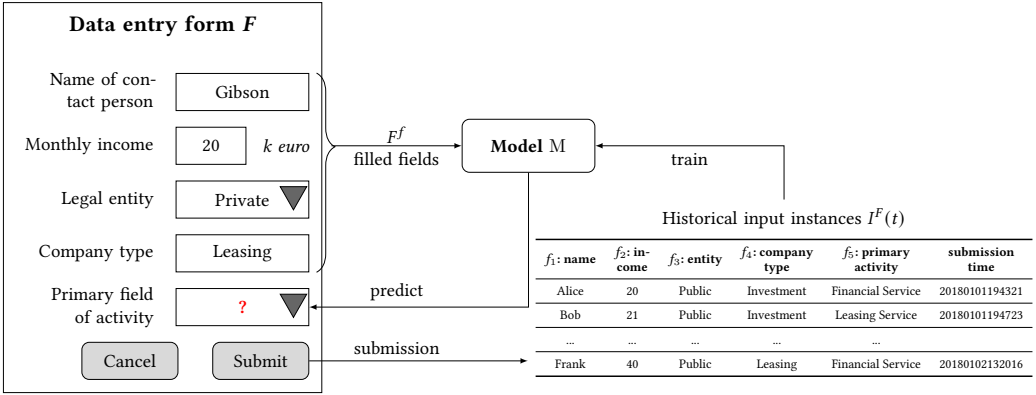

\textbf{Example 2}: \textit{Users require cognitive effort to match options with the actual value they plan to fill in}.
The second example is inspired
by a use case of our industrial partner. The use case refers to the opening of a bank account for
business customers;
a simplified data entry form for this activity is shown on the
left-hand side of \figurename~\ref{fig:def}.
We will use this data entry form as a running example 
to explain the form filling problem and illustrate our solution.
For simplicity, let us assume that the data entry
form contains only two fields: ``company type'' (textual) and
``(company) primary field of activity'' (categorical).  When a company called
\emph{SmartLease} requests to open a bank account, the bank clerk Bill
(who is the end-user interfacing with the data entry form) asks the
\emph{SmartLease} representative about the company type and inputs
``leasing (company)'' in the corresponding field; then, Bill
selects ``other financial services'' for the ``primary field of activity'' field.  Several weeks after the account opening, the data quality
division of the bank detects a potential data quality issue regarding
\emph{SmartLease}, in the form of a mismatch between the actual
activity implied by the company's type and operations, and the company
activity recorded when the account was open. This issue can be quickly
solved by checking with \emph{SmartLease} and amending the ``primary field of
activity'' field with the correct value: ``leasing services''. Nevertheless, such
an issue can cause a business loss: for example, by knowing the actual
``primary field of activity'' of the company, the bank could have offered targeted products to its
customer since the beginning of the business relation.
Further investigation reveals that this issue
occurred because, as a new employee in the bank, 
Bill was not familiar with all the 75 options defined in the ``primary field of
activity'' field. 
He browsed the list for a limited time, compared the candidate values with the actual value he intended to fill in and finally selected an inappropriate value.

The aforementioned problems cannot be solved satisfactorily by
existing solutions that support filling categorical fields,
including those based on the search-by-keyword functionality, web
browser plugins for autofill, as well as approaches that build field ontologies.

First, some solutions help users locate the candidate values in categorical fields with the search-by-keyword functionality.
This functionality cannot solve the issues in the two examples, since users need to carefully compare textually similar values (as shown in Example 1) 
and have the burden to remember all the options to avoid searching an inappropriate value (as shown in Example 2).

Second, web browser plugins such as Chrome Autofill
Forms~\cite{chrome} provide automatic form filling, but they simply
reuse the inputs provided in past forms to
automatically fill out fields in different forms with the same
information (for example ``zip code''). They do not leverage the
knowledge provided by already filled fields to provide ``intelligent''
suggestions. Moreover, these tools are usually personalized for a
single user, and cannot be used in the context of enterprise software
systems, in which the end-user fills the same form with different
information.  As shown in Example~2, a bank clerk works daily with several bank
accounts of different customers and cannot directly reuse the input
instance of a customer to pre-fill an input form for another customer.

Third, some approaches automatically build ontologies for form filling~\cite{an2012learning,winckler2011approach,araujo2010carbon}.
They map a `target' field (e.g., ``zip code'') in a form to `source' fields (e.g., ``postal code'' or ``postcode'') in other filled forms 
to support data exchange across software systems.
However, for domain-specific software systems (e.g., biology information management platforms), 
many fields are domain-specific (such as ``tail shape'' and ``species
name'' in Example 1) and cannot be easily mapped to fields in other forms.
In addition, due to legal or security policies, 
software systems for governments and enterprises have constraints on sharing records across systems (as the banking system in Example~2).

For all the above reasons, there is a need to design a semi-automated method
that developers can adopt
to
support and guide users during the form filling activity.

\subsection{Problem Definition}
\label{sec:prob_def}

In this paper, we deal with the automated form filling problem, which
can be informally defined as the problem of suggesting possible values
for the form fields a user is about to fill in, based on the values of
the other fields and on the values provided as input in previous data
entry sessions.  We target \emph{categorical fields} for automated filling
since they require cognitive effort and focused attention for users
to choose among the (typically) large set of options. The task of
filling categorical fields may be slow and frustrating to
users~\cite{gutwin2006improving}, and may lead to data quality issues,
as shown in the examples above.  We define the automated form filling
problem as follows.

Let $F$
be a data entry form with $n$ fields
$F=\{f_1, f_2, \dots, f_n\}$; each field $f_i$ can take values from a
domain $V_i$ (which always includes a special element $\bot$ representing
an empty  field); let $F^c \subseteq F$ be the set of categorical fields.

When a form $F$ is being filled, at any time the fields can be partitioned in two
groups: fields that have been already filled (denoted by $F^{f}$) and
unfilled fields (denoted by $F^{u}$); we have that $F^{f} \cup
F^{u}=F$ and $F^{f} \cap F^{u}=\emptyset$.

When a filled form $F$ is about to be submitted (e.g., to be stored in
a database), we define an \emph{input instance} of $F$
$I^F=\{\langle f_1, v_1 \rangle, \dots, \langle f_n, v_n\rangle\}$ with
$f_i \in F$ and $v_i \in V_i$, as the set of pairs
$\langle \mathit{field}, \mathit{value}\rangle$ from $F$; we use the
subscript $t_j$ as in $I^F_{t_j}$ to denote that the input instance $I^F$ was
submitted at time $t_j$.
We use the notation $I^F(t)$ to represent the set of \emph{historical
  input instances} of form $F$ that have been submitted up to a certain time
instant $t$;  $I^F(t)=\{I^F_{t_{i}}, I^F_{t_{j}}, \dots,
I^F_{t_{k}}\}$, where $t_i < t_j < t_k < t$. Hereafter, we drop the
superscript $F$ when it is clear from the context.

The automated form filling problem can be defined as follows. Given a
(partially filled) form $F=F^{f} \cup F^{u}$, a set of historical
input instances $I^F(t)$, and a target field $f_p \in (F^u \cap F^c)$ to fill, we
want to build a model $M$ that at time $t$ can predict a value $v_p$
for $f_p$ based on $F^{f}$ and $I^F(t)$. 
Notice that in this problem definition the
filling  order of the fields in $F$ is unrestricted.

\subsubsection*{Application to the running example}
\figurename~\ref{fig:def} shows an example explaining the automated form filling problem.
We have a data entry form $F$ for a banking system with five fields: 
$f_1$:``name of contact person'', $f_2$:``monthly income'',
$f_3$:``legal entity'', $f_4$:``company type'', and $f_5$:``primary field of activity''.
Among them, fields ``legal entity'' and ``primary field of activity'' are categorical (i.e., $F^c=\{f_3, f_5\}$).
During the data entry session, users provide values for these fields,
which are stored in a database upon submission of the form.
The table on the right-hand side of \figurename~\ref{fig:def}
shows some historical input instances filled by the bank customers
through the data entry form;
the submission timestamp $t$ of these input instances is automatically recorded.
In the table, each row represents an input instance 
(e.g., $I^F_{20180101194321}=\{\langle \textit{``name''}, \textit{Alice} \rangle, \dots, \langle \textit{``primary activity''}, \textit{Financial Service}\rangle\}$);
the column names correspond to the field names in the data entry
form\footnote{In the case of this illustrative example, we assume that
the mapping between field names and column names can be retrieved in
some way, for
example by manually analyzing the existing software design documentation or software implementation. 
We provide more explanations on identifying such a mapping in section~\ref{sec:implications:sw-dev}.}.
With these historical input instances, 
we can build a model $M$ to learn the relationships of values filled in fields $f_1$ to $f_5$ by different customers.
Notice the submission timestamp is not used for model building;
it is only used for distinguishing different input instances.

Continuing the example, let us assume that, at a certain point of the
data entry session, a customer has provided the values \emph{Gibson},
\emph{20}, \emph{Private}, and \emph{Leasing} for fields $f_1$ to
$f_4$ respectively, as shown on the left-hand side of
\figurename~\ref{fig:def}; the unfilled field $f_5$ (``primary field
of activity'') is the next (categorical) field to fill in.  Our goal
is to use the model $M$ to predict the value of $f_5$ based on the
values of the filled fields $f_1$ to $f_4$.

\subsection{Challenges of Automated Form Filling}
Several
automated form filling approaches have been
proposed~\cite{martinez2019using, hermens1994machine,troiano2017modeling}; the basic idea is to mine dependencies among
fields from the values recorded in previous form filling sessions, to
build recommendation models. These models can then be used to suggest
possible values on a target field based on the filled fields in the
current form.  Nevertheless, state-of-the-art approaches exhibit some
limitations when dealing
with
\begin{inparaenum}[(1)]
\item forms filled following an arbitrary order, and 
\item partially filled forms.
\end{inparaenum}

First, while filling in a data entry form, it is very frequent to have
little or no restriction on the order of user' inputs. A user may
select any field as the next target or even go back and modify already
filled fields. In other words, the set of filled fields ($F^f$) and the target
field to suggest ($f_p \in F^u$) keep changing.  This scenario is different from the
one considered by many recommender systems in the software engineering
domain~\cite{mcintosh2017fix, menzies2011local},
in which models are trained on predefined features/attributes (e.g.,
code metrics) to predict a specific target (e.g., source code
defects).  Some automated form filling
approaches~\cite{hermens1994machine} require a fixed
form filling order before building the recommendation models. However,
this assumption is unrealistic from a practical standpoint.
Although some approaches~\cite{martinez2019using,troiano2017modeling} are insensitive to the form filling order (e.g., suggesting possible values of a target field based on the frequency of values in historical inputs), they may not provide accurate suggestions due to their limitations in accurately mining dependencies among fields (as discussed in section~\ref{sec:rq1}).
Hence, one
of the challenges in automated form filling is how to build
recommendation models (e.g., by mining dependencies among fields)
without making any assumption on the order in which fields are filled.

Second, at a certain time during the data entry session, a form is
usually partially filled: this means that a recommender system for
automated form filling can only use the knowledge in currently filled
fields ($F^f$).  However, when the filled fields do not provide enough
knowledge to predict the target---based on our preliminary
experiments---existing approaches~\cite{martinez2019using,
  hermens1994machine} yield inaccurate suggestions.
The challenge, when dealing with partially filled forms, is how to
discard low-confidence suggestions, in order to make suggestions only when a high degree of confidence is achieved.

 \section{Preliminaries}
\label{sec:preliminaries}

Before illustrating our approach, we first briefly introduce two basic
machine learning algorithms we rely on.

\subsection{Bayesian Networks}
\label{sec:bayesian-networks}

Bayesian networks (BNs) are probabilistic graphical models (PGM) in
which a set of random variables and their conditional dependencies are
encoded as  a directed acyclic graph: nodes corresponds to random
variables and edges correspond to conditional probabilities.

The use of BNs for supervised learning~\cite{friedman1997bayesian}
typically consists of two phases: structure learning and variable inference.

During \emph{structure learning}, the graphical structure of the BN is
automatically learned from a training set.  First, the conditional
probability between any two random variables is computed.  
Based on
these probabilities, optimization-based search (e.g., hill
climbing~\cite{gamez2011learning}) is applied to search the graphical structure.
The search algorithm initializes a random structure,
and then iteratively adds/deletes its nodes and edges to generate new structures.
For each new structure, the search algorithm calculates a fitness
function (e.g., Bayesian information criterion,
BIC~\cite{raftery1995bayesian}) based on the nodes' conditional
probabilities and on Bayes' theorem~\cite{friedman1997bayesian}.
Structure learning stops when it finds a graphical structure that minimizes the fitness function.

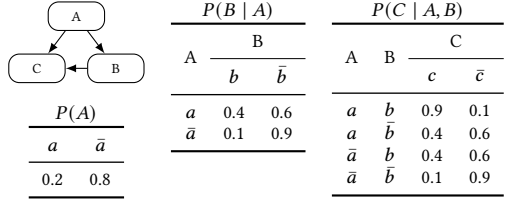
\begin{figure}[tb]
	
\tikzstyle{round_rect} = [rectangle, rounded corners, minimum width=1cm, minimum height=0.5 cm,text centered, draw=black, fill=white!30]
\tikzstyle{rect} = [rectangle, minimum width=1.5cm, minimum height=0.5cm, text centered, text width=1.5cm, draw=black, fill=white!30]
\tikzstyle{text_node} = [rectangle, minimum width=0.5cm, minimum height=0.35cm, text centered, text width=4cm, draw=white, fill= white!30]

\begin{minipage}[t]{\linewidth}
  \centering
  \scriptsize
\begin{tabular}{ccc}
\begin{tikzpicture}[baseline,scale=0.7, every node/.style={scale=0.75},node distance=3mm, >=latex]

\node (A) [round_rect] {A};
\node (B) [round_rect, below= of A,xshift= 0.7cm] {B};
\node (C) [round_rect, below= of A,xshift=-0.7cm] {C};

\draw [arrow,->] (A) -- (C);
\draw [arrow,->] (A) -- (B);
\draw [arrow,->] (B) -- (C);

\end{tikzpicture}

&\multirow{3}{*}{\begin{tabular}[b]{ c c c}
          \multicolumn{3}{c}{$P(B \mid A)$}\\
          \toprule
        \multirow{2.5}{*}{A} & \multicolumn{2}{c}{B} \\ 
	\cmidrule{2-3}
	&$b$&$\Bar{b}$ \\
	\midrule
	$a$  & 0.4 & 0.6 \\
	$\Bar{a}$ & 0.1 & 0.9 \\
	\bottomrule
	\end{tabular}}
    &\multirow{3}{*}{\begin{tabular}[b]{c  c c  c}
          \multicolumn{4}{c}{$P(C \mid A, B)$} \\
          \toprule
	\multirow{2.5}{*}{A} & \multirow{2.5}{*}{B} & \multicolumn{2}{c}{C} \\             
	\cmidrule{3-4}
	&&$c$&$\Bar{c}$\\
	\midrule
	$a $& $b $& 0.9 & 0.1  \\
	$a $& $\Bar{b}$ & 0.4 & 0.6 \\
	$\bar{a}$&$b$& 0.4 & 0.6 \\
	$\Bar{a}$ &$\Bar{b}$ & 0.1 & 0.9 \\
	\bottomrule
	\end{tabular}}\\ &&\\ 
   \begin{tabular}[b]{ c c}
    \multicolumn{2}{c}{$P(A)$}\\
	\toprule
	$a$&$\Bar{a}$ \\
	\midrule
	0.2 & 0.8 \\
	\bottomrule
   \end{tabular}
&&\\    
\end{tabular}
\end{minipage}

 	\caption{An Example of BN and the Probability
          Functions of its Nodes}
	\label{fig:bayesian_network}
\end{figure}

\figurename~\ref{fig:bayesian_network} shows an
example of the BN structure with three random variables: variable $B$
depends on variable $A$; variable $C$ depends on variables $A$ and
$B$.  In the PGM, each node is associated with a probability function
(in this case encoded as a table), which represents the conditional
probability between the node and its parent(s).  For example, in
\figurename~\ref{fig:bayesian_network} each variable has two values;
the probability table for $B$ reflects the conditional probability
$P(B\mid A)$ between $A$ and $B$ on these values.

\emph{Variable inference} infers unobserved variables from the
observed variables and the graphical structure of the BN using Bayes' theorem~\cite{friedman1997bayesian}.
For example, we can infer the probability of $C=c$ when the value of $A$ is $a$ (denoted as $P(c \mid a)$) as follows:
\begin{equation*}
\begin{aligned}
P(c \mid a) &= \frac{P(a, c)}{P(a)} = \frac{P(a, b, c) + P(a, \overline{b}, c)}{P(a)} \\
&= \frac{P(c \mid a, b) P(b \mid a)  P(a) + P(c \mid a, \overline{b}) P(\overline{b} \mid a)  P(a)}{P(a)} \\
&= \frac{0.9*0.4*0.2 + 0.4*0.6*0.2}{0.2} =0.6
\end{aligned}
\end{equation*}

BNs have been initially proposed for learning dependencies among discrete
random variables.  They are also robust when dealing with missing
observed variables; more specifically, variable inference can be
conducted when some conditionally independent observed variables are
missing~\cite{friedman1997bayesian}.

\subsection{K-modes }
\emph{K}-modes is a clustering algorithm that extends the
\emph{k}-means one to enable clustering of categorical
data~\cite{huang1998extensions}.  The algorithm first randomly selects
\emph{k} instances in the data set as the initial centroids. Each
instance is represented with a vector of categorical attributes.  The
algorithm clusters the instances in the dataset by calculating the
distances between the instances and each centroid.  The distance, also
called dissimilarity measure, is defined as the total mismatches of
the corresponding categorical attributes of an instance and of a centroid.
Based on the clustering results, new centroids are selected, which
represent the modes of categorical attributes in each cluster.  The
algorithm then re-clusters the instances according to the new
centroids.  This process is repeated until the centroids remain
unchanged or until it reaches a certain number of iterations.

 \section{Approach}
\label{sec:approach}

\begin{figure}[tb]
  \centering
  
\tikzstyle{rounded_rect} = [rectangle, rounded corners, minimum width=2.5cm, minimum height=1cm,text centered, draw=black, fill=white!30]

\tikzstyle{rect} = [rectangle, minimum width=2.5cm, minimum height=1cm,  text width=2.5cm, draw=black,align=center, fill=white!30]

\tikzstyle{text_node} = [rectangle, minimum width=0.5cm, minimum height=0.5mm, text centered, text width=1cm, draw=white, fill= white!30]
  
\tikzstyle{list_box} = [rect,align=left ,minimum width=3.2cm,text width=3.2cm, minimum height= 0.5cm]

\resizebox{0.66\textwidth}{!}{
\begin{tikzpicture}[node distance=7mm, >=latex]
  \node (nd1) [rect,align=center] {Historical input instances};
  \node (nd2) [rounded_rect, below= of nd1, yshift=-0.5cm] {Pre-processing};
  \node (nd3) [rounded_rect, below= of nd2] {Model building};
  \node (big_box1)[draw=black, fit={(nd2)(nd3)},minimum width=2.9cm,minimum height=3.2cm,line width=0.4mm]{};
  
  \node (nd4) [rounded_rect, right= of nd2,xshift= 1.2cm] {Pre-processing};
  \node (nd5) [rounded_rect, below= of nd4] {Model selection};
  \node (nd6) [rounded_rect, right= of nd5, xshift=1.45cm] {Endorsing}; 
  \node (big_box2) [draw=black, line width=0.4mm,fit={(nd4)(nd5)(nd6)},minimum width=7.7cm, minimum height=3.2cm]{};
  
  \node (user)[inner sep=0pt, above= of nd4,xshift=-0.9cm,yshift=-0.3cm]
  {\includegraphics[width=.03\textwidth]{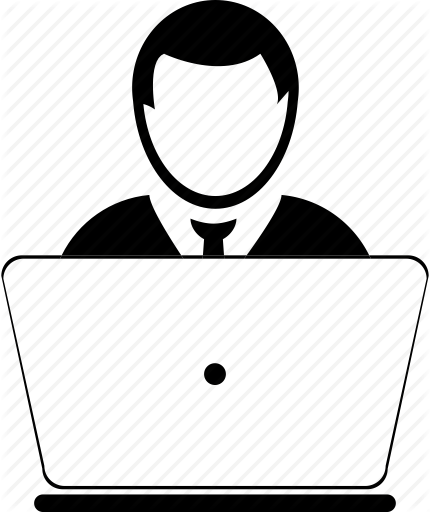}};
  \node (current_input) [rect] at (nd4 |- nd1) {Current input};
  
  \node (list_box_p1)[list_box] at (nd6 |- current_input) {Leasing Serv.   \hfill 0.70\\ Financial  Serv. \hfill \ \ \ 0.15\\ Accommoda-  \hfill \ \ \ 0.05 \\ tion Serv.};
  \node(list_box_p2)[list_box,below=of list_box_p1, yshift=0.8cm]{Other values};

  \node(MB_phase) [text_node, below=of big_box1,text width=2.8cm,yshift=0.5cm]{\textbf{Model Building Phase}};
  \node(FFS_phase) [text_node, text width=3cm] at (big_box2|-MB_phase){\textbf{Form Filling Suggestion Phase}};
  
  \draw[arrow,->] (nd1)--(nd2);
  \draw[arrow,->] (nd2)--(nd3);
  \draw[arrow,->] (nd3)--node [anchor=south] {BNs}(nd5);
  \draw[arrow,->] (nd4)--(nd5);
\draw[arrow,->] (nd5) --node [anchor=south,text width = 2cm,align=center] {Probabilistic distribution}(nd6);
\draw [arrow,->] (current_input) -- (nd4);
\draw [arrow,->] (nd6)-- node [anchor=east,text width = 3cm,xshift=1.2cm]{Suggestions}(list_box_p2);

\end{tikzpicture}
}

   \caption{Main Steps of the \approach Approach}
  \label{fig:laf-framework}
\end{figure}
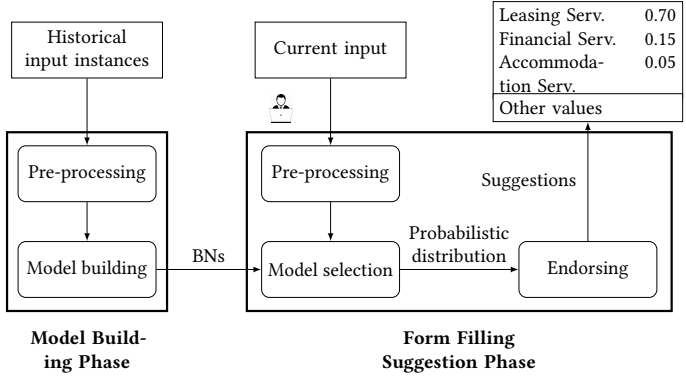

In this section, we present our machine-learning based approach for form filling, named
\approach (\underline{L}earning-based \underline{A}utomated \underline{F}orm
\underline{F}illing).

\approach includes two phases: \emph{model building} and \emph{form
  filling suggestion}, whose main steps are shown in
\figurename~\ref{fig:laf-framework}. In the former, \approach analyzes
historical input instances of a data entry form and uses dependency
analysis to build BNs that represent the fields in
the form and their conditional dependencies. This phase
occurs offline, before deploying \approach as an assistive technology
for data entry. The \emph{form filling suggestion} phase occurs during
the data entry session: given a target field, \approach selects a
BN among those built in the \emph{model building}
phase and predicts possible values based on the values in the
already-filled fields and their conditional dependencies; the
predicted values and the corresponding predicted probability (endorsed  based on field dependencies and prediction confidence) are
then provided to the end-user as suggestions.

\subsection{Pre-processing}
\label{sec:pre-processing}
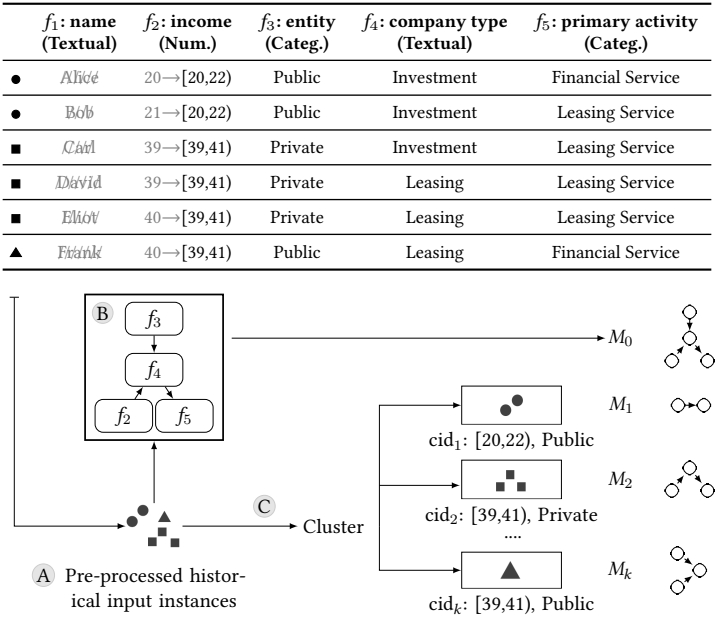
\begin{figure}[tb]
  \centering

\newcommand{\tikzmark}[2]{
	\tikz[overlay,remember picture,baseline] 
	\node [anchor=base] (#1) {$#2$};}

\newcommand{\DrawVLine}[3][]{\begin{tikzpicture}[overlay,remember picture]
	\draw[shorten <=0.3ex, #1] (#2.north) -- (#3.south);
	\end{tikzpicture}
}

\newcommand{\DrawCircle}[1][]{\begin{tikzpicture}[overlay,remember picture]
	\draw[black,fill=black] (0,0.5ex) circle (0.5ex);
	\end{tikzpicture}
}

\newcommand{\DrawRec}[1][]{\begin{tikzpicture}[overlay,remember picture]
	\draw[black,fill=black] (-0.5ex,0) rectangle ++(1ex, 1ex);
	\end{tikzpicture}
}

\newcommand{\DrawTriangle}[1][]{\begin{tikzpicture}[overlay,remember picture]
	\filldraw[fill=black] (90:1.2ex) -- (170:0.7ex) -- (10:0.7ex) --cycle;
	\end{tikzpicture}
}

\tikzstyle{round_rect} = [rectangle, rounded corners, minimum width=1.5cm, minimum height=0.5 cm,text centered, draw=black, fill=white!30]

\tikzstyle{rect} = [rectangle, minimum width=1.5cm, minimum height=0.8cm, text centered, text width=1.5cm, draw=black, fill=white!30]

\tikzstyle{text_node} = [rectangle, minimum width=0.5cm, minimum height=0.35cm, text centered, text width=4cm, draw=white, fill= white!30]

\begin{tikzpicture}

\node (tab1) [shape=rectangle,draw=white] {
	\scriptsize
	\begin{tabular}{cccccc}
	\toprule
	
	&\textbf{$f_1$: name}&\textbf{$f_2$: income}&\textbf{$f_3$: entity}&\textbf{$f_4$: company type}&\textbf{$f_5$: primary activity}\\
	&\textbf{(Textual)}&\textbf{(Num.)}&\textbf{(Categ.)}&\textbf{(Textual)}&\textbf{(Categ.)}\\
	\midrule
	\DrawCircle
	&\tikzmark{topC1}{\color{gray}\text{\xout{Alice}}} &{\color{gray}20$\rightarrow$}[20,22) &Public &Investment&Financial Service\\
	\midrule
	\DrawCircle&\color{gray}{\xout{Bob}}&{\color{gray}21$ \rightarrow$}[20,22) & Public &Investment&Leasing Service \\ 
	\midrule
	\DrawRec&\color{gray}{\xout{Carl}}&{\color{gray}39$\rightarrow$}[39,41)& Private &Investment&Leasing Service  \\ 
	\midrule
	\DrawRec&\color{gray}{\xout{David}}&{\color{gray}39$\rightarrow$}[39,41)&Private &Leasing &Leasing Service  \\ 
	\midrule
	\DrawRec&\color{gray}{\xout{Eliot}}&{\color{gray}40$\rightarrow$}[39,41)&Private &Leasing&Leasing Service  \\ 
	\midrule
	\DrawTriangle&\tikzmark{bottomC1}{\color{gray}\text{\xout{Frank}}}&{\color{gray}40$\rightarrow$}[39,41)&Public &Leasing &Financial Service\\ 
	\bottomrule
	\end{tabular}
};
\end{tikzpicture}
	
\resizebox{0.572\textwidth}{!}{	
\hspace*{-3em}{
	\begin{tikzpicture} [node distance=5mm, >=latex]
	\begin{scope}[transform canvas={xshift=0em},node distance=3mm]
	\node (f3) [round_rect, minimum width=1cm] {$f_3$};
	\node (f4) [round_rect,minimum width=1cm, below= of f3] {$f_4$};
	\node (f2) [round_rect, minimum width=1cm,below left= of f4,xshift=0.7cm] {$f_2$};
	\node (f5) [round_rect, minimum width=1cm,below right= of f4,xshift=-0.7cm] {$f_5$};
	\draw [arrow,->] (f3) -- (f4);
	\draw [arrow,->] (f2) -- (f4);
	\draw [arrow,->] (f4) -- (f5);
	\end{scope}
	\node (big_box)[draw=black, fit={(f3)(f4)(f2)(f5)},minimum width=2.4cm,minimum height=2.5cm,line width=0.3mm]{};
	
	\node[shape=circle, draw, fill=gray, opacity=.2, text opacity=1, inner sep=0.5pt, left=of f3, xshift=0.3cm, yshift=0.1cm] (charB) {B};
		
	\node (shapes)[inner sep=0pt, below=of f4, yshift= -1.5cm]
	{\includegraphics[width=.07\textwidth]{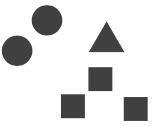}};
	\node (text3) [text_node, below= of shapes, yshift= 0.3cm] {Pre-processed historical input instances};
	\node (text4) [text_node, right= of shapes] {Cluster};
	
	\node[shape=circle, draw, fill=gray, opacity=.2, text opacity=1, inner sep=0.5pt, left=of text3, xshift=0.9cm, yshift=0.25cm] (charA) {A};
	
	\node[shape=circle, draw, fill=gray, opacity=.2, text opacity=1, inner sep=0.5pt, above=of text4, xshift=-1.2cm, yshift=-0.6cm] (charC) {C};
	
	\begin{scope}[transform canvas={xshift=0em},node distance=6mm]
	\draw [arrow,->] (shapes) -- (big_box);
	\draw [arrow,->]  (shapes) --++(2.5,0)(text4) ;
	\node (cluster2) [rect, right=of text4, xshift=-0.5cm,yshift=0.8cm,minimum height= 0.4cm]{\includegraphics[width=0.4\textwidth]{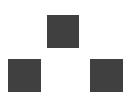}};
	\node (textc2) [text_node, below= of cluster2,text width=4cm,yshift=0.6cm] {cid$_2$: [39,41), Private};
	\node (cluster1) [rect, above=of cluster2,minimum height= 0.4cm]{\includegraphics[width=0.35\textwidth]{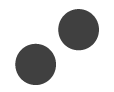}};
	\node (textc1) [text_node, below= of cluster1,text width=4cm,yshift=0.6cm] {cid$_1$: [20,22), Public};
	\node (textpoints) [text_node, below= of textc2,text width=4cm,yshift= 0.7cm] {....};
	\node (cluster3) [rect, below=of cluster2, yshift=-0.3cm,minimum height= 0.4cm]{\includegraphics[width=.3\textwidth]{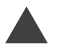}};
	\node (textc3) [text_node, below= of cluster3,text width=4cm,yshift=0.6cm] {cid$_k$: [39,41), Public};
\draw[arrow,->] (text4)++(0.8,0) |- (cluster1);
	\draw[arrow,->] (text4) ++(0.8,0.7) --++(1.4,0) (cluster2);
	\draw[arrow,->] (text4) ++(0.8,0)|- (cluster3);
	\end{scope}
		
	\node (M1) [text_node, right = of cluster1,text width=0.8cm] {{$M_1$}};
	\node (M2) [text_node, right = of cluster2,text width=0.8cm] {{$M_2$}};
	\node (Mk) [text_node, right= of cluster3,text width=0.8cm] {{$M_k$}};
	\node (M0) [text_node, text width=0.8cm,yshift=0.5cm] at (M1 |- big_box) {{$M_0$}};
	
	\draw [arrow,->] (big_box) ++(1.35,0.5)--++(6.5,0)  (M0);
	
	\node (anchor_table) [left=of f3, xshift=-1.3cm, yshift=0.5cm]{};
	\draw [arrow,|->] (anchor_table) |- (shapes);

	\begin{scope}[transform canvas={xshift=1em}, node distance=2mm]
	\node (m0f4) [round_rect, minimum width=0.2cm, minimum height=0.08cm, right= of M0] { };
	\node (m0f3) [round_rect, minimum width=0.2cm, minimum height=0.08cm, above=of m0f4] { };
	\node (m0f2) [round_rect, minimum width=0.2cm, minimum height=0.08cm, below left= of m0f4, xshift=0.7cm] { };
	\node (m0f5) [round_rect, minimum width=0.2cm, minimum height=0.08cm, below right= of m0f4, xshift=-0.7cm] { };
	\draw [arrow,->] (m0f3) -- (m0f4);
	\draw [arrow,->] (m0f5) -- (m0f4);
	\draw [arrow,->] (m0f4) -- (m0f2);
	\end{scope}

	\begin{scope}[transform canvas={xshift=1.7em}, node distance=2mm]
	\node (m1f4) [round_rect, minimum width=0.2cm, minimum height=0.08cm, right= of M1] { };
	\node (m1f3) [round_rect, minimum width=0.2cm, minimum height=0.08cm, left=of m1f4] { };
	\draw [arrow,->] (m1f3) -- (m1f4);
	\end{scope}
	
	\begin{scope}[transform canvas={xshift=1em}, node distance=2mm]
	\node (m2f4) [round_rect, minimum width=0.2cm, minimum height=0.08cm, yshift=2mm, right= of M2] { };
	\node (m2f2) [round_rect, minimum width=0.2cm, minimum height=0.08cm, below left= of m2f4, xshift=0.7cm] { };
	\node (m2f5) [round_rect, minimum width=0.2cm, minimum height=0.08cm, below right= of m2f4, xshift=-0.7cm] { };
	\draw [arrow,->] (m2f5) -- (m2f4);
	\draw [arrow,->] (m2f4) -- (m2f2);
	\end{scope}
	
	\begin{scope}[transform canvas={xshift=1.5em}, node distance=2mm]
	\node (mkf4) [round_rect, minimum width=0.2cm, minimum height=0.08cm, right= of Mk] { };
	\node (mkf3) [round_rect, minimum width=0.2cm, minimum height=0.08cm, above left=of mkf4, , yshift=-0.1cm] { };
	\node (mkf2) [round_rect, minimum width=0.2cm, minimum height=0.08cm, below left= of mkf4, yshift=0.1cm] { };
	\draw [arrow,->] (mkf3) -- (mkf4);
	\draw [arrow,->] (mkf2) -- (mkf4);
	\end{scope}
	
	\end{tikzpicture}
}} 	\caption{Example of Model Building on Pre-processed Historical Input Instances}
	\label{fig:dependency}
\end{figure}

Both phases of \approach include a pre-processing step to improve
the quality of the form filling data; this step is based on best
practices for predictive data
mining~\cite{Alexandropoulos2019DataPI}.

Typically, historical input instances have many \emph{missing values}
due to the presence of optional fields in input forms. Fields that
have a high number of missing values do not provide representative
information for model building; hence, they can be removed. We remove
fields for which $T_p^v\%$ or more of the values are missing, where
$T_p^v$ is a user-configurable threshold (with default value
equal to $90$).

We also remove file and textual fields that have a high number of
unique values, since they typically correspond to form fields for
which users frequently provide new string values (e.g., the textual
field ``name''). To identify such fields, we compute the ratio of
unique values of a field in the historical input instances; if the
ratio is larger than a user-configurable threshold $T_p^u$ (with
default value equal to $0.9$), the corresponding field is removed.

Furthermore, we delete a historical input instance if more than
$T_p^m\%$ of its field values are missing, where $T_p^m$ is a
user-configurable threshold (with default value equal to $50$).  After
deletion, we perform data imputation~\cite{jing2016missing} on the
remaining data that exhibit missing values. Numerical fields are
imputed using the mean value of this field; categorical and textual
fields are imputed using a default label ``UNKNOWN''.

We also apply data discretization to numerical fields to reduce the
number of unique values. Numerical values are transformed into
discrete intervals based on information gain analysis, a widely used
discretization method first proposed in decision
trees~\cite{breiman1984classification}.

During the data entry session, we ignore values in the fields that were
removed in historical input instances, and map numbers onto intervals.

\subsubsection*{Application to the running example}
The table at the top of \figurename~\ref{fig:dependency} shows an example of historical
input instances filled through the data entry form in \figurename~\ref{fig:def}.
Each row is a historical input instance filled by a user.
During pre-processing, \approach removes the field
``name'' since all its values are unique (we crossed out the text of
these values with a hatch pattern to represent the removal). 
Also, the values of field ``income'' are discretized into intervals.
During the data entry session, as shown in \figurename~\ref{fig:def},
a user fills the fields ``name'' with \textit{Gibson}, ``income'' with
\textit{20}, ``legal entity'' with \textit{Private}, and ``company type'' with ``Leasing''; 
``primary activity'' is the next field to be filled. 
Through the application of the pre-processing steps, 
\approach ignores the value for field ``name'' and maps the value
\textit{20} of field ``income'' to the interval \emph{[20, 22)}.

\subsection{Model building}
\label{sec:model-building}

\begin{algorithm}[tb]
	\footnotesize
	\caption{Model Building}
	\label{alg:LAF-training}
	\KwIn{Pre-processed historical input instances $I^F(t)^\prime$}
	\KwOut{List of probabilistic graphical models $\mathcal{M}$ \newline
		Historical input instance clusters $C$}
	$\mathcal{M}\gets$empty list\;
	$M_0\gets \mathit{trainBayesianNetwork}(I^F(t)^\prime)$\;\label{line:train-m0}
	$\mathcal{M}.\mathit{append}(M_0)$\;
	
	independent field set $F^I\gets \emptyset$\; \label{line:train-s-fields}
	\ForEach{field $f_i \in M_0$}
	{
		\If{$\mathit{getParents}(M_0, f_i) = \emptyset$}{
			$F^I\gets \{f_i\} \cup F^I$\;
		}
	}	\label{line:train-e-fields}

	number of cluster $k \gets \mathit{elbowMethod}(I^F(t)^\prime, F^I)$\;\label{line:train-s-cluster}
	$C = \{I^F(t)^\prime_1, \dots, I^F(t)^\prime_k\} \gets  \mathit{kModes}(I^F(t)^\prime, F^I, k)$\;\label{line:train-e-cluster}
	
	\For{$i\gets1$ \KwTo $k$} {\label{line:train-mk:begin}
		$M_i\gets \mathit{trainBayesianNetwork}(I^F(t)^\prime_i)$\;
		$\mathcal{M}.\mathit{append}(M_i)$;
	}\label{line:train-mk:end}
	\KwRet{$\mathcal{M}$, $C$}\;
\end{algorithm}

The goal of the \emph{model building} phase is to mine dependencies from historical input instances of a data entry form.

Due to the arbitrary order for filling the form,
the filled fields and the target field keep changing.
When we take the filled fields as features to predict the target field,
the arbitrary form filling order results into a large set of feature-target combinations.
For example, let us consider a data entry form with $n$ fields, 
with $t \leq n$ of them being categorical and thus representing the possible targets.
When we take one of the categorical fields as the target, 
assuming that a random order is used for form filling,
in principle users may fill any subset of the remaining $n-1$ fields, 
resulting in a total of $2^{n-1}-1$ possible combinations of filled fields (i.e., features). The total number of feature-target combinations is equal to $t*(2^{n-1}-1)$.
Normally, a model would need to be trained on each target-features combination 
to ensure the assumption of identical features and
target~\cite{dekel2010learning} in the model building and form filling
suggestion phases.
As we will show through our evaluation in
section~\ref{sec:evaluation},
adopting such an approach would require to train more than
\num{220000} models on one of our datasets. The total time required to
train this large number of models would be impractical for a production-grade system.

To solve this problem, we capture dependencies with BNs,
in which variables correspond to form fields. 
By using BNs, 
we can analyze the dependency between filled fields and target
fields without training models on specific combinations of features (i.e., filled fields) and target field.
In addition, as mentioned in section~\ref{sec:bayesian-networks}, BNs are robust when dealing with missing values.
This means that even when a data entry form is partially filled,
BNs can still infer the probability distribution of target fields
using only the information in the filled fields and the underlying PGM.

In this work, we learn the structure of BN from the pre-processed historical input instances. 
Following the workflow of BN presented in section~\ref{sec:bayesian-networks},
we represent each field in the historical input instances as a random variable.
BN computes the conditional probability between any two fields 
and uses a search-based optimizer to automatically optimize 
the structure of BN based on the conditional probability of fields and the fitness function.
In this study, we use hill climbing as the optimizer,
because it shows a good trade-off between computational demands and the quality of the models learned~\cite{gamez2011learning}.
We define the fitness function in terms of BIC~\cite{raftery1995bayesian}, 
which aims at best fitting the data, while avoiding over-fitting by complex structures.
  The element denoted with \myblock{B} in \figurename~\ref{fig:dependency} shows an
example of the BN structure learned from the data in block
\myblock{A} (where the different black shapes correspond to
the various rows in the table at the top of  \figurename~\ref{fig:dependency}).

Algorithm~\ref{alg:LAF-training} illustrates the main steps of this
phase. \approach takes as input the pre-processed historical input
instances $I^F(t)^\prime$ as the training data to mine field
dependencies.
Initially,
we learn the BN over the entire training data
(line~\ref{line:train-m0}).  
This global model, denoted as $M_0$, represents
the general dependencies among fields.
However, historical input instances may form different groups that
share similar characteristics.
For example, in the historical input instances of
\figurename~\ref{fig:dependency}, users having the same value for
fields ``income'' and ``legal entity'' may share specific values for
fields ``company type'' and ``primary activity''.
The global model $M_0$ may
not learn the fine-grained field dependencies for specific values of
``income'' and ``legal entity''
due to the influence of input instances with other values for those
fields.
For example, using the entire dataset in
\figurename~\ref{fig:dependency}, one could determine that the conditional
probability of having ``primary activity'' equal to \emph{Leasing Service} when ``Company type'' is \emph{Leasing} is $66.7\%$. However, 
this conditional probability increases to $100\%$,
if we only
consider the input instances where ``income'' is in the range \emph{[39,
  41)} and ``legal entity'' is equal to \emph{Private}.
Hence, \approach trains local models
on subsets of $I^F(t)^\prime$ to learn fine-grained
dependencies. 

More specifically, \approach first selects 
a set of fields $F^I$ that are independent from other fields in the
probabilistic graph of $M_0$
(lines~\ref{line:train-s-fields}--\ref{line:train-e-fields}).  
For
example, in block \myblock{B} of \figurename~\ref{fig:dependency}, fields $f_2$ and $f_3$
are selected as they do not depend on other parent nodes (fields).
We use the fields in $F^I$ as the main fields to form
partitions of $I^F(t)^\prime$ having similar characteristics for two reasons.
First, these fields are not intercorrelated since they do not directly and strongly depend on each other.
Second,  these fields are root nodes and influence the values of other fields in the BN;
when the values on these fields are similar, we are likely to obtain a similar
probability distribution for the other fields.

\approach produces
(lines~\ref{line:train-s-cluster}--\ref{line:train-e-cluster}) a set
$C$ of clusters of $I^F(t)^\prime$ based on the fields in $F^I$.
We assume that the clustering process reduces the data variation of $I^F(t)^\prime$:
models trained on these data, which show less statistical variation, may provide more accurate suggestions 
even when the size of each cluster is smaller than that of $I^F(t)^\prime$.
This process is called local modeling; it has
been already applied in software engineering, e.g., to cluster software projects and mine project-specific relationships of software metrics~\cite{menzies2011local}.

To extract clusters from $I^F(t)^\prime$,
\approach represents each historical input instance as a tuple of the form $\langle \text{values in } F^I, \text{input instance} \rangle$.
It clusters these tuples based on the values in $F^I$
using the \emph{k}-modes algorithm.
The optimal number of clusters  $k$ is automatically determined with the elbow method. 
\approach runs \emph{k}-modes
within a range of $k$  values (e.g., from 1 to 100) and determines the value of $k$ that minimizes the average within-cluster distance with the
cluster centroids (denoted with ``cid'' in block \myblock{C} of
\figurename~\ref{fig:dependency}).  
After clustering, \approach
trains a local BN model $M_i$
(lines~\ref{line:train-mk:begin}--\ref{line:train-mk:end}) based on the input instances in each cluster. These
\emph{local} models, denoted $M_1, \dots, M_k$, capture specialized field dependencies on
partitions of $I^F(t)^\prime$.

The algorithm ends by producing a list $\mathcal{M}$ of BNs, where $\mathcal{M}=[M_0,
M_1, \ldots M_k]$, and the set $C$ of clusters of the historical
input instances.

\subsubsection*{Application to the running example}
Initially, \approach trains a global model $M_0$ with the 
historical input instances in block \myblock{A} in  \figurename~\ref{fig:dependency}. 
Block \myblock{B} of \figurename~\ref{fig:dependency} shows an example of the learned field dependencies. 
Based on $M_0$, \approach selects fields $f_2$:``income'' and
$f_3$:``legal entity'' as the main fields for local modeling
since they do not depend on other parent nodes (fields).
\approach clusters the historical input instances according to fields
``income'' and ``legal entity'' (block \myblock{C} of \figurename~\ref{fig:dependency}).
Three clusters are automatically identified with centroids ``\textit{[20, 22), Public'}', ``\textit{[39, 41), Private}'', and ``\textit{[39, 41), Public}'' (\textit{k}=3).
We use circular, rectangular, and triangular icons to represent the historical input instances belonging to different clusters.
\approach trains three local models $M_1$, $M_2$, and $M_3$, based on
these clusters; these three models are three distinct BNs capturing specialized field dependencies (as shown on the right of \figurename~\ref{fig:dependency}).
After the model building phase, \approach outputs four models: a global
BN model $M_0$ and the three local BN models $M_1$, $M_2$, and $M_3$.

\subsection{Form Filling Suggestion}
\label{sec:filling-suggestion}

\begin{algorithm}[tbp]
	\footnotesize
	\caption{Form Filling Suggestion}
	\label{alg:LAF-prediction}
	\KwIn{Models $\mathcal{M}=[M_0, M_1, \dots, M_k]$ \newline
		Clusters $C=\{I^F(t)'_1, I^F(t)'_2, \dots, I^F(t)'_k\}$ \newline
		Filled fields $F^{f}=\{\langle f^f_1, v^f_1 \rangle, \dots, \langle f^f_m, v^f_m\rangle\}$ \newline
                Target field $f_p$ \newline
                Number of suggested values $n_r$ \newline
		Threshold $\theta$
	}
	\KwOut{List of predicted values $V_p$ for $f_p$}
	
	$F^{f'} \gets \mathit{getPreprocessedData}(F^{f})$ \;
	$D=\{d_1, \dots, d_k\} \gets \mathit{calcClusterDistance}(C, F^{f'})$\;\label{line:pre-distance}
	
	Model $M_{cur} \gets \mathcal{M}.\mathit{get}(M_0)$\;\label{line:pre-s-selection}
	\If{$\mathit{getNumOfMinDistance}(D) = 1$}{
		$i \gets \mathit{getMinDistanceID}(D)$\;
		$M_{\mathit{cur}} \gets \mathcal{M}.\mathit{get}(M_i)$\;
	}\label{line:pre-e-selection}

	List of Pairs $\langle v_{p}, \mathit{pr}\rangle$ of candidate values and probability distribution 
	${\mathit{Candidates}} =
        \mathit{predictCandidates}(M_{\mathit{cur}}, F^{f'}, f_p)$\;\label{line:pre-s-topn}
	${\mathit{Candidates}}^{R} \gets
        \mathit{getTopRanked}(\mathit{Candidates},
        n_r)$\;\label{line:pre-e-topn}
        Bool $\mathit{checkDep} \gets \mathit{isMember}(
        \mathit{getParents}(M_{cur}, f_p),F^f)$\;\label{line:pre-flag-dd}
        Bool $\mathit{checkProb} \gets (\mathit{getSumProb}(\mathit{Candidates}^R) >  \theta)$\;       \label{line:pre-flag-prob}
	\If{$ \mathit{checkProb} \lor  \mathit{checkDep} $} {\label{line:pre-b-filter}
          \ForEach{$v_{p_i} \text{ s.\ t. }  \langle v_{p_i}, {\mathit{pr}}_i\rangle\ \in {\mathit{Candidates}}^{R}$}
         { $V_p.\mathit{append}(v_{p_i})$;}\label{line:pre-e-filter}}
	\KwRet{$V_p$}\;
\end{algorithm}

The \emph{form filling suggestion} phase occurs during the data
entry session and assumes that the models in $\mathcal{M}$, built in the
\emph{model building} phase, are available. Given a target field
$f_p$, \approach selects a BN model $M \in \mathcal{M}$ and predicts possible
values of $f_p$ based on the already-filled fields $F^f$ and
their conditional dependencies captured in $M$.	 The main steps
of the \emph{form filling suggestion} phase are shown in
Algorithm~\ref{alg:LAF-prediction}.

The algorithm takes as input a list of models $\mathcal{M}$, a set of
clusters $C$, a set $F^f$ of already-filled fields with their
values, a target field $f_p$, and some auxiliary parameters
representing the number of expected suggestions for $f_p$ and an
endorsing threshold.  After pre-processing the filled fields in $F^f$ using the
techniques discussed in \S~\ref{sec:pre-processing} and obtaining the
new set $F^{f^\prime}$, \approach computes the distance between the
filled fields $F^{f^\prime}$ and each cluster in $C$
(line~\ref{line:pre-distance}).  The distance is defined as the
dissimilarity measure adopted in the $k$-modes clustering algorithm
used in the model building phase; it is the total
number of mismatches between $F^{f^\prime}$ and the centroid of each
cluster on the corresponding fields.

\approach attempts to select a local model from
$M_1, \ldots M_k$, corresponding to the cluster with minimal distance, to predict the target
field, since this model may capture the fine-grained characteristics
of $F^{f^\prime}$. In our example, this could be a model trained on
the instances that have the same values for the fields ``legal entity'' and
``income'' in $F^{f^\prime}$.  However, a unique and optimal local model cannot
always be found: given a partially filled data entry form, there could
be cases for which the distance of the filled fields to different
centroids is equal.  For example, in
\figurename~\ref{fig:dependency}, local models $M_2$ and $M_k$ are
specialized for different values of the field ``legal entity'' (i.e.,
``Private'' and ``Public'') but assume the same value for the
field ``income'' ($[39, 41)$). Let us consider the case in which the set $F^f$
contains only the field ``income'' (with a value equal to ``40'') and the
target field is ``primary activity''.  In this case, we cannot reliably select
between the two models $M_2$ and $M_k$ for prediction, since we have
insufficient information to decide which model is ``more local''
(i.e., specialized) for this
input (i.e., the distance to the two centroids is the same).
One possible solution for this problem is ensemble learning, which considers the predictions of both models jointly 
(e.g., bagging)~\cite{zhou2021ensemble}.
However, this solution could significantly increase prediction
time. Specifically, depending on the deployment configuration, in the worst case, 
the ensemble prediction time
would be the sum of the prediction time of all $k$ local models 
(i.e., all the local models are not specialized for the current input), 
which may exceed the acceptable response time for a practical
application, as presented in section~\ref{sec:rq:performance} 
and further discussed in section~\ref{sec:implications:sw-dev}. 
Given the interactive nature of data-entry applications, 
having a short prediction time is important.
Hence, when \approach finds more than one minimal distance and no single cluster is particularly suited for the current input,
it selects $M_0$ for prediction, since it is trained with the entire
set of historical input instances (lines~\ref{line:pre-s-selection}--\ref{line:pre-e-selection}).

After selecting the most appropriate model for prediction, \approach
predicts the candidate values for the target field
(line~\ref{line:pre-s-topn}) and ranks the topmost $n_r$ values,
according to their probability distribution
(line~\ref{line:pre-e-topn}).

\subsubsection*{Endorsing}

During the data entry session, the filled fields in the current input
instance do not always provide enough information to predict values for
the target field, leading to inaccurate suggestions.  
Such a situation can occur because of two reasons.  One reason is that the filled fields may not have explicit
dependencies with the target field, according to the
probabilistic graph.  For example, in
\figurename~\ref{fig:dependency}, $f_2$ is independent from $f_3$; \approach will not accurately infer $f_3$ merely
with the knowledge of $f_2$.  Another reason is that there may not be enough
historical input instances to learn the conditional probability
between two fields for specific values.  For example, in the example in
\figurename~\ref{fig:dependency}, we have no historical input
instance with values of field ``income'' greater than 41; such value
provides limited information to infer other fields.

In the context of automated form filling, users might be reluctant to
use an automated form filling tool, if the tool provides many inaccurate
suggestions which users can hardly find the correct value they intend to fill in. To avoid such a situation, \approach includes a
\emph{heuristic-based endorser},
which decides whether the suggestions determined in the previous
step are accurate enough to be returned to the user.

To deal with the first cause of inaccurate suggestions, \approach
analyzes the dependency between the filled fields and the target field
(line~\ref{line:pre-flag-dd}), to check whether the target field
directly depends on one of the filled fields in the BN.  More
precisely, function \textit{getParent} computes a list of parent
fields the target field directly depends on; function
\textit{isMember} checks whether any of the filled fields is in the
parent field list.  The result of this check is saved in the Boolean
flag $\mathit{checkDep}$, which is true when the target field directly
depends on one of the filled fields.  A direct dependency indicates
that the filled fields can reliably determine the value of the target
field.

To deal with the second cause of inaccurate suggestions, \approach analyzes the
predicted probability distribution of the values for the target field.
For a probability model like BN,
the probability of each value is inferred based on the information from the filled fields.
\approach computes the sum of the top-$n_r$ probability values in the
distribution through function \textit{getSumProb}.
If this value is larger than a user-defined threshold $\theta$,
it means \approach may have enough information for variable inference;
the result of this check is saved in the Boolean flag  $\mathit{checkProb}$
(line~\ref{line:pre-flag-prob}). From a practical standpoint,
threshold $\theta$ reflects how much uncertainty users are willing to accept regarding the suggestions provided by \approach.

If one of the flags $\mathit{checkProb}$ and $\mathit{checkDep}$ evaluates
to \textit{true}, \approach populates the list of suggested values to be
returned to the user based on the top-ranked candidate values;
otherwise, \approach returns an empty list
(lines~\ref{line:pre-b-filter}--\ref{line:pre-e-filter}).  For
example, assuming $\theta=0.70$, $n_r=3$, and the probability for the
top-3 candidate values as shown in the top right corner of \figurename~\ref{fig:laf-framework}, the sum of the top-3 probability values (returned by $\mathit{getSumProb}$) is $0.70+0.15+0.05=0.90$; $\mathit{checkProb}$
corresponds to the evaluation of $0.90 > 0.70$, which is \textit{true};
hence, \approach decides to yield the list of suggestions to the user.

\begin{figure}[tb]
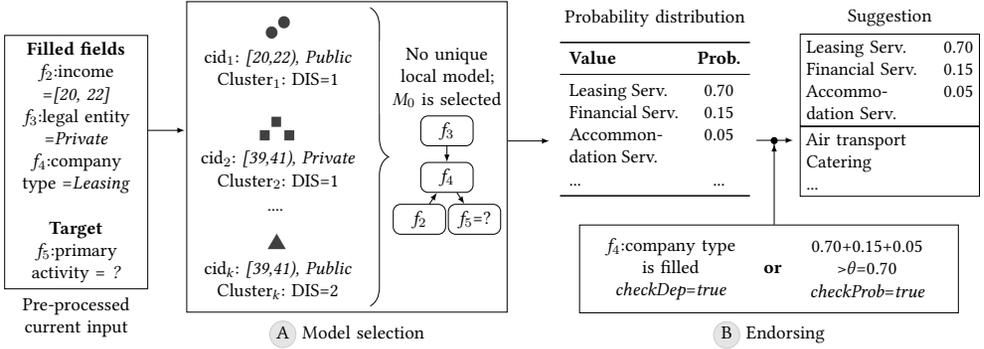

	\centering
	
\tikzstyle{round_rect} = [rectangle, rounded corners, minimum width=1.5cm, minimum height=0.5 cm,text centered, draw=black, fill=white!30]
\tikzstyle{rect} = [rectangle, minimum width=1.5cm,  text centered, text width=1.5cm, draw=black, fill=white!30]
\tikzstyle{text_node} = [rectangle, minimum width=0.5cm, text centered, text width=4cm, draw=white, fill= white!30]
\tikzstyle{text_model} = [rectangle, minimum width=0.8cm, text centered, text width=0.8cm, draw=white, fill= white!30]

\tikzstyle{list_box} = [rect,align=left ,minimum width=2.39cm,text width=3.19cm, minimum height= 0.5cm]

\resizebox{1\textwidth}{!}{
	\begin{tikzpicture} [node distance=0.3mm, >=latex]
	
	\node (input_val) [rect, text
        width=2.5cm,xshift=0.5cm]{\textbf{Filled fields}\\
          $f_2$:income =\emph{[20, 22]}\\$f_3$:legal entity =\emph{Private}\\ $f_4$:company type =\emph{Leasing}\\$ $\\ \textbf{Target}\\ $f_5$:primary activity = \emph{?}};		
	\node (current_input)[text_node, below = of input_val]
	{Pre-processed \\current input}; 
	
	\node (cluster1) [rect, draw=none, right =of input_val.north, minimum height= 0.4cm, yshift=-0.3cm, xshift=2.2cm, text width=3cm]{ \includegraphics[width=0.2\textwidth]{pics/cr.png}\\cid$_1$: \emph{[20,22), Public}\\Cluster$_1$: DIS=1};
	
	\node (cluster2) [rect, draw=none, below=of cluster1, xshift=-0cm,yshift=-0.2cm, minimum height= 0.4cm, text width=3cm]{\includegraphics[width=0.25\textwidth]{pics/rect.png}\\cid$_2$: \emph{[39,41), Private}\\Cluster$_2$: DIS=1};
	
	\node (textpoints) [text_node, below= of cluster2, text width=0cm, xshift=-0.2cm, yshift= -0.1cm] {....};
	
	\node (cluster3) [rect, draw=none, below=of textpoints, xshift=0.2cm, yshift=-0.1cm, minimum height= 0.4cm, text width=3cm]{\includegraphics[width=.15\textwidth]{pics/tr.png}\\cid$_k$: \emph{[39,41), Public}\\Cluster$_k$: DIS=2};

	\draw [decorate, decoration={brace, aspect=0.5, amplitude=10pt, raise=4pt}, yshift=0pt] (cluster1.north east) -- (cluster3.south east)  node [black, xshift=0cm] { };
	
	\begin{scope}[transform canvas={xshift=0em}, node distance=3mm]
		\node (f4) [round_rect, minimum width=1cm, right= of cluster2, xshift=0.8cm, yshift=-0.5cm] {$f_4$};
		\node (f3) [round_rect, minimum width=1cm, above= of f4] {$f_3$};
		\node (f2) [round_rect, minimum width=1cm, below left= of f4,xshift=0.7cm] {$f_2$};
		\node (f5) [round_rect, minimum width=1cm, below right= of f4,xshift=-0.7cm] {$f_5$=?};
		\draw [arrow,->] (f3) -- (f4);
		\draw [arrow,->] (f2) -- (f4);
		\draw [arrow,->] (f4) -- (f5);
	\end{scope}

	\node (sel_explain) [text_node, above= of f3, yshift= 0cm, text width=2.3cm] {No unique local model;\\$M_0$ is selected};
	
	\node (big_box_selection)[draw=black, fit={(cluster1)(cluster2)(cluster3)(f2)(f3)(f4)(f5)},minimum width=1.4cm,minimum height=2.5cm]{};
	
	\draw [arrow,->] (input_val.east) ++(0, 0.6)--++(0.7, 0) (big_box_selection);
	
	\node (sel_explain2) [text_node, below= of big_box_selection, yshift= 0cm, text width=3.5cm] {\myblock{A} Model selection};
	
	\node (table) [shape=rectangle,draw=white, right = of f4, yshift=1.1cm, xshift=1.5cm] {
		\begin{tabular}{p{2.1cm}c}
		\toprule
		\textbf{Value}& \textbf{Prob.}\\
		\midrule
		Leasing Serv.& 0.70  \\
		Financial Serv. & 0.15  \\
		Accommon-dation Serv.&0.05  \\
		... & ... \\
		\bottomrule
		\end{tabular}
	};

	\node(text_distribution) [text_node, above = of  table,minimum width=0.8cm,text width=3.5cm]{Probability distribution};

	\node(list_box_p2)[list_box, right=of table, yshift=-0.8cm, xshift=0.8cm]{Air transport\\Catering\\...};
	\node (list_box_p1)[list_box, above=of list_box_p2, yshift=-0.3mm] {Leasing Serv.  \hfill 0.70\\ Financial Serv. \hfill \ \ \ 0.15\\ Accommo-  \hfill \ \ \ 0.05 \\dation Serv.};

	\node (text_sug) [above=of list_box_p1, text_node]{Suggestion};
	\draw [arrow,->] (table.east) ++(0, -0.4)-- ++(0.7, 0)(list_box_p1);

	\node (big_box_suggestion)[draw=none, fit={(table)(list_box_p2)},minimum width=1.4cm,minimum height=2.5cm]{};

	\draw [arrow,->] (big_box_selection.east) ++(0, 0.3)--++(0.8, 0) (big_box_suggestion);

	\node (check_dep)[ text_node, below= of table, text width=3cm, draw=none, yshift=-0.5cm, xshift=0.3cm] {$f_4$:company type\\ is filled \\ \emph{checkDep}=\emph{true}};
	
	\node (text_or)[ text_node, right= of check_dep, minimum width=0.1cm, text width=0.2cm, xshift=0cm] {\textbf{or}};
	
	\node (check_prob)[ text_node, right= of text_or, text width=3cm, draw=none, yshift=0cm, xshift=-0cm] {0.70+0.15+0.05\\ >$\theta$=0.70 \\ \emph{checkProb}=\emph{true}};
	
	\node (big_box_endorser)[draw=black, fit={(check_dep)(check_prob)(text_or)},minimum width=1.4cm,minimum height=1.5cm]{};
	
	\node (endorser_explain) [text_node, below= of big_box_endorser, yshift= 0cm, text width=2.5cm] {\myblock{B} Endorsing};
	
	\draw [arrow, ->{Circle}] (big_box_endorser.north) ++(0.1,0)--++(0,1.65) (big_box_suggestion);

	\end{tikzpicture}
} 	\caption{Workflow for Form Filling Suggestion}
	\label{fig:selection-filter}
\end{figure}

\subsubsection*{Application to the running example}
Given the new input instance  shown on the left side of \figurename~\ref{fig:selection-filter}
(i.e., the instance ``income''=\emph{[20, 22)}, ``legal
entity''=\emph{Private}, and ``company type''=\emph{Leasing}, as obtained after pre-processing), 
\approach suggests the possible values of ``primary activity''.
As shown in block \myblock{A} of \figurename~\ref{fig:selection-filter},
\approach first attempts to select a unique local model by 
calculating the distance between the current input instance and the centroids of the three clusters 
generated in block \myblock{C} of \figurename~\ref{fig:dependency};
however, such a local model cannot be found because the distances with ``cid$_1$'' and ``cid$_2$'' are both 1.
Hence, \approach uses $M_0$ for prediction.
According to the variable inference method in BN (explained in section~\ref{sec:bayesian-networks}),
\approach outputs the probability distribution of the candidate values for the field ``primary activity''. 
Let us assume, as an example, that the probability distribution is ``\emph{Leasing
  Service}=0.70, \emph{Financial Service}=0.15, \emph{Accommodation
  Service}=0.05, \dots''.
By means of the endorser module (block \myblock{B} of
\figurename~\ref{fig:selection-filter}), \approach uses this probability distribution to decide whether to present the suggestions to the user.
For example, let us further assume the data quality engineers in the bank set $\theta$ to 0.70 and configures \approach to suggest three values. 
On the one hand, \approach finds that the target field $f_5$:``primary activity'' directly depends on $f_4$:``company type'', 
which was already filled by the user; 
the \emph{checkDep} flag is \emph{true}. 
On the other hand, the sum of the top-3 probability values is 0.90, which is higher than $\theta$; 
the \emph{checkProb} flag is \emph{true}.
Since the endorser module endorses a suggestion when one of these two flags evaluates to true, 
\approach provides a suggestion to the user: the three values above are put to the top of the list while the other candidate values are presented in their original order (e.g., alphabetically).

\section{Evaluation}
\label{sec:evaluation}

In this section, we report on the evaluation of our approach (\approach)
for automated form filling.  First, we assess the overall accuracy of
\approach in suggesting appropriate values to automatically fill in
the fields of data entry forms, and compare it with state-of-the-art
form filling algorithms. We also assess the performance of \approach,
in terms of training time and prediction time, for practical applications.
Then, we evaluate how the use of
local modeling (in the \emph{model building} phase) and
heuristic-based endorser (in the \emph{form filling suggestion} phase) affect the
accuracy of \approach.
Last, we assess the impact of the number of filled fields and the size of the training set on the effectiveness of \approach.

More specifically, we evaluated \approach by answering the following
research questions:
\begin{compactenum}[RQ1]
\item \emph{Can \approach provide accurate suggestions for automated
  form filling, and how does it compare with state-of-the-art
  algorithms?}
\item \emph{Is the performance of \approach (in terms of training time
  and prediction time) suitable for practical application in
  data-entry scenarios?}
\item \emph{What is the impact of using local modeling and
    heuristic-based endorser on the effectiveness of \approach?}
\item \emph{What is the impact of the number of filled fields on the effectiveness of \approach?}
\item \emph{What is the impact of the size of the training set on the effectiveness of \approach?}
\end{compactenum}

\subsection{Dataset and Settings}
\label{sec:dataset_and_setting}

\begin{table}[tbp]
	\caption{Information about the Fields in the Datasets}
	\centering
	\scriptsize
	\begin{filecontents*}{data/dataset-info.csv}
Dataset; 			     Fields; 				 Instances; 				Categorical;										 Freq1;						Freq5
\multirow{2}{*}{NCBI};	 \multirow{2}{*}{26};    \multirow{2}{*}{74105};    sex(3), tissue(68), cell-line(50), cell-type(63),	;\multirow{2}{*}{40.8\%};\multirow{2}{*}{59.4\%}
					 ;						;						   ;	disease(84), ethnicity(40)							;						;
\multirow{5}{*}{PROP} ;	 \multirow{5}{*}{33};	 \multirow{5}{*}{174446};	title(18), sex(3), legal capacity(7), country(208),	;\multirow{5}{*}{48.4\%};\multirow{5}{*}{65.6\%}
					 ;						;							;	first nationality(206), civil status(8),			;						;
					 ;						;							;	matrimonial regime(6), activity(13), status(15),	;						;
					 ;						;							;	function(41), contract(8), field of activity(75),	;						;
					 ;						;							;	primary activity(3), country of activity(198)		;						;
\end{filecontents*}
\pgfplotstabletypeset[
	every head row/.style={
		before row={\toprule
			\multirow{2}{*}{Dataset}& \# of& \# of & Name of categorical fields&\multicolumn{2}{c}{Value frequency} \\ 
		},
		after row=\midrule,
	},
	every last row/.style={
		after row=\bottomrule},
	every nth row={7[+2]}{before row=\midrule},
	columns/Dataset/.style ={column name=}, 
	columns/Fields/.style ={column name=fields}, 
	columns/Instances/.style ={column name=instances},
	columns/Categorical/.style={column name=(\# of candidate values)},	
	columns/Freq1/.style={column name=top-1},
	columns/Freq5/.style={column name=top-5\%},
	col sep=semicolon,
	string type,
	]{data/dataset-info.csv}
	\label{tab:dataset-info}
\end{table}

\subsubsection*{Datasets}
We evaluated \approach using a public dataset in the biomedical
domain (dubbed NCBI) and a proprietary dataset, extracted from a production-grade enterprise information system, provided by our industrial partner (dubbed PROP).

The NCBI dataset contains the metadata for diverse types of
biological samples from multiple species~\cite{barrett2012bioproject}.
We selected this dataset because it has been used in a previous study
on metadata suggestion for biomedical
datasets~\cite{martinez2019using}, which provided also the design of
the corresponding data entry form in the CEDAR
workbench~\cite{goncalves17:_cedar_workb}.  More specifically,
following the evaluation methodology described
in~\cite{martinez2019using}, we considered the subset of the NCBI
dataset related to the species ``Homo sapiens'' and the corresponding
data entry form based on the specification of the BioSample Human
package
v1.0\footnote{\url{https://submit.ncbi.nlm.nih.gov/biosample/template/?package=Human.1.0&action=definition}}.
We downloaded the dataset  
from the official NCBI website\footnote{\url{https://ftp.ncbi.nlm.nih.gov/biosample/}}.
In the dataset, the data is organized as a table.
Each row is an input instance filled by a user.
Retrieving the mapping between column names and field names was trivial 
since the column names in the dataset are the same as the field
names.
As shown in Table~\ref{tab:dataset-info}, the NCBI dataset has 26
fields, six of which are categorical. These categorical fields have between
3 and 84 candidate values to be selected by users.
We calculated the frequency by which users select different values during form filling: the most frequent (i.e., top-1)  and the
top-5\% most frequent values are selected, on average, respectively in
40.8\% and 59.4\% of the instances for different categorical fields.
Given the sparseness of the
dataset (caused by the optional fields), as suggested in~\cite{martinez2019using}, we identified the
empty values (e.g., ``n/a'', ``null''), and only retained records with at
least three fields (out of six) with non-empty values; in total, the NCBI
dataset contains \num{74105} input instances.  

The PROP dataset contains customer data that are provided through a
web-based data entry form, which is filled out upon creation of a new
customer account.  
We extracted the dataset from the distributed database of our industrial partner,
where all the input instances of a certain form are organized as a database table.
Each row in the table is an input instance 
and each column represent a form field.
We identified the mapping between the column names in the table
  and the field names in the data entry form by consulting the available software documentation.
As shown in Table~\ref{tab:dataset-info}, the PROP
dataset has 33 fields, 14 of which are categorical (with the number of
candidates values ranging from \numrange{3}{206}).
In terms of frequency according to which users select different candidate
values, the top-1 and the top-5\% most frequent values are selected,
on average, respectively in 48.4\% and 65.6\% of the instances.
According to the form design, eight of the categorical fields are mandatory to be filled; hence, we do not remove spare records as done for the
other dataset; in total, the PROP dataset contains \num{174446} input
instances.

We remark that both datasets represent the input instances from real-world data entry forms 
(i.e., the NCBI platform and a production-grade enterprise information system).
The number of fields in these systems is comparable with or larger
than the data entry forms used in the related work. For example, 
we calculated the average number of fields of data entry forms in the TEL-8 dataset, 
a manually collected dataset with 447 web forms (with no input instances),
which is used in the literature on form filling~\cite{araujo2010carbon,jou2019schema}.
In this dataset, each form has 6.39 fields on average.
The data entry forms in our study are more
complex, ranging from 26 to 33 fields of different types.

\subsubsection*{Dataset Preparation}
For the two datasets,
as discussed in section~\ref{sec:prob_def},
all the categorical fields are the targets for automated form
filling. However, we excluded the fields with less than 10 candidate
values (e.g., ``sex'', which has only three values in both datasets)
as users may easily browse all the values in these fields, without the
need for form-filling automation. The threshold for excluding fields
was determined together with the data quality engineers and some data entry
operators of our partner.
We find the majority of categorical fields we evaluated are related to certain domains or business processes;
they include fields ``tissue'', ``cell-line'', ``cell-type'', ``disease'' and ``ethnicity'' for the biological domain,
and fields ``activity'', ``status'', ``function'', ``field of activity'', and ``country of activity'' for the financial domain.
These fields are more difficult to fill than basic user information (e.g., name, sex, and age), since users need to understand the meaning of candidate values.

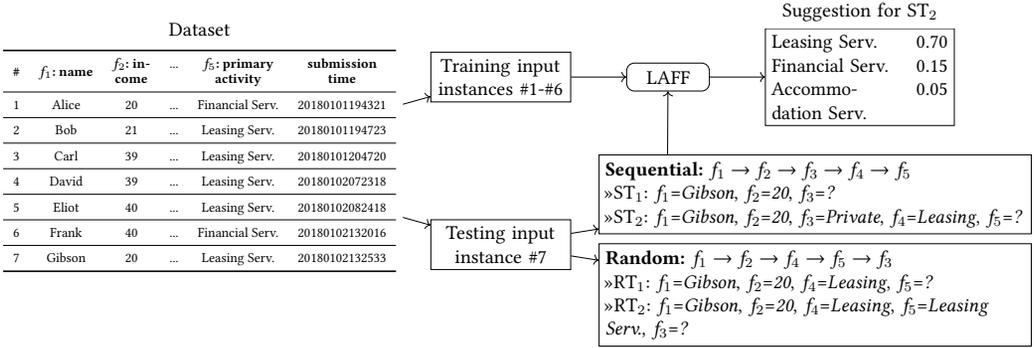
\begin{figure}[tb]
	\centering
	
\newcommand{\tikzmark}[2]{
	\tikz[overlay,remember picture,baseline] 
	\node [anchor=base] (#1) {$#2$};}

\newcommand{\DrawVLine}[3][]{\begin{tikzpicture}[overlay,remember picture]
	\draw[shorten <=0.3ex, #1] (#2.north) -- (#3.south);
	\end{tikzpicture}
}

\newcommand{\DrawCircle}[1][]{\begin{tikzpicture}[overlay,remember picture]
	\draw[black,fill=black] (0,0.5ex) circle (0.5ex);
	\end{tikzpicture}
}

\newcommand{\DrawRec}[1][]{\begin{tikzpicture}[overlay,remember picture]
	\draw[black,fill=black] (-0.5ex,0) rectangle ++(1ex, 1ex);
	\end{tikzpicture}
}

\newcommand{\DrawTriangle}[1][]{\begin{tikzpicture}[overlay,remember picture]
	\filldraw[fill=black] (90:1.2ex) -- (170:0.7ex) -- (10:0.7ex) --cycle;
	\end{tikzpicture}
}

\tikzstyle{round_rect} = [rectangle, rounded corners, minimum width=1.5cm, minimum height=0.5 cm,text centered, draw=black, fill=white!30]

\tikzstyle{rect} = [rectangle, minimum width=1.5cm, minimum height=0.8cm, text centered, text width=1.5cm, draw=black, fill=white!30]

\tikzstyle{text_node} = [rectangle, minimum width=0.5cm, minimum height=0.35cm, text centered, text width=4cm, draw=white, fill= white!30]

\tikzstyle{list_box} = [rect,align=left ,minimum width=3.19cm,text width=3.19cm, minimum height= 0.5cm]

\resizebox{1\textwidth}{!}{
\begin{tikzpicture}

\node (tab1) [shape=rectangle,draw=white] {
	\scriptsize
	\begin{tabular}{cccccc}
	\toprule
	
	\multirow{2}{*}{\textbf{\#}}&\multirow{2}{*}{\textbf{$f_1$: name}}&\textbf{$f_2$: in-}&...&\textbf{$f_5$: primary}&\textbf{submission}\\
	& & \textbf{come}& &\textbf{activity}&\textbf{time}\\
	\midrule
	1&Alice& 20 &...&Financial Serv.&20180101194321\\
	\midrule
	2&Bob&21 & ...&Leasing Serv.& 20180101194723\\ 
	\midrule
	3&Carl& 39 & ...&Leasing Serv.&  20180101204720\\ 
	\midrule
	4&David&39 &... &Leasing Serv.&  20180102072318\\ 
	\midrule
	5&Eliot& 40 &...&Leasing Serv.&  20180102082418\\ 
	\midrule
	6&Frank&40 &... &Financial Serv.& 20180102132016\\ 
	\midrule
	7&Gibson&20 & ... & Leasing Serv.& 20180102132533\\
	\bottomrule
	\end{tabular}
};

\node (dataset) [ text_node, above=of tab1, yshift=-1cm] {Dataset};

\node (train) [ text_node, draw=black, text width=2.3cm, right= of tab1, xshift=-0.5cm, yshift= 1.5cm] { Training input instances \#1-\#6};
\node (test) [ text_node, draw=black, text width=2.3cm, right= of tab1, xshift=-0.5cm, yshift= -1.5cm] { Testing input instance \#7};

\draw [arrow,->] (tab1) -- (train);
\draw [arrow,->] (tab1) -- (test);

\node(laff)[ round_rect, draw=black, text width=1.2cm, right= of train, xshift=-0.0cm] { \approach};

\draw [arrow,->] (train) -- (laff);

\node (seq) [ text_node, align= left, draw=black, text width=7.6cm, right= of test, xshift=-0.5cm, yshift= 0.9cm] { \textbf{Sequential:} $f_1\rightarrow f_2 \rightarrow f_3 \rightarrow f_4 \rightarrow f_5$\\
>>ST$_1$: $f_1$=\emph{Gibson}, $f_2$=\emph{20}, $f_3$=\emph{?}\\
>>ST$_2$: $f_1$=\emph{Gibson}, $f_2$=\emph{20}, $f_3$=\emph{Private}, $f_4$=\emph{Leasing}, $f_5$=\emph{?}};

\node (rand) [ text_node, align= left, draw=black, text width=7.6cm, right= of test, xshift=-0.5cm, yshift= -0.9cm] { \textbf{Random:} $f_1\rightarrow f_2 \rightarrow f_4 \rightarrow f_5 \rightarrow f_3$\\
	>>RT$_1$: $f_1$=\emph{Gibson}, $f_2$=\emph{20}, $f_4$=\emph{Leasing}, $f_5$=\emph{?}\\
	>>RT$_2$: $f_1$=\emph{Gibson}, $f_2$=\emph{20}, $f_4$=\emph{Leasing}, $f_5$=\emph{Leasing Serv.}, $f_3$=\emph{?}};

\draw [arrow,->] (test) -- (seq);
\draw [arrow,->] (test) -- (rand);
\draw [arrow,->] (seq) ++(-2.675, 0.7)-- (laff);

\node (list_box_p1)[list_box, right=of laff, yshift=-0.0mm] {Leasing Serv.   \hfill 0.70\\ Financial  Serv. \hfill \ \ \ 0.15\\ Accommo-  \hfill \ \ \ 0.05 \\dation Serv.};
\node (text_sug) [above=of list_box_p1, yshift=-1cm, text_node]{Suggestion for ST$_2$};

\draw [arrow,->] (laff) -- (list_box_p1);

\end{tikzpicture}
}
 	\caption{Example of Dataset Preparation (Training and Testing Sets)}
	\label{fig:eval}
\end{figure}

Since both datasets automatically recorded the submission time of each input instance.
we split the dataset into two subsets containing 80\% and 20\% of
input instances based on their submission time, used respectively for training and testing.  
The input instances (excluding the information of the submission time) in the training set are used to train \approach.  
As for the testing input instances, since
there is no information on the actual filling order used to input the
data, we considered two form filling orders to simulate the data entry
session.  More specifically, we simulated two types of filling
scenarios: ``sequential filling'' and ``random filling''.  The former
corresponds to filling data entry forms in the default order, as
determined by the form \emph{tab sequence}, e.g., the navigation order
determined by the HTML attribute \texttt{tabindex} in web UI
designs~\cite{fowler2004web}.
It simulates the logical order many users follow to fill out forms, 
especially when they use a keyboard to navigate form fields~\cite{change2013microsoft}.
The latter represents the scenario when
users may select any field as the next target, and even go back to
modify already filled fields.
These two form filling orders represent two opposite extremes\footnote{For some large data entry forms, 
UI designers can semantically partition related fields into sections.
Users can then move between sections in sequential order, while using
the random order to fill fields within a section.
This is a ``middle-ground filling'' order, which sits between 
``sequential filling'' and ``random filling''. We have not
evaluated this scenario since it 
requires additional knowledge about the partitioned sections, which was
not available for the two datasets we have considered. } of user behavior
during a real data-entry session.
We simulated random filling by randomly
generating an order for each testing input instance.  In both form
filling scenarios, the filled fields considered by \approach are the
fields that precede each target.
For each target field, we consider the actual value filled by the user as the ground truth.

\subsubsection*{Dataset Preparation - Example of Application}
\figurename~\ref{fig:eval} shows an example of application of our
dataset preparation method for the training and testing sets.
Let us consider a dataset containing seven input instances submitted
through a data entry form, shown on the left-hand side of
\figurename~\ref{fig:eval}.
Following the running example introduced in
section~\ref{sec:prob_def}, the form has five fields, two of which are categorical (e.g., $f_3$:
``legal entity'' and $f_5$:``primary activity'').
We split the dataset  into the training and testing sets according to the submission time: we take 80\% of input instances (i.e., \#1-\#6) to train \approach;
the testing set contains the remaining 20\% of the instances, in this case input instance \#7.
The testing set is further processed to simulate the two types of
filling scenarios.
When using  the sequential filling order, 
users fill the data entry form following the \texttt{tabindex} of fields in the form 
(e.g., from $f_1$ to $f_5$ sequentially): starting from the input
instance \#7, we generate test instances ST$_1$ and ST$_2$ for categorical fields $f_3$ and $f_5$, respectively.
For each categorical field (i.e., the target),
the actual value filled by the user is the ground truth 
(e.g., the ground truth for the field $f_5$ is `Leasing Serv.').
When using the random filling order, 
we randomly generate a field order for each input instance 
(e.g., $f_1\rightarrow f_2\rightarrow f_4\rightarrow f_5\rightarrow
f_3$ for the input instance \#7); based on this order, we then generate test instances RT$_1$ and RT$_2$.

\subsubsection*{Implementation and Settings}
We implemented \approach as a Python program; we used the open-source
library \texttt{pgmpy}~\cite{ankan2015pgmpy} for working with Bayesian
networks.

We configured \approach (through parameter $n_r$ in Algorithm~\ref{alg:LAF-prediction}) to suggest the top 5\%, most likely values for each target field.
Based on the number of candidate values for each field in the datasets
(indicated in parentheses in the rightmost column of Table~\ref{tab:dataset-info}),
suggesting the top 5\% values means showing between one (for field
``activity'' in the PROP dataset) and ten (for field ``country'' in
the PROP dataset) suggested values to users.
This is in accordance with other recommender systems in software engineering,
in which only a list of few candidates is suggested for
consideration~\cite{yang2016combining,pearson2017evaluating}.
We set the threshold $\theta$ to 0.7 based on the feedback received by data entry operators and data quality engineers of our
partner.

We performed the experiments on the NCBI dataset with a computer running macOS 10.15.5 with a \SI{2.30}{\giga\hertz}
Intel Core i9 processor with \SI{32}{\giga\byte} memory. As for the
experiments on the PROP dataset\footnote{Due to the data protection
  policy of our partner, we were obliged to run the experiments on the
  PROP dataset using an on-premise, dedicated server that, however,
  could not be used to store external data (like the NCBI dataset).}, we performed them on a server running CentOS 7.8 on a \SI{2.60}{\giga\hertz} Intel Xeon E5-2690 processor with \SI{125}{\giga\byte} memory.

\subsection{Effectiveness (RQ1)}
\label{sec:rq1}

To answer RQ1, we assessed the effectiveness of \approach to
suggest appropriate values for each of the target fields in the
dataset.
We compared \approach with MFM (most frequent model), ARM (association
rule mining)~\cite{martinez2019using}, Na{\"i}veDT (na{\"i}ve
application of decision trees), and FLS (first letter search),
which are able to provide suggestions under different form filling orders: 
\begin{enumerate}
	\item MFM is a widely-used form filling
	algorithm, which suggests possible values of a
	target field based on their frequency in historical input instances.
	\item ARM
	is a state-of-the-art algorithm for form filling. ARM uses historical input instances to mine
	association rules with a minimal level of support and confidence;
	it matches the filled fields with mined association rules,
	and suggests the consequents of the matched rules to users.
        
      \item Na{\"i}veDT is a na{\"i}ve application of decision trees
        for form filling.  We use decision trees because this type of
        model has been already used in the form filling
        literature~\cite{hermens1994machine}.  Given a target field,
        this approach takes a subset of the remaining fields as filled
        fields (i.e., features); it then trains a decision tree for
        each feature-target combination.  During form filling, based
        on the filled fields and the target field, Na{\"i}veDT selects
        the decision tree trained on the same feature-target
        combination in order to predict the target field.
We considered this na{\"i}ve application of decision
          trees because previous work~\cite{hermens1994machine} has
          shown that the effectiveness of a single decision tree trained
          for each target field is poor (see also
          section~\ref{sec:related_work}); our preliminary evaluation
          has also confirmed this.
        
      \item
        FLS simulates form filling in categorical fields through a ``typing'' function. 
	This function allows users to type the first letter of the candidate value they intend to fill 
	(i.e., the first letter of the ground truth in this study). 
	FLS filters the list of candidate values based on this letter 
	and presents the refined candidate values as suggestions.
    \end{enumerate}

We did not consider other approaches for automated form filling, since
they rely on additional information
beyond the input values provided in the past for the same form.
For example, they reuse the values filled in other software systems~\cite{araujo2010carbon}, 
extract information from text files 
(e.g., a CV file to fill job search sites)~\cite{toda2010probabilistic}, 
or refactor forms for effective form filling~\cite{chen2011usher}.
An empirical comparison with these techniques is not feasible, 
since such additional knowledge is not always available during form filling;
moreover, \approach does not assume the existence of such knowledge.
We discuss the differences between \approach and these related approaches
in section~\ref{sec:related_work}.

\subsubsection*{Choosing effectiveness metrics}

We reviewed the
main metrics used for evaluating the effectiveness of building a suggestion list.  More specifically, we investigated the
metrics used in the recommender systems area because, similar to
automated form filling, many software applications use recommender
systems to support software stakeholders in their decision-making
while interacting with large information
spaces~\cite{robillard2009recommendation} (e.g., locating faulty code
snippets in software projects~\cite{ye2014learning}).
Table~\ref{tab:metrics} shows, for each metric we reviewed, its dimension, description, and rationale.

\begin{table}[tb]
	\caption{Main metrics used in the  recommender system
		area for evaluating the effectiveness of building a list of
		suggestions }
	\centering
	\scriptsize
    \begin{tabular}{l p{5.8cm }p{5.7cm}}
	\toprule
	Dimension&Description&Intuition\\
	\midrule
    
    Diversity 
    &Diversity metrics generally measure the average dissimilarity between all pairs of items in the suggestion list~\cite{kunaver2017diversity}.
    &The suggested items can cover a broad area of the information space to increase the chance of satisfying the user’s information need. For example, with a movie recommender system, users may hope to get relevant items from different genres (e.g., ``comedy'', ``romance'') \\
	\midrule

	Novelty
	&The novelty of an item is typically estimated by the inverse of its popularity (e.g., measured by the number of ratings it has received); novelty metrics measure the ratio of the suggested relevant items that have  low popularity~\cite{kaminskas2016diversity}.
	&A tool has the ability to suggest relevant items that are unknown to users. For example, a movie recommender system should have the ability to recommend ``new'' movies that users did not watch or know before.  \\ 
	\midrule

	\multirow{4}{*}{Accuracy}
	&\emph{Precision} measures the fraction of suggested relevant items among all the suggested items~\cite{herlocker2004evaluating}.
	&The suggestion list only contains relevant items.\\
	\cmidrule{2-3}
	&\emph{Recall} measures the fraction of suggested relevant items among all the relevant items~\cite{herlocker2004evaluating}.
	&The suggestion list contains all the relevant items regarding a target.\\
	\cmidrule{2-3}
	 &\emph{MAP (Mean Average Precision)} measures the mean of the average precision at the rank of each relevant item~\cite{herlocker2004evaluating}.
	 &All the relevant items can be ranked at the top of a suggestion list. \\ 
	 \cmidrule{2-3}
	 &\emph{MRR (Mean Reciprocal Rank)} measures the mean of the reciprocal rank at which the first relevant item was suggested in a list~\cite{martinez2019using}.
	 &The relevant item can be ranked at the top of a suggestion list.\\ 
	\midrule

	\multirow{2}{*}{Coverage}
	&\emph{Catalog coverage} measures the ratio of the items suggested to users over the total number of candidate items of a given target~\cite{ge2010beyond}.
	&A tool can avoid suggesting a long list of items to users (e.g., only the top 5\% suggested items are presented to the users)\\
	\cmidrule{2-3}
	&\emph{Prediction coverage} measures the ratio of suggestions provided by a tool over the total number of targets requiring suggestions~\cite{ge2010beyond}.
	&A tool can avoid making ``useless'' suggestions to users. Low-confidence suggestions with many unrelated items should be filtered out to fit the user's interests.\\
	\midrule

	\multirow{2}{*}{Combined}
	&\emph{F$_1$-score} is the harmonic mean of the precision and recall~\cite{avazpour2014dimensions}.
	&A tool can suggest (recall) and only suggest (precision) all the relevant items to users.\\
	\cmidrule{2-3}
	&\emph{Quality} of suggested items is the product between their similarity to the user's query and the diversity of the items~\cite{avazpour2014dimensions}. 
	&The items suggested in the suggestion list are relevant to the user's query (similarity) and at the same time different from each other (diversity).\\
	\bottomrule
\end{tabular} 	\label{tab:metrics}
\end{table}

Metrics for evaluating recommender systems span over four dimensions,
including diversity~\cite{kunaver2017diversity},
novelty~\cite{castells2011novelty,kaminskas2016diversity},
accuracy~\cite{herlocker2004evaluating}, and
coverage~\cite{ge2010beyond,herlocker2004evaluating}.  As shown in
Table~\ref{tab:metrics}, diversity and novelty focus on the
dissimilarity among the suggested items: the former by looking at 
pairwise dissimilarity, and the latter by
determining the difference between the currently and previously
suggested items.  Both metrics can be applied in contexts where more
than one relevant item can be suggested.  As for assessing accuracy,
precision and recall are the most common
metrics~\cite{herlocker2004evaluating}; they measure the ratio of
correctly suggested items. However, precision and recall 
ignore the exact ranking of items as only the correct or incorrect
classification is measured~\cite{schroder2011setting}.  Other common
accuracy metrics include MRR and
MAP~\cite{schroder2011setting,karimi2018news,herlocker2004evaluating},
which are designed to evaluate a list of suggested items. MRR
calculates the rank of the first relevant item, and MAP measures the
average precision of relevant items at different positions.  Regarding
coverage, two definitions have been proposed in the
literature~\cite{ge2010beyond}: catalog coverage and prediction
coverage. Catalog coverage measures the length of a list of suggested items relative to its maximum length; prediction coverage calculates the ratio of targets for which an algorithm provides suggestions over the total number of targets requiring suggestions.  Finally, in the literature, some metrics are also
proposed to combine different metrics from the same dimension to
evaluate the trade-off between them.  For example, metrics like
F$_1$-score and the quality metric~\cite{avazpour2014dimensions} have
been proposed to balance the weight of precision and recall (accuracy
dimension) or similarity and diversity (diversity dimension),
respectively.

According to a previous study~\cite{robillard2009recommendation}, an
effective recommender system in software engineering (e.g., a form
filling system) is expected to avoid ``helpless'' suggestions that may
be ignored by users, but provide users a large number of ``helpful''
suggestions.  Considering the dimensions in
Table~\ref{tab:metrics}, metrics for diversity and novelty are not
applicable for form filling, as most categorical fields only contain a
single correct value for a user to select.  The accuracy metrics can
be used to assess the ``helpfulness'' of suggestions.  Since \approach
suggests a list of values for users to select the correct one, MRR is
the most appropriate metric in our context, since it evaluates if the
correct value is ranked at the top of a suggestion list.  Regarding
coverage, we select the prediction coverage to evaluate the extent to which the endorser module of \approach can avoid ``helpless'' suggestions;
it calculates the frequency of suggestions made by \approach when required to make one.  Since MRR and prediction coverage belong to different dimensions, we separately evaluate the two metrics instead of combining their results with a single score.

In the following, we provide the definition of MRR and prediction coverage. 

MRR (Mean Reciprocal Rank) is defined as:
\begin{equation}
\mathit{MRR} = \frac{1}{|S|}\sum_{i=1}^{S} \frac{1}{k_i},
\end{equation}
where $|S|$ is the number of target fields that the algorithm provides suggestions, 
and $k_i$ is the position of the first correct value in the $i$-th suggestion.

Prediction coverage rate is defined as:
\begin{equation}
\mathit{PCR} =
\frac{|S|}{|S_{\mathit{all}}|}, 
\end{equation}
where $|S|$ is again the number of target fields that the algorithm provides suggestions, and $|S_{\mathit{all}}|$ is the total number of target fields in all the testing input instances~\cite{herlocker2004evaluating,ge2010beyond}.

\subsubsection*{Methodology}

To assess the effectiveness of the various form filling algorithms, we
computed $\mathit{MRR}$ and the prediction coverage rate
$\mathit{PCR}$.

  We remind the reader that \approach uses the filled fields (i.e.,
  features) of each test instance to predict the value of a target
  field.  In our datasets, all the target categorical fields only have
  a single correct value (e.g., in \figurename~\ref{fig:eval},
  \emph{Leasing Serv.} is the ground truth for the field $f_5$ of the
  input instance \#7).

For each test instance, we checked the position of the correctly suggested
value (i.e., the value that corresponds to the ground truth) in each suggestion list 
and computed the reciprocal rank of the correct value. 
If no correct value was found, we set the reciprocal rank to zero. 
For example, in \figurename~\ref{fig:eval}, \approach suggested three values; 
the reciprocal rank of the suggestion for ST$_2$ is 1 
since the user can find the correct value to fill in first
position (i.e., $k_i = 1$).
The $\mathit{MRR}$ value was computed as the mean of the reciprocal
ranks for all the suggestion lists.
Given a test set, we calculated the average $\mathit{MRR}$ value for different targets.
Concerning $\mathit{PCR}$, we counted the number of target fields for which an algorithm provided suggestions (i.e., a list of suggested values) and the number of target fields for which no suggestion was provided. $\mathit{PCR}$ was computed as the percentage of target fields receiving suggestions over the total number of target fields.

  As there is no publicly available
implementation of the baselines (ARM, MFM,  Na{\"i}veDT, and FLS) for form filling, we implemented them from
scratch.

In the case of ARM, we set the minimum acceptable support and
confidence to 5 and 0.3, respectively, consistent with previous
work~\cite{martinez2019using}.

We implemented Na{\"i}veDT using the open-source library
\texttt{scikit-learn}~\cite{scikit-learn}.
For the NCBI dataset, after the preprocessing step (see section~\ref{sec:pre-processing}) we obtained six fields; all of them are categorical.
According to the discussion in section~\ref{sec:model-building}, we have  $n=6$ and $t=6$, 
resulting in $6*(2^{6-1}-1)=186$ feature-target combinations (i.e.,
decision trees to train).
For the PROP dataset, after the preprocessing step, we obtained 15
fields, among which 14 are categorical (i.e., $n=15$ and $t=14$). This leads to $14 * (2^{15-1}-1)=\num{229362}$ decision trees to train.

Regarding FLS,
we ranked the candidate values for a given target field alphabetically 
and refined these values by the first letter the user intends to fill. 
We assume that the first letter of the value of the ground truth is what the user intends to fill. 
Based on this assumption, FLS suggests the candidate values that start with the same letter as the ground truth.

  We set a timeout of 24 hours to train each algorithm. This timeout
  value reflects the realistic situation in which the algorithm gets
  daily updates of its models, empowered with the information derived
  from new input instances collected throughout the day. 

\subsubsection*{Results}
\begin{table}[tb]
	\caption{$\mathit{MRR}$  and $\mathit{PCR}$  of Form Filling
          Algorithms (t/o: timeout; N/A: not applicable)}
	\centering
	\scriptsize
	\pgfplotstabletypeset[
every head row/.style={
	before row={\toprule
		\multirow{2.5}{*}{}&\multirow{2.5}{*}{Alg.}&\multicolumn{2}{c}{Sequential}&\multicolumn{2}{c}{Random}&\multirow{1}{*}{Train}&\multicolumn{2}{c}{Predict (\si{\milli\s})}\\
		\cmidrule(r){3-4}
		\cmidrule(l){5-6}
	},
	after row=\midrule,
},
every last row/.style={
	after row=\bottomrule
},
every nth row={5}{before row=\midrule},
columns/Dataset/.style ={column name=}, 
columns/Algorithm/.style ={column name=}, 
columns/MRR-S1/.style={column name=MRR}, 
columns/Covrg-S1/.style ={column name=PCR},
columns/MRR-R1/.style={column name=MRR},	 
columns/Covrg-R1/.style ={column name=PCR},
columns/Train/.style ={column name=(\si{\s})},
columns/P-Avg/.style ={column name=avg},
columns/P-Range/.style ={column name=min--max},
col sep=comma,
string type,
]{data/rq-applicability.csv}

 	\label{tab:rq1}
\end{table}

Table~\ref{tab:rq1} shows the effectiveness of the various algorithms
for the two form filling scenarios.
  We remark that Na{\"i}veDT timed out during the training phase on
  the PROP dataset\footnote{For completeness, based on our
      preliminary evaluation, we note that using
      a single decision tree trained for each target field (as done in
      previous work~\cite{hermens1994machine}) would have
      avoided the time-out on the PROP dataset but would have  yielded worse effectiveness
      results than the lowest values reported in
      Table~\ref{tab:rq1}, confirming the trend already reported
      by~\citet{hermens1994machine} (see also
section~\ref{sec:related_work}).}

In terms of coverage rate (columns $\mathit{PCR}$), MFM, ARM, Na{\"i}veDT, and FLS
provide suggestions on almost all the target fields ($\mathit{PCR}
\approx 1$),  while \approach achieves a
$\mathit{PCR}$ value ranging from \numrange{\LAFFRandCov}{\LAFFRandCovBGL} on the two datasets.
The lower $\mathit{PCR}$ value of \approach is ascribable to its
endorsing module, which discards  low-confidence suggestions.

As to the accuracy of these suggestions,
MFM, ARM, and  Na{\"i}veDT achieve
$\mathit{MRR}$ values ranging from \numrange{\MFSeqMrr}{\ARMRandMrr}
on the NCBI dataset;
MFM and ARM achieve $\mathit{MRR}$ values ranging
 from \numrange{\MFSeqMrrBGL}{\ARMSeqMrrBGL} on the PROP dataset.
\approach substantially outperforms MFM, ARM, and Na{\"i}veDT, achieving a
$\mathit{MRR}$ value above {\LAFFRandMrr} in both datasets.  The
improvement in $\mathit{MRR}$ value obtained by \approach over Na{\"i}veDT is
\SI{+22}{\pp} on the NCBI dataset; the one over ARM is \SI{+11}{\pp} on the PROP dataset for the sequential filling
scenario. For the random filling scenario, the improvement over
Na{\"i}veDT and ARM is
\SI{+20}{\pp} on the NCBI dataset; the improvement over ARM is \SI{+16}{\pp} on the PROP
dataset.
According to Table~\ref{tab:rq1}, \approach also outperforms the interaction-based approach FLS by \SIrange{+19}{+25}{\pp} 
on the two datasets for different filling scenarios.
The reason is that, after refining the candidate values by typing letters, 
users could still have dozens of candidate values with the same initial letter to check. 
For example, when users type ``L''  to find the country ``Luxembourg'', 
the form returns 9 results (e.g., ``Laos'', ``Latvia'', ``Lebanon'') where ``Luxembourg'' is the last one in the refined list;
in contrast, \approach can rank ``Luxembourg'' as the first country
for users to choose, leading to much higher $\mathit{MRR}$ values.
Compared with FLS, \approach has two advantages.
First, many users lack a detailed conceptual model of
the software system~\cite{horsky2003framework} 
defined by the requirements analysts and domain experts. 
They may not remember all the candidate values predefined in a categorical field, 
leading to a potential lengthy search process~\cite{horsky2003framework}.
In this case, \approach can directly provide the most-likely suggestions for users to choose based on the filled fields.
Second, \approach is compatible with FLS. 
Based on the highly accurate suggestions made by \approach,
users can continue refining the suggested list with FLS when needed
to further accelerate their form filling process. 
For example, when users type ``L'' after  \approach's suggestion, only country names starting with ``L'' can be retained (e.g., ``Luxembourg'', ``Laos'', ``Latvia'', ``Lebanon'').

We use the Mann-Whitney U test to assess the statistical significance of the
difference between the MRR values of \approach and the baselines, 
with a level of significance $\alpha$ = 0.05. The results show that \approach
always achieves a statistically higher MRR value than the baselines for the two 
form filling scenarios on both the NCBI and PROP datasets ($p$-value < 0.01).

These results have to be interpreted with the usage scenarios of a
recommender system.  Previous studies show that, for a recommender
system, inaccurate suggestions increase users' decision time and the
risk of making wrong decisions~\cite{ore2018assessing}.  
The $\mathit{MRR}$ and
$\mathit{PCR}$ values achieved by \approach show that 
the suggestions provided by \approach allow users to find the correct
value among the top-ranked suggested values.

\subsubsection*{Error analysis}\label{sec:rq1:error}
We further analyzed the suggestions made by \approach,
to identify the cases in which it does not perform well.  We recall
that in our experiments, \approach suggested the top 5\% most likely
values for each target field.  On the NCBI dataset, \approach captures
the correct value in the top 5\% suggested values for 79.0\% of the
suggestions when using sequential filling and for 79.9\% of the
suggestions when using random filling.  On the PROP dataset, the
correct value is in the top 5\% suggested values for 89.5\%
(sequential filling) and 90.1\% (random filling) of the suggestions.
Overall, 79.0\% to 90.1\% of the suggestions made by \approach allow
users to find the correct value among the 5\% top-ranked suggested
values.  For the remaining incorrect suggestions (around 21\% on the NCBI dataset and 9.9\% on the PROP dataset), in
which the correct value is not in the top 5\% suggested values, we
identified the following main reasons.

First, \approach tends to provide incorrect suggestions when the
number of filled fields used for prediction is small.
Specifically,  on the NCBI dataset,
33.1\% of the incorrect suggestions when using sequential filling and
53.1\%  of the incorrect suggestions when using random filling  were made when there was only one filled field;
on the PROP dataset, the ratio is 17.4\% and 23.5\%, respectively.
With few filled fields, \approach may not get enough knowledge (i.e., the information of dependent field values) for prediction. 
This affects both the variable inference step within BNs and the
behavior of the endorser module.
In this case, \approach tends to use the most frequent value in the target field for prediction, 
since this value has a higher prior probability.
One possible way to mitigate this issue, to be investigated as part of
future work, could be to define form refactoring
techniques that allow users to first fill fields that
provide additional knowledge used to predict the values of other fields.
We  analyze the impact of the number of filled fields on the
effectiveness of \approach as part of RQ4 (\S~\ref{sec:rq4}).

Second, incorrect suggestions are caused by the number of training input instances. 
Due to optional fields, users may not provide values for all the fields. 
For example, for the field ``ethnicity'' in the NCBI dataset, 
only 15.6\% of input instances contain a non-empty value. 
The sparseness of the filled values for a field leads to a small
number of training input instances to learn the corresponding dependency. 
Continuing the example, the MRR values for the field ``ethnicity'' are 0.608 and 0.553 
for the sequential filling and random filling scenarios, respectively, 
thus leading to many incorrect suggestions.
To mitigate this problem, as part of future work,
instead of using a single threshold for endorsing suggestions, we
could modify the endorser used  in \approach to support field-specific
thresholds.
We analyze the impact of the size of the training set on the
effectiveness of \approach as part of RQ5 (\S~\ref{sec:rq5}).

Third, the number of options (i.e., candidate values) for a field \emph{may} affect the
effectiveness of \approach.  To investigate this, we computed the correlation between the
number of options and the MRR value for a field,
considering  the MRR values achieved in the random filling
scenario\footnote{We did not consider the sequential filling scenario,
  since it could introduce some bias in our analysis: The number of
  filled fields to predict for each target field is different, as it
  depends on the tabindex order of the field.}.  The resulting Pearson
correlation coefficient is -0.09 on the NCBI dataset ($p$-value=0.722) and  -0.477 on the PROP dataset ($p$-value=0.001),
thus showing no correlation for NCBI and a moderate but significant correlation for PROP~\cite{emam1999benchmarking}. The difference in results between datasets can be easily explained by the fact that the variance in number of options is very low for NCBI (\num{228.8}), while it is much larger for PROP \num{6713.25}.  These results therefore suggest that, when a field has more options, \approach tends to provide
more incorrect suggestions.

To conclude, \emph{
the answer to RQ1 is that \approach can yield a large number (with a
$\mathit{PCR}$ value ranging from \numrange{\LAFFRandCov}{\LAFFRandCovBGL}) of accurate suggestions, with
a $\mathit{MRR}$ value above {\LAFFRandMrr}, significantly outperforming
state-of-the-art approaches.
}

 \subsection{Performance (RQ2)}
\label{sec:rq:performance}

To answer RQ2, we measured the execution time required to perform the
model building phase of \approach (i.e., training time), as well as
the time to predict a target field (i.e., prediction time). The
training time indicates the feasibility of using \approach in contexts
where the training set (i.e., the set of historical input instances)
is updated often as new input instances are recorded in the
system. The prediction time indicates how fast \approach can provide
form filling suggestions during a data entry session.

\subsubsection*{Methodology}
We used the same settings (i.e., form filling scenarios) as in RQ1. We
computed the training time as the time to build all BN models over the
historical input instances. The prediction time is the average time
(over the various target fields) taken to provide a suggestion for one
input instance using locally deployed models. 
We also compared \approach with the MFM, ARM, and Na{\"i}veDT algorithms.
Notice
that FLS does not require any training and the prediction can be
considered instantaneous.

\subsubsection*{Results}
The results are shown in the last two columns in Table~\ref{tab:rq1},
in term of training time (column \textit{Train}) and the prediction time
(column \textit{Predict}, with sub-columns indicating the average,
minimum, and maximum values, when applicable) for the two datasets. 

The training time of \approach is much higher than the one of MFM,
ARM, FLS (and Na{\"i}veDT on the NCBI dataset);
\approach takes \SI{\LAFFTrain}{\s} and \SI{\LAFFTrainBGL}{\s} to
train models on the NCBI and PROP datasets, respectively.
This
can be easily explained since \approach trains several models, as explained
in section~\ref{sec:model-building}.
Although Na{\"i}veDT took only \SI{\DTTrain}{\s} to train all the decision trees on the NCBI dataset, 
it timed out on the PROP dataset. 
As for prediction time, that of \approach is higher than that of
MFM, Na{\"i}veDT, and FLS.
MFM and FLS directly suggest the frequency-based (for MFM)
or matching-based (for FLS) value list to users.
For Na{\"i}veDT, the prediction time includes the time to select the
appropriate model (based on the feature-target combination)
among the trained models. 
However, the prediction time of \approach is, on average, faster than the one of
ARM by \SI{105}{\milli\s} (\SI{\LAFFPredict}{\milli\s} vs
\SI{\ARMPredict}{\milli\s}) over the NCBI dataset and by \SI{740}{\milli\s} (\SI{\LAFFPredictBGL}{\milli\s} vs \SI{\ARMPredictBGL}{\milli\s}),
over the PROP dataset.
The Mann-Whitney U test also confirms that the differences in prediction time
between \approach and the baselines are statistically significant
($p$-value < 0.01 for the two datasets).

These results have to be interpreted taking into account the usage
scenarios of our approach. The training time has to be considered
when training the models with new data, i.e., the new input
instances recorded in the system since the last execution of the model
building phase. This task is performed \emph{offline} and
\emph{periodically} (e.g., once a day), so a training
time of the order of one hour is acceptable from a practical
standpoint.

Given the interactive nature of data-entry applications, having a
short prediction time is much more important for an automated form
filling approach.  According to human-computer interaction
principles~\cite{heeter2000interactivity}, users feel a system reacts
instantaneously when its response time is within \SI{100}{\milli\s}
and feel they are seamlessly interacting with the software system when the
response time is within \SI{1}{\s}.  In our context, the prediction
time of \approach depends on the computational power at our disposal
and the complexity of the trained BN models (i.e., the number of nodes
and the size of probability tables).  The experiments using the
proprietary dataset (extracted from a production-grade system) show
that \approach is fast enough (requiring at most \SI{\LAFFPredictMax}{\milli\s}) to
provide real-time suggestions during data entry sessions.

\emph{The answer to RQ2 is that the performance of \approach, with a
  training time of about one hour (or less) and a prediction time of at
  most \SI{\LAFFPredictMax}{\milli\s}, is suitable for practical
  application in data-entry scenarios.}

 \subsection{Impact of Local Modeling and Endorser (RQ3) }
\label{sec:rq3}

\approach has two important modules: the local modeling module, which
builds local models based on local field dependencies of partitions of
historical input instances (section~\ref{sec:model-building}); the
endorsing module, which uses heuristic rules to remove possibly
inaccurate suggestions (section~\ref{sec:filling-suggestion}).  To
answer RQ3, we assessed the impact of these two modules on the
effectiveness of \approach.

\subsubsection*{Methodology}
As shown in Table~\ref{tab:component}, in the two sub-columns of
column \emph{Module}, we
considered four variants of \approach, to reflect possible
configurations with the two modules. In the table,
\emph{L} refers to the local modeling module and \emph{E}
refers to the endorsing module; symbols `\ding{51}' and `\ding{55}'
indicate whether the variant of \approach includes or not a certain
module, respectively.  When the local modeling module is disabled (denoted by \approach-L), it
means \approach only uses the global model for prediction; when the
endorsing module is disabled (denoted by \approach-E), it means we do not scrutinize (and
possibly discard) the suggestions
provided by \approach.  If both modules are disabled (denoted by \approach-LE), \approach
becomes a plain BN model trained on the entire set of historical input
instances.  We ran the vanilla version of \approach (i.e., the one
presented in section~\ref{sec:approach}) and the additional variants
using the same settings as in RQ1, and measured effectiveness in
terms of $\mathit{MRR}$ and $\mathit{PCR}$.

\subsubsection*{Results}

\begin{table}[tb]
	\centering
		\scriptsize
	\caption{Effectiveness of \approach with Different Modules}
	\label{tab:component}
		\pgfplotstabletypeset[
	every head row/.style={
		before row={\toprule
			\multirow{4.5}{*}{ID}&\multicolumn{2}{c}{\multirow {2.5}{*}{Module}}&\multicolumn{4}{c}{NCBI}&\multicolumn{4}{c}{PROP}\\
			\cmidrule(lr){4-7}
			\cmidrule(lr){8-11}
			&&&\multicolumn{2}{c}{Sequential}&\multicolumn{2}{c}{Random}&\multicolumn{2}{c}{Sequential}&\multicolumn{2}{c}{Random}\\
			\cmidrule(lr){2-3}
			\cmidrule(lr){4-5}
			\cmidrule(lr){6-7}
			\cmidrule(lr){8-9}
			\cmidrule(l){10-11}
		},
		after row=\midrule,
	},
	every last row/.style={
		after row=\bottomrule
	},
	columns/ID/.style = {column name=},
	columns/Local/.style = {column name=L},
	columns/Filter/.style = {column name=E},
	columns/MRR-S1/.style ={column name=MRR}, 
	columns/Covrg-S1/.style={column name=PCR},  
	columns/MRR-R1/.style ={column name=MRR}, 
	columns/Covrg-R1/.style={column name=PCR},  
	columns/MRR-S2/.style={column name=MRR},  
	columns/Covrg-S2/.style={column name=PCR},  
	columns/MRR-R2/.style={column name=MRR},  
	columns/Covrg-R2/.style={column name=PCR},  
	col sep=comma,
	string type,
	]{data/rq-component.csv}

 \end{table}

As shown in Table~\ref{tab:component}, each module impacts the
effectiveness of \approach. 
The local modeling module improves
the ability of BNs in ranking the correct values ahead of the
incorrect ones, leading to a higher MRR value (while the PCR value
remains equal to 1).
The endorsing module mainly reduces the quantity of inaccurate suggestions made by
different BN models (and therefore reduces the $\mathit{PCR}$ value). 

When we compare \approach (with both modules enabled) with a plain BN
(i.e., \approach-LE) on the
NCBI dataset, \approach improves the $\mathit{MRR}$ value by
\SI{+20}{\pp} ({\LAFFSeqMrr} vs {\BNSeqMrr}) for the sequential
filling scenario and by \SI{+17}{\pp} ({\LAFFRandMrr} vs {\BNRandMrr})
for the random filling scenario; on the PROP dataset the improvement
is smaller (\SI{+4}{\pp} for both scenarios).
Hence, the integration of the local modeling and endorsing modules 
positively affect the effectiveness of \approach.
When we apply the endorsing module on a plain BN, we get an
MRR improvement ranging from \SIrange{+1}{+5}{pp} on the
two datasets (see \approach-LE vs \approach-L).
Even though the local modeling module alone does not affect the number of suggestions, 
the integration of the local modeling in \approach-L leads to a
further reduction of the number of inaccurate suggestions (with PCR dropping by \SIrange{0}{13}{pp});
it improves the $\mathit{MRR}$ value by \SIrange{0}{+19}{pp} on the two 
datasets (see \approach-L vs \approach).

These results can be explained as follows.
As mentioned in section~\ref{sec:filling-suggestion}, the endorsing module 
endorses suggestions based on two heuristics: $\mathit{checkDep}$ (which 
checks if the filled fields are parents of the target field) and $\mathit{sumProb}$ 
(which checks whether the sum of probabilities for the top-$n_r$ suggested values 
is higher than a threshold). 
When the local modeling module is enabled, the local models can learn 
fine-grained dependencies for a field and exclude some useless dependencies 
that are present in a plain BN.
As a result of the absence of the dependencies between filled fields and the target 
field, the $\mathit{checkDep}$ heuristic may endorse more suggestions (i.e., the 
$\mathit{PCR}$ value decreases from \approach-L to \approach), in order to retain 
high-confidence suggestions (i.e., the $\mathit{MRR}$ value increases).
On the PROP dataset, the improvement is not obvious because of the high quality 
of its input instances. As a proprietary dataset from the banking domain, the data 
quality division usually double-checks the data entry to minimize the effect of data 
errors and data conflicts on financial software systems. Compared to a public 
dataset, BNs trained on the PROP dataset can find more meaningful field 
dependencies and also achieve high $\mathit{sumProb}$ values for most 
suggestions; hence the differences in both $\mathit{MRR}$ and $\mathit{PCR}$ values 
are relatively small when different modules are enabled.

\emph{The answer to RQ3 is that  the local modeling module and the
  endorsing module improve the effectiveness of \approach.}

 \subsection{Impact of the Number of Filled Fields (RQ4)}
\label{sec:rq4}

\approach takes as input a set of already-filled fields with their values to suggest the value of the target field.
To answer RQ4, we assessed the impact of the number of filled fields
on the effectiveness of \approach as well as of the ARM and MFM baselines.
We did not compare to the Na{\"i}veDT and FLS baselines. 
Na{\"i}veDT is impractical to be used in a production-grade system due
to the timeout issue during the training phase.
FLS does not use the
information contained in the already-filled fields.

\subsubsection*{Methodology}
We generated new test sets by varying the number of filled fields on
the testing input instances obtained as described in section~\ref{sec:dataset_and_setting}.
To generate a test set with $i$ filled fields for a target $t$, for
each testing input instance, we set the field $t$ as the target and
randomly selected $i$ non-empty fields as filled fields. The
unselected fields were considered as unfilled and  their values  replaced with a dummy value representing empty fields.
Given a data entry form with $t$ targets for automated form filling, we can generate $t$ new test sets with $i$ filled fields, each of which has a different target.
We ran \approach on these new test sets and computed the 
$\mathit{MRR}$ and $\mathit{PCR}$ values for predicting different targets. The results indicate the effectiveness of \approach when $i$ fields are filled.

Following this strategy, we assessed the effectiveness of \approach and the baselines on
the NCBI dataset with one, two, and three filled fields. We discarded
the configuration with four filled fields, since we could only
generate 505 new testing input instances due to the optional fields;
this number is significantly smaller than the number of testing input
instances we obtained for the configurations with one/two/three filled
fields (which have more than \num{17000} new testing input instances)
and might have introduced  bias.

For the PROP dataset, we generated
16 test sets representing the configurations with one to 16 filled
fields; each of them has more than \num{300000} testing input
instances.
However, running the experiment for all testing input instances would be
infeasible on the dedicated server provided by our industrial partner,
which caps the duration of any job to 168 hours.
The issue of the execution time was mainly introduced when evaluating ARM.
As mentioned in section \ref{sec:rq:performance}, ARM may take more than \SI{3}{\s} to provide a suggestion.
This means ARM could take, in the worst case,  $\approx \num{300000}
\times 16 \times \SI{3}{\s} \approx 166\;\text{days}$ to execute on the
testing input instances for all the 16 generated test sets in the PROP
dataset.
Hence, to assess the impact of the number of filled fields on all the
algorithms when using the PROP dataset,
we randomly sampled \num{12600} testing input instances from each of the
16 newly generated test sets.
We chose this number because, in the worst
case, the experiments for all the algorithms could be finished within the 168 hours limit
(e.g., $\num{12600} \times 16 \times\SI{3}{\s}=168\;\text{hours}$ for ARM).

\begin{figure}[tb]\centering
        \pgfmathdeclarefunction{fpumod}{2}{\pgfmathfloatdivide{#1}{#2}\pgfmathfloatint{\pgfmathresult}\pgfmathfloatmultiply{\pgfmathresult}{#2}\pgfmathfloatsubtract{#1}{\pgfmathresult}\pgfmathfloatifapproxequalrel{\pgfmathresult}{#2}{\def\pgfmathresult{4}}{}}

\subfigure[MRR on the NCBI test sets]{
	\label{fig:rq4-ncbi-mrr}
	\begin{tikzpicture}[scale=0.8]
	\begin{axis}[
		legend entries = {MFM, ARM, \approach},
		legend to name={legend},
		legend style={font=\fontsize{6.5}{5}\selectfont, mark=*, mark size = 1pt, nodes={scale=0.8, transform shape}},
		name=border,
boxplot/draw direction=y,
		ylabel={MRR},
		height=6cm,
		ymin=0,ymax=1.05,
		ytick={0, 0.1,0.2,0.3,0.4,0.5,0.6,0.7,0.8,0.9,1,1.10},
		cycle list={{green!60!black},{blue},{red}},
		boxplot={
draw position={1/4+ floor(\plotnumofactualtype/3) + 1/4*mod(\plotnumofactualtype,3)},
box extend=0.2
		},
x=2cm,
xlabel={Number of filled fields},
		xtick={0,1,2,...,10},
		x tick label as interval,
		xticklabels={1, 2, 3},
		x tick label style={
			text width=2.5cm,
			align=center
		},
		scatter/classes={
			laff={red},
			arm={blue},
			mfm={green!60!black}
		}
		]
		
		\addplot table [y index=1, col sep=comma] {data/rq-fields-ncbi-mfm-mrr.csv};
		\addplot table [y index=1, col sep=comma] {data/rq-fields-ncbi-arm-mrr.csv};
		\addplot table [y index=1, col sep=comma] {data/rq-fields-ncbi-laff-mrr.csv};
		\addplot table [y index=2, col sep=comma] {data/rq-fields-ncbi-mfm-mrr.csv};
		\addplot table [y index=2, col sep=comma] {data/rq-fields-ncbi-arm-mrr.csv};
		\addplot table [y index=2, col sep=comma] {data/rq-fields-ncbi-laff-mrr.csv};
		\addplot table [y index=3, col sep=comma] {data/rq-fields-ncbi-mfm-mrr.csv};
		\addplot table [y index=3, col sep=comma] {data/rq-fields-ncbi-arm-mrr.csv};
		\addplot table [y index=3, col sep=comma] {data/rq-fields-ncbi-laff-mrr.csv};
\addplot [scatter, mark=*, only marks, mark size = 0.75pt, scatter src=explicit symbolic] 
		table [
			x expr={1/4 + floor(\coordindex/3) + 1/4*mod(\coordindex,3)},
			y=mean,
			meta=label,col sep=comma
		]{data/rq-fields-ncbi-mean-mrr.csv};
\addplot [scatter, mark=+, only marks, mark size = 1.25pt, scatter src=explicit symbolic] 
		table [
			y=outlier,
			meta=label,col sep=comma
		]{data/rq-fields-ncbi-outlier-mrr.csv};
	\end{axis}
	\node[below right] at (0.0, 4.4) {\ref{legend}};
	\end{tikzpicture}
}
\quad
\subfigure[PCR on the NCBI test sets]{
	\label{fig:rq4-ncbi-pcr}
	\begin{tikzpicture}[scale=0.8]
	\begin{axis}[
		legend entries = {MFM, ARM, \approach},
		legend to name={legend},
		legend style={font=\fontsize{6.5}{5}\selectfont, mark=*, mark size = 1pt, nodes={scale=0.8, transform shape}},
		name=border,
boxplot/draw direction=y,
		ylabel={PCR},
		height=6cm,
		ymin=0,ymax=1.05,
		ytick={0, 0.1,0.2,0.3,0.4,0.5,0.6,0.7,0.8,0.9,1,1.10},
		cycle list={{green!60!black},{blue},{red}},
		boxplot={
			draw position={1/4+ floor(\plotnumofactualtype/3) + 1/4*mod(\plotnumofactualtype,3)},
			box extend=0.2
		},
		x=2cm,
		xlabel={Number of filled fields},
		xtick={0,1,2,...,10},
		x tick label as interval,
		xticklabels={1, 2, 3},
		x tick label style={
			text width=2.5cm,
			align=center
		},
		scatter/classes={
			laff={red},
			arm={blue},
			mfm={green!60!black}
		}
		]	
		
		\addplot table [y index=1, col sep=comma] {data/rq-fields-ncbi-mfm-pcr.csv};
		\addplot table [y index=1, col sep=comma] {data/rq-fields-ncbi-arm-pcr.csv};
		\addplot table [y index=1, col sep=comma] {data/rq-fields-ncbi-laff-pcr.csv};
		\addplot table [y index=2, col sep=comma] {data/rq-fields-ncbi-mfm-pcr.csv};
		\addplot table [y index=2, col sep=comma] {data/rq-fields-ncbi-arm-pcr.csv};
		\addplot table [y index=2, col sep=comma] {data/rq-fields-ncbi-laff-pcr.csv};
		\addplot table [y index=3, col sep=comma] {data/rq-fields-ncbi-mfm-pcr.csv};
		\addplot table [y index=3, col sep=comma] {data/rq-fields-ncbi-arm-pcr.csv};
		\addplot table [y index=3, col sep=comma] {data/rq-fields-ncbi-laff-pcr.csv};
\addplot [scatter, mark=*, only marks, mark size = 0.75pt, scatter src=explicit symbolic] 
		table [
			x expr={1/4 + floor(\coordindex/3) + 1/4*mod(\coordindex,3)},
			y=mean,
			meta=label,col sep=comma
		]{data/rq-fields-ncbi-mean-pcr.csv};
\addplot [scatter, mark=+, only marks, mark size = 1.25pt, scatter src=explicit symbolic] 
		table [
			y=outlier,
			meta=label,col sep=comma
		]{data/rq-fields-ncbi-outlier-pcr.csv};
	\end{axis}	
	\node[below right] at (0.0, 1.5) {\ref{legend}};
	\end{tikzpicture}
}

\subfigure[MRR on the sampled PROP test sets]{
	\label{fig:rq4-prop-small-mrr}
	\begin{tikzpicture}[scale=0.8]
	\begin{axis}[
		legend entries = {MFM, ARM, \approach},
		legend to name={legend},
		legend style={font=\fontsize{6.5}{5}\selectfont, mark=*, mark size = 1pt, nodes={scale=0.8, transform shape}},
		name=border,
boxplot/draw direction=y,
		ylabel={MRR},
		height=6cm,
		ymin=0,ymax=1.05,
		ytick={0, 0.1,0.2,0.3,0.4,0.5,0.6,0.7,0.8,0.9,1,1.10},
		cycle list={{green!60!black},{blue},{red}},
		boxplot={
			draw position={1/4 + floor(\plotnumofactualtype/3) + 1/4*fpumod(\plotnumofactualtype,3)},
			box extend=0.15
		},
		x=0.8cm,
		xlabel={Number of filled fields},
		xtick={0,1,2,...,16},
		x tick label as interval,
		xticklabels={1, 2, 3,...,16},
		x tick label style={
			text width=2.5cm,
			align=center
		},
		scatter/classes={
			laff={red},
			arm={blue},
			mfm={green!60!black}
		}
		]
			
		\addplot table [y index=1, col sep=comma] {data/rq-fields-prop-mfm-mrr.csv};
		\addplot table [y index=1, col sep=comma] {data/rq-fields-prop-arm-mrr.csv};
		\addplot table [y index=1, col sep=comma] {data/rq-fields-prop-laff-mrr.csv};
		\addplot table [y index=2, col sep=comma] {data/rq-fields-prop-mfm-mrr.csv};
		\addplot table [y index=2, col sep=comma] {data/rq-fields-prop-arm-mrr.csv};
		\addplot table [y index=2, col sep=comma] {data/rq-fields-prop-laff-mrr.csv};
		\addplot table [y index=3, col sep=comma] {data/rq-fields-prop-mfm-mrr.csv};
		\addplot table [y index=3, col sep=comma] {data/rq-fields-prop-arm-mrr.csv};
		\addplot table [y index=3, col sep=comma] {data/rq-fields-prop-laff-mrr.csv};
		\addplot table [y index=4, col sep=comma] {data/rq-fields-prop-mfm-mrr.csv};
		\addplot table [y index=4, col sep=comma] {data/rq-fields-prop-arm-mrr.csv};
		\addplot table [y index=4, col sep=comma] {data/rq-fields-prop-laff-mrr.csv};
		\addplot table [y index=5, col sep=comma] {data/rq-fields-prop-mfm-mrr.csv};
		\addplot table [y index=5, col sep=comma] {data/rq-fields-prop-arm-mrr.csv};
		\addplot table [y index=5, col sep=comma] {data/rq-fields-prop-laff-mrr.csv};
		\addplot table [y index=6, col sep=comma] {data/rq-fields-prop-mfm-mrr.csv};
		\addplot table [y index=6, col sep=comma] {data/rq-fields-prop-arm-mrr.csv};
		\addplot table [y index=6, col sep=comma] {data/rq-fields-prop-laff-mrr.csv};
		\addplot table [y index=7, col sep=comma] {data/rq-fields-prop-mfm-mrr.csv};
		\addplot table [y index=7, col sep=comma] {data/rq-fields-prop-arm-mrr.csv};
		\addplot table [y index=7, col sep=comma] {data/rq-fields-prop-laff-mrr.csv};
		\addplot table [y index=8, col sep=comma] {data/rq-fields-prop-mfm-mrr.csv};
		\addplot table [y index=8, col sep=comma] {data/rq-fields-prop-arm-mrr.csv};
		\addplot table [y index=8, col sep=comma] {data/rq-fields-prop-laff-mrr.csv};
		\addplot table [y index=9, col sep=comma] {data/rq-fields-prop-mfm-mrr.csv};
		\addplot table [y index=9, col sep=comma] {data/rq-fields-prop-arm-mrr.csv};
		\addplot table [y index=9, col sep=comma] {data/rq-fields-prop-laff-mrr.csv};
		\addplot table [y index=10, col sep=comma] {data/rq-fields-prop-mfm-mrr.csv};
		\addplot table [y index=10, col sep=comma] {data/rq-fields-prop-arm-mrr.csv};
		\addplot table [y index=10, col sep=comma] {data/rq-fields-prop-laff-mrr.csv};
		\addplot table [y index=11, col sep=comma] {data/rq-fields-prop-mfm-mrr.csv};
		\addplot table [y index=11, col sep=comma] {data/rq-fields-prop-arm-mrr.csv};
		\addplot table [y index=11, col sep=comma] {data/rq-fields-prop-laff-mrr.csv};
		\addplot table [y index=12, col sep=comma] {data/rq-fields-prop-mfm-mrr.csv};
		\addplot table [y index=12, col sep=comma] {data/rq-fields-prop-arm-mrr.csv};
		\addplot table [y index=12, col sep=comma] {data/rq-fields-prop-laff-mrr.csv};
		\addplot table [y index=13, col sep=comma] {data/rq-fields-prop-mfm-mrr.csv};
		\addplot table [y index=13, col sep=comma] {data/rq-fields-prop-arm-mrr.csv};
		\addplot table [y index=13, col sep=comma] {data/rq-fields-prop-laff-mrr.csv};
		\addplot table [y index=14, col sep=comma] {data/rq-fields-prop-mfm-mrr.csv};
		\addplot table [y index=14, col sep=comma] {data/rq-fields-prop-arm-mrr.csv};
		\addplot table [y index=14, col sep=comma] {data/rq-fields-prop-laff-mrr.csv};
		\addplot table [y index=15, col sep=comma] {data/rq-fields-prop-mfm-mrr.csv};
		\addplot table [y index=15, col sep=comma] {data/rq-fields-prop-arm-mrr.csv};
		\addplot table [y index=15, col sep=comma] {data/rq-fields-prop-laff-mrr.csv};
		\addplot table [y index=16, col sep=comma] {data/rq-fields-prop-mfm-mrr.csv};
		\addplot table [y index=16, col sep=comma] {data/rq-fields-prop-arm-mrr.csv};
		\addplot table [y index=16, col sep=comma] {data/rq-fields-prop-laff-mrr.csv};
		
\addplot [scatter, mark=*, only marks, mark size = 0.75pt, scatter src=explicit symbolic] 
		table [
		x expr={1/4 + floor(\coordindex/3) + 1/4*fpumod(\coordindex,3)},
		y=mean,
		meta=label,col sep=comma
		]{data/rq-fields-prop-mean-mrr.csv};
\addplot [scatter, mark=+, only marks, mark size = 1.25pt, scatter src=explicit symbolic] 
		table [
		y=outlier,
		meta=label,col sep=comma
		]{data/rq-fields-prop-outlier-mrr.csv};
	\end{axis}
	\node[below right] at (13.7, 1.5) {\ref{legend}};
	\end{tikzpicture}
}

\subfigure[PCR on the sampled PROP test sets]{
	\label{fig:rq4-prop-small-pcr}\begin{tikzpicture}[scale=0.8]
	\begin{axis}[
	legend entries = {MFM, ARM, \approach},
	legend to name={legend},
	legend style={font=\fontsize{6.5}{5}\selectfont, mark=*, mark size = 0.75pt, nodes={scale=0.8, transform shape}},
	name=border,
boxplot/draw direction=y,
	ylabel={PCR},
	height=6cm,
	ymin=0,ymax=1.05,
	ytick={0, 0.1,0.2,0.3,0.4,0.5,0.6,0.7,0.8,0.9,1,1.10},
	cycle list={{green!60!black},{blue},{red}},
	boxplot={
		draw position={1/4 + floor(\plotnumofactualtype/3) + 1/4*fpumod(\plotnumofactualtype,3)},
		box extend=0.15
	},
	x=0.8cm,
	xlabel={Number of filled fields},
	xtick={0,1,2,...,16},
	x tick label as interval,
	xticklabels={1, 2, 3,...,16},
	x tick label style={
		text width=2.5cm,
		align=center
	},
	scatter/classes={
		laff={red},
		arm={blue},
		mfm={green!60!black}
	}
	]

	\addplot table [y index=1, col sep=comma] {data/rq-fields-prop-mfm-pcr.csv};
	\addplot table [y index=1, col sep=comma] {data/rq-fields-prop-arm-pcr.csv};
	\addplot table [y index=1, col sep=comma] {data/rq-fields-prop-laff-pcr.csv};
	\addplot table [y index=2, col sep=comma] {data/rq-fields-prop-mfm-pcr.csv};
	\addplot table [y index=2, col sep=comma] {data/rq-fields-prop-arm-pcr.csv};
	\addplot table [y index=2, col sep=comma] {data/rq-fields-prop-laff-pcr.csv};
	\addplot table [y index=3, col sep=comma] {data/rq-fields-prop-mfm-pcr.csv};
	\addplot table [y index=3, col sep=comma] {data/rq-fields-prop-arm-pcr.csv};
	\addplot table [y index=3, col sep=comma] {data/rq-fields-prop-laff-pcr.csv};
	\addplot table [y index=4, col sep=comma] {data/rq-fields-prop-mfm-pcr.csv};
	\addplot table [y index=4, col sep=comma] {data/rq-fields-prop-arm-pcr.csv};
	\addplot table [y index=4, col sep=comma] {data/rq-fields-prop-laff-pcr.csv};
	\addplot table [y index=5, col sep=comma] {data/rq-fields-prop-mfm-pcr.csv};
	\addplot table [y index=5, col sep=comma] {data/rq-fields-prop-arm-pcr.csv};
	\addplot table [y index=5, col sep=comma] {data/rq-fields-prop-laff-pcr.csv};
	\addplot table [y index=6, col sep=comma] {data/rq-fields-prop-mfm-pcr.csv};
	\addplot table [y index=6, col sep=comma] {data/rq-fields-prop-arm-pcr.csv};
	\addplot table [y index=6, col sep=comma] {data/rq-fields-prop-laff-pcr.csv};
	\addplot table [y index=7, col sep=comma] {data/rq-fields-prop-mfm-pcr.csv};
	\addplot table [y index=7, col sep=comma] {data/rq-fields-prop-arm-pcr.csv};
	\addplot table [y index=7, col sep=comma] {data/rq-fields-prop-laff-pcr.csv};
	\addplot table [y index=8, col sep=comma] {data/rq-fields-prop-mfm-pcr.csv};
	\addplot table [y index=8, col sep=comma] {data/rq-fields-prop-arm-pcr.csv};
	\addplot table [y index=8, col sep=comma] {data/rq-fields-prop-laff-pcr.csv};
	\addplot table [y index=9, col sep=comma] {data/rq-fields-prop-mfm-pcr.csv};
	\addplot table [y index=9, col sep=comma] {data/rq-fields-prop-arm-pcr.csv};
	\addplot table [y index=9, col sep=comma] {data/rq-fields-prop-laff-pcr.csv};
	\addplot table [y index=10, col sep=comma] {data/rq-fields-prop-mfm-pcr.csv};
	\addplot table [y index=10, col sep=comma] {data/rq-fields-prop-arm-pcr.csv};
	\addplot table [y index=10, col sep=comma] {data/rq-fields-prop-laff-pcr.csv};
	\addplot table [y index=11, col sep=comma] {data/rq-fields-prop-mfm-pcr.csv};
	\addplot table [y index=11, col sep=comma] {data/rq-fields-prop-arm-pcr.csv};
	\addplot table [y index=11, col sep=comma] {data/rq-fields-prop-laff-pcr.csv};
	\addplot table [y index=12, col sep=comma] {data/rq-fields-prop-mfm-pcr.csv};
	\addplot table [y index=12, col sep=comma] {data/rq-fields-prop-arm-pcr.csv};
	\addplot table [y index=12, col sep=comma] {data/rq-fields-prop-laff-pcr.csv};
	\addplot table [y index=13, col sep=comma] {data/rq-fields-prop-mfm-pcr.csv};
	\addplot table [y index=13, col sep=comma] {data/rq-fields-prop-arm-pcr.csv};
	\addplot table [y index=13, col sep=comma] {data/rq-fields-prop-laff-pcr.csv};
	\addplot table [y index=14, col sep=comma] {data/rq-fields-prop-mfm-pcr.csv};
	\addplot table [y index=14, col sep=comma] {data/rq-fields-prop-arm-pcr.csv};
	\addplot table [y index=14, col sep=comma] {data/rq-fields-prop-laff-pcr.csv};
	\addplot table [y index=15, col sep=comma] {data/rq-fields-prop-mfm-pcr.csv};
	\addplot table [y index=15, col sep=comma] {data/rq-fields-prop-arm-pcr.csv};
	\addplot table [y index=15, col sep=comma] {data/rq-fields-prop-laff-pcr.csv};
	\addplot table [y index=16, col sep=comma] {data/rq-fields-prop-mfm-pcr.csv};
	\addplot table [y index=16, col sep=comma] {data/rq-fields-prop-arm-pcr.csv};
	\addplot table [y index=16, col sep=comma] {data/rq-fields-prop-laff-pcr.csv};
	
\addplot [scatter, mark=*, only marks, mark size = 0.75pt, scatter src=explicit symbolic] 
	table [
	x expr={1/4 + floor(\coordindex/3) + 1/4*fpumod(\coordindex,3)},
	y=mean,
	meta=label, col sep=comma
	]{data/rq-fields-prop-mean-pcr.csv};
\addplot [scatter, mark=+, only marks, mark size = 1.25pt, scatter src=explicit symbolic] 
	table [
	y=outlier,
	meta=label,col sep=comma
	]{data/rq-fields-prop-outlier-pcr.csv};
	\end{axis}
	\node[below right] at (13.7, 1.5) {\ref{legend}};
	\end{tikzpicture}
}
         \caption{Effectiveness of \approach with different number of filled fields on the NCBI  and PROP datasets}\label{fig:rq4-number-filled-fields}\end{figure}
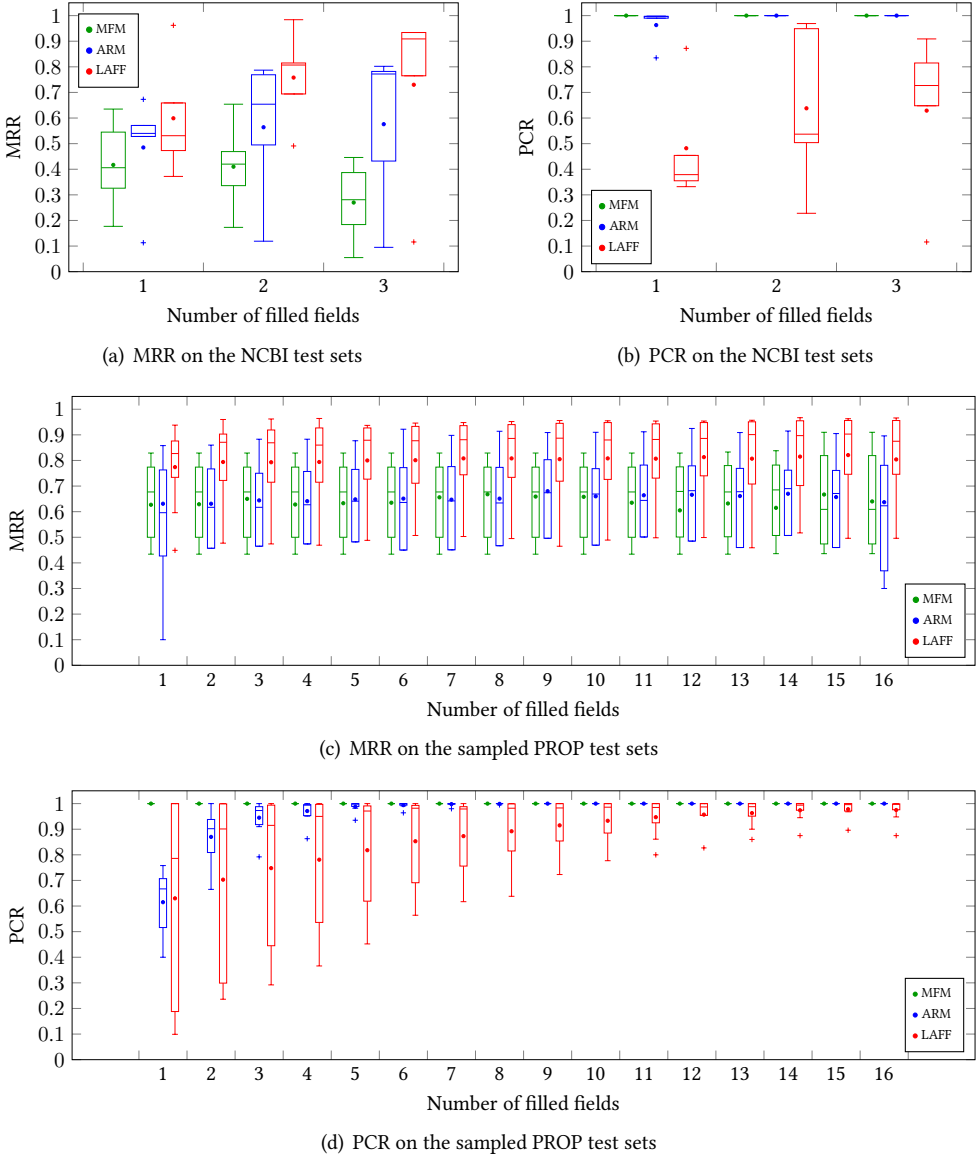

\subsubsection*{Results}
Figure~\ref{fig:rq4-number-filled-fields}  shows the results of
running the form filling algorithms on the NCBI and PROP datasets with
different numbers of filled fields.
The x-axis of the figures represents the number of filled fields and the y-axis shows the $\mathit{MRR}$ or $\mathit{PCR}$ value of each form filling algorithm.

In terms of accuracy (in
Figure~\ref{fig:rq4-ncbi-mrr} and Figure~\ref{fig:rq4-prop-small-mrr}), the $\mathit{MRR}$ values of \approach and ARM increase when more fields are filled.
For example, the average $\mathit{MRR}$ value of \approach increases from
\LAFFOneFFNcbiMRR $\pm 0.227$ to \LAFFAllFFNcbiMRR $\pm 0.350$ on the NCBI dataset
and from \LAFFOneFFPropMRR $\pm 0.160$ to \LAFFAllFFPropMRR $\pm 0.190$ on the PROP dataset as the number of filled fields increases from one to three and
from one to 16, respectively.
In contrast, MFM shows a different trend. In theory, MFM is not
affected by the number of filled fields, as it always provides the
same suggestion for a target field, based on the most frequent values
filled in the target field in the past.
However, our experiments show a variation in $\mathit{MRR}$ as
the number of fields increases. 
More specifically, since the testing input instances generated for each number of filled fields are different, 
the $\mathit{MRR}$ value of MFM decreases when the correct values for
a target in the generated testing input instances are not the most
frequent historical values.
Overall, according to the boxplots, when varying the number of filled fields, the $\mathit{MRR}$ values of \approach for predicting different targets remain higher than those of MFM and ARM.
Given a partially filled form with only one filled field,
\approach outperforms MFM and ARM by \SI{+12}{\pp} and \SI{+11}{\pp}, 
respectively, on the NCBI dataset; it also outperforms both MFM and ARM by
\SI{+14}{\pp} on the PROP dataset.

As for coverage, the $\mathit{PCR}$ value of the baselines is 1 for
the majority of the numbers of filled fields (as shown in
Figure~\ref{fig:rq4-ncbi-pcr} and Figure~\ref{fig:rq4-prop-small-pcr}).
For \approach, the PCR value increases as more fields are filled.
However, we find that the PCR values of \approach for different target fields vary significantly, especially with a small number of filled fields. 
For example, with one filled field, the PCR values of \approach are \LAFFOneFFNcbiPCR $\pm 0.22$ and \LAFFOneFFPropPCR $\pm 0.41$ respectively on the NCBI and PROP datasets.
This is because, when the number of filled fields is small, the filled fields may not always provide enough knowledge (i.e., the information of dependent field values) for \approach to predict all the target fields. 
For some targets, the endorsing module of \approach may filter out many suggestions.
When the number of filled fields increases,  \approach gets more
information from the input instances to provide suggestions that can
be endorsed.
For example, the PCR value of \approach increases to \LAFFAllFFPropPCR $\pm 0.04$ on the PROP dataset with 16 filled fields.
Overall, The PCR values show that \approach could correctly endorse the
suggestions when the form is partially filled, which helps \approach
achieve higher $\mathit{MRR}$ values than those of the baselines.

\emph{The answer to RQ4 is that the effectiveness (in terms of
  $\mathit{MRR}$ and $\mathit{PCR}$) of \approach increases as more
  fields are filled. Further, \approach can better handle partially filled
  forms than state-of-the-art algorithms.}

 \subsection{Impact of the Size of the Training Set (RQ5)}
\label{sec:rq5}

\approach is a learning-based approach that requires a training set (i.e., historical input instances) to train machine learning models. To answer RQ5, we assessed the impact of the size of the training set on the effectiveness of \approach.

\subsubsection*{Methodology}
We evaluated \approach by varying the size of the training set from
\SIrange{10}{100}{\percent} of the historical input instances included in
the training set, with a step of 10\%.
For the NCBI dataset, the size of the sampled training set ranged from
\numrange{5928}{59284} historical input instances (where \num{59284}
is 80\% of the \num{74105} input instances in the dataset).
For the PROP dataset, the size of the sampled training set ranged from
\numrange{13955}{139557} historical input instances (where \num{139557}
is 80\% of the \num{174446} input instances in the dataset).

For a given percentage value $p$, we randomly sampled $p$\% of the
historical input instances in the training set to form a smaller training set.
We trained \approach on the sampled training set and used the trained
model to conduct automated form filling on the testing input instances
obtained as described in section~\ref{sec:dataset_and_setting}.

As for RQ1, we measured the effectiveness of
\approach in terms of $\mathit{MRR}$ and $\mathit{PCR}$.

\subsubsection*{Results}

\begin{figure}[tb]\centering
	\subfigure[Sequential  filling scenario on NCBI dataset ]{\label{fig:rq5-ncbi-seq}\begin{tikzpicture}[scale=0.7]
		\begin{axis}[
			xlabel={Number of historical input instances},
			ylabel={MRR and PCR},
			ylabel near ticks,
			xmin=0, xmax=61,
			ymin=0, ymax=1.05,
			xtick={0, 6,12,18,24,30,36,42,48,54,60},
			xticklabel={$\pgfmathprintnumber{\tick}k$},
			legend pos=north east,
			ymajorgrids=true,
			grid style=dashed,
			]
			
			\addplot [color=blue ,mark=x] table [x=Size, y=MRR, col sep=comma] {data/rq-sample-seq-ncbi.csv};\addlegendentry{MRR};
			\addplot [color=red ,mark=triangle]  table [x=Size, y=PCR, col sep=comma] {data/rq-sample-seq-ncbi.csv};\addlegendentry{PCR}
		\end{axis}
\end{tikzpicture}}
\quad
\subfigure[Random filling scenario  on NCBI dataset]{\label{fig:rq5-ncbi-rand}\begin{tikzpicture}[scale=0.7]
		\begin{axis}[
			xlabel={Number of historical input instances},
			ylabel={MRR and PCR},
			ylabel near ticks,
			xmin=0, xmax=61,
			ymin=0, ymax=1.05,
			xtick={0, 6,12,18,24,30,36,42,48,54,60},
			xticklabel={$\pgfmathprintnumber{\tick} k$},
			legend pos=south east,
			ymajorgrids=true,
			grid style=dashed,
			]
			
			\addplot  [color=blue ,mark=x] table [x=Size, y=MRR, col sep=comma] {data/rq-sample-rand-ncbi.csv};\addlegendentry{MRR};
			\addplot [color=red ,mark=triangle] table [x=Size, y=PCR, col sep=comma] {data/rq-sample-rand-ncbi.csv};\addlegendentry{PCR}
		\end{axis}
\end{tikzpicture}}
\quad

\subfigure[ Sequential  filling scenario on PROP dataset]{\label{fig:rq5-prop-seq}\begin{tikzpicture}[scale=0.7]
		\begin{axis}[
			xlabel={Number of historical input instances},
			ylabel={MRR and PCR},
			ylabel near ticks,
			xmin=0, xmax=145,
			ymin=0.5, ymax=1.05,
			xtick={0,14,28,42,56,70,84,98,112,126,140},
			xticklabel={$\pgfmathprintnumber{\tick}k$},
			legend pos=north east,
			ymajorgrids=true,
			grid style=dashed,
			]
			
			\addplot [color=blue ,mark=x] table [x=Size, y=MRR, col sep=comma] {data/rq-sample-seq-prop.csv};\addlegendentry{MRR};
			\addplot [color=red ,mark=triangle]  table [x=Size, y=PCR, col sep=comma] {data/rq-sample-seq-prop.csv};\addlegendentry{PCR}
		\end{axis}
\end{tikzpicture}}
\quad
\subfigure[ Random filling scenario on PROP dataset]{\label{fig:rq5-prop-rand}\begin{tikzpicture}[scale=0.7]
		\begin{axis}[
			xlabel={Number of historical input instances},
			ylabel={MRR and PCR},
			ylabel near ticks,
			xmin=0, xmax=145,
			ymin=0.5, ymax=1.05,
			xtick={0,14,28,42,56,70,84,98,112,126,140},
			xticklabel={$\pgfmathprintnumber{\tick} k$},
			legend pos=south east,
			ymajorgrids=true,
			grid style=dashed,
			]
			
			\addplot  [color=blue ,mark=x] table [x=Size, y=MRR, col sep=comma] {data/rq-sample-rand-prop.csv};\addlegendentry{MRR};
			\addplot [color=red ,mark=triangle] table [x=Size, y=PCR, col sep=comma] {data/rq-sample-rand-prop.csv};\addlegendentry{PCR}
		\end{axis}
\end{tikzpicture}}

 	\caption{Effectiveness of \approach with different training set sizes on the NCBI and PROP datasets}\label{fig:rq5-size-of-data}\end{figure}
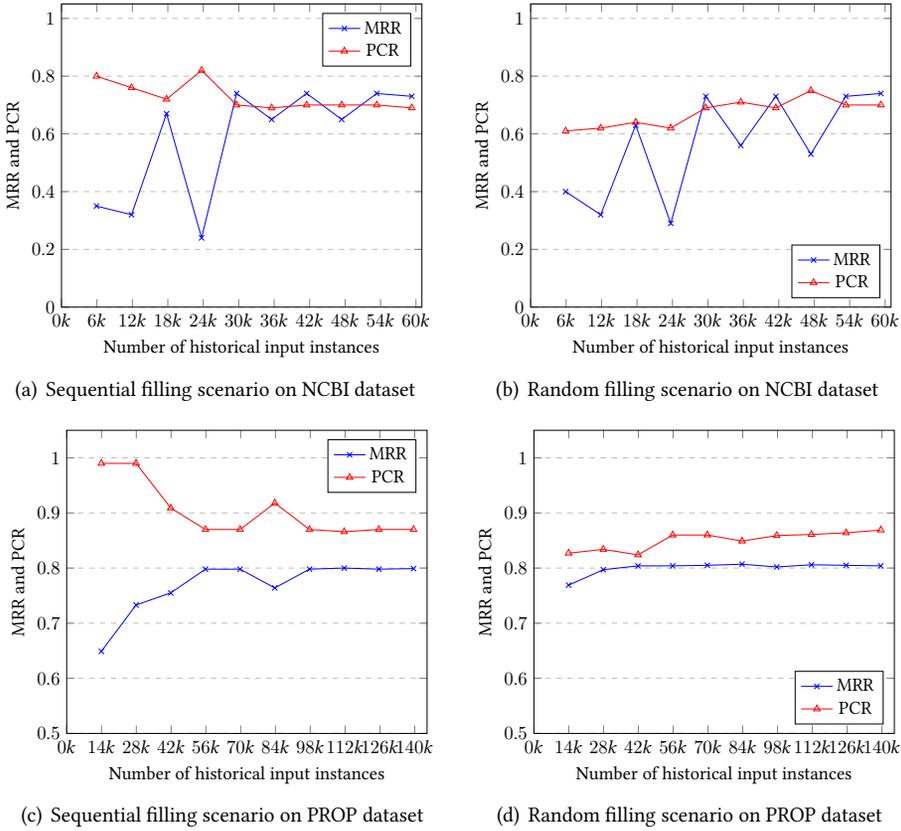

Figure~\ref{fig:rq5-size-of-data}  shows the
results of \approach on the NCBI and PROP datasets with different training set sizes.
The x-axis of the figure represents the number of historical input
instances; 
the y-axis shows the $\mathit{MRR}$ and $\mathit{PCR}$ values of
\approach under different filling scenarios.

As shown in the figure, the value of $\mathit{MRR}$ increases and gradually becomes stable on the two datasets when the size of the training set increases. 
For the NCBI dataset, the value of $\mathit{MRR}$ is low and
fluctuates significantly with a training set of less than \num{30000}
historical input instances; the majority of $\mathit{MRR}$ values are
lower than 0.40 for the two form filling scenarios. When we have more
historical input instances (between \num{30000} and \num{60000}), the
value of $\mathit{MRR}$ increases and becomes more stable, ranging
from \numrange{\NCBISeqMRRLow}{\NCBISeqMRRHigh} for the sequential
filling scenario and from \numrange{\NCBIRandMRRLow}{\NCBIRandMRRHigh}
for the random filling scenario. The value of $\mathit{MRR}$ has a
similar trend on the PROP dataset; it gradually increases and then becomes stable (between {\PROPSeqMRRLow} and {\PROPRandMRRHigh}) when the number of historical input instances is higher than \num{56000}.

In terms of $\mathit{PCR}$, the endorser module of \approach works
poorly with a small training set. \approach either keeps the majority
of suggestions (for the sequential filling scenario) or wrongly
removes many suggestions (for the random filling scenario), leading to
low $\mathit{MRR}$ values. As the size of the training set increases
to more than \num{30000} (for the NCBI dataset) and  \num{56000} (for
the PROP dataset), \approach is able to remove more inaccurate suggestions, achieving a $\mathit{PCR}$ value between {\NCBISeqPCRLow} and {\NCBIRandPCRHigh} and between {\PROPRandPCRLow} and {\PROPSeqPCRHigh}, respectively on the two datasets.

When comparing the results of \approach across the two datasets,
we observe a more significant fluctuation of $\mathit{MRR}$ values on the NCBI dataset than that on the PROP dataset as the size of the training set increases.
This is caused by the data quality of the NCBI dataset. In contrast to
the proprietary dataset PROP, there is no field constraint or
additional check on the values filled in the NCBI data entry
form. This means that, when more historical input instances are added
to the training set, one may also introduce many conflicting or erroneous field values, which increase the uncertainty of \approach when predicting on the NCBI dataset.

\emph{
The answer to RQ5 is that the size of the training set affects the effectiveness of  \approach. $\mathit{MRR}$ values increase on both datasets when the size of the training set increases; more suggestions are also correctly endorsed. With more than \num{56000} historical input instances, \approach achieves accurate suggestions on both datasets}

 \subsection{Threats to Validity}

The size of the pool of historical input instances can affect the
effectiveness of \approach, a common issue among learning 
algorithms. Nevertheless, we do not expect this to be a strong
limitation since it targets data entry functionalities in 
enterprise software, in which one can expect thousands of input
instances per day, as it is the case for the system used by our
industrial partner.

Another threat to the validity is the choice of the
value of the endorser threshold $\theta$. With a higher threshold,
\approach will filter out more suggestions (resulting in a lower PCR
value), only keeping the ones with a high predicted probability
(resulting in higher MRR value). Hence, this threshold reflects the
degree of uncertainty users are willing to accept regarding the
suggestions provided by \approach. To mitigate this threat, we selected a
threshold value based on the feedback received by data entry operators
and data quality engineers of our partner.

The choice of the deployment method for \approach can impact its
performance in terms of prediction time. In our experiments, we
deployed \approach locally; using a different deployment (e.g.,
cloud-based) could lead to different results, since the prediction
time would be  affected by
many other factors, such as the DNS lookup time, the connection time, and the
data transmission time~\cite{cardellini2000geographic}.
Since the prediction time
of \approach is less than \SI{\LAFFPredictMax}{\milli\s}, an
application using a non-local deployment would have
enough leeway to optimize these factors and provide seamless interactions for
users, complying with human-computer interaction
principles~\cite{heeter2000interactivity} (i.e., with a response time
less than \SI{1}{\s}).
As part of future work, we plan to assess the performance of \approach
under different deployment configurations.

To increase the external validity of our results, \approach should be
further studied on other datasets, possibly from other domains.  To
partially mitigate this threat, we use two different types of datasets
to evaluate \approach, including both a public and a proprietary
dataset, and the corresponding data entry forms. Meanwhile, we
simulated two form filling scenarios (sequential and random filling)
that are plausible during a real data-entry session.  As part of
future work, we plan to conduct a user study using different datasets
and data entry forms, to analyze the effect of \approach on reducing
form filling time and input errors.
Another external threat is our implementation of the four algorithms
(MFM, ARM, Na{\"i}veDT, and FLS) to which we compared, 
which may be different from the original definitions; 
to mitigate this threat, two of the authors cross-reviewed the
implementations, taking into account the relevant literature (when available).

\subsection{Data Availability}\label{sec:data-availability}

The implementation of \approach,
the  NCBI dataset, and the scripts used for the evaluation are available at \url{https://figshare.com/s/b191fcb5ee7f3ad634bf}; \approach is distributed under the MIT license. The PROP dataset cannot be
distributed due to an NDA.

\section{Discussion}
\label{sec:discussion}

\subsection{Usefulness}
\label{sec:usefulness}

The fundamental question we seek to answer is whether \approach can help users fill forms.
To answer this question, we evaluated \approach with two real-world form filling
datasets, one from the biomedical domain and another from the banking domain. 
The results show that \approach
outperforms state-of-the-art form filling algorithms in providing a
larger number (with a $\mathit{PCR}$ value over {\LAFFRandCov}) of accurate
suggestions (with a $\mathit{MRR}$ value over  {\LAFFRandMrr}).
The $\mathit{MRR}$ value reflects the ability of \approach in avoiding inaccurate suggestions.
For example, considering a list with three suggested values, 
if the correct value is in the top-1, top-2, and top-3 of the list, 
the corresponding $\mathit{MRR}$ value is 1, 0.5, and 0.33, respectively; 
when all the values are incorrect, the $\mathit{MRR}$ value is 0.
In the context of our experiments, an $\mathit{MRR}$ value of
at least {\LAFFRandMrr} indicates that, for more than 73\% of suggestions, 
\approach can help users find the correct value from the top-ranked ones.
The $\mathit{PCR}$ value indicates the number of suggestions made by \approach. 
A $\mathit{PCR}$ value over {\LAFFRandCov} indicates that 
\approach can confidently make suggestions for more than 70\% of target fields, 
where the correct values are usually ranked before the incorrect ones.
According to a previous study~\cite{chen2011usher}, 
as users are presented with more candidate values for selection, 
they tend to make more mistakes, which makes form filling a
frustrating activity. 
Hence, we speculate \approach reduces the mental load of users in filling forms 
by helping them go through fewer candidate values (top-k\%) before finding the
correct one; further user studies are required to corroborate this hypothesis.

  We can interpret the above results from a point of view of
  usefulness as follows.
  On the one hand, as shown in RQ1, with an $\mathit{MRR}$ value over
  {\LAFFRandMrr}, \SIrange{79.0}{90.1}{\percent} of the suggestions made by
  \approach allow users to find the correct value among the 5\%
  top-ranked suggested values on the two datasets.  This means that
  when \approach makes suggestions, in \SIrange{79.0}{90.1}{\percent} of the
  cases, it leads to at least 95\% effort saving when browsing
  possible values, since users need only to check the top 5\% most
  likely items recommended by \approach.  For example, on our
  datasets, users need to check between one (for field ``activity''
  having 13 candidate values) and ten (for field ``country'' having
  206 candidate values) values before finding the correct one.
  On the other hand,
according to our analysis in RQ1, a frequency-based method like MFM
  achieves $\mathit{MRR}$ values ranging from
  \numrange{\MFSeqMrr}{\MFSeqMrrBGL} on the NCBI and PROP datasets.
  Hence, when comparing \approach with the widely-used data entry
  solution MFM, an $\mathit{MRR}$ improvement of 
  \SIrange{+14}{+32}{\pp} suggests that \approach can better help
  users select candidate values in data entry forms.

 \subsection{Practical Implications}
\label{sec:implication}

This subsection discusses the practical implications of \approach for
its different stakeholders: software developers, end users, system
administrators, and researchers.

\subsubsection{Software Developers}\label{sec:implications:sw-dev}
Automated form filling is a common requirement for software systems.
Some popular languages and APIs (e.g., HTML~\cite{w3chtml2021html} and
Android APIs~\cite{android2021android}) also provide pre-fill or
auto-completion frameworks, for which customized form filling
strategies can be implemented.  In its current version, \approach is a
stand-alone tool that developers can integrate into their data entry
form implementations as an effective and efficient strategy for
filling categorical fields (e.g., Listboxes and Dropdown Lists).
Since \approach shows a higher accuracy than state-of-the-art
approaches, it can be used in data-reliant enterprise systems across 
different domains, especially when there are constraints on sharing or
accessing data of other software systems.

As for adopting \approach in production, in terms of execution time, a training time of about one hour allows \approach to compute
daily updates for its models, empowered with the information derived from
new input instances.
Moreover, 
considering the interactive nature of data-entry applications, 
we use a model selection strategy to select the suitable local or global model for prediction from multiple models. 
Compared with many learning strategies that consider the predictions of multiple models jointly (e.g., ensemble learning),
our strategy could significantly reduce prediction time in
  specific deployment configurations.
For example, the worst-case prediction time of ensemble learning is,
when using a
single-thread configuration, the sum of the prediction
time of all local models. This could be reduced to the maximum
of the prediction time of all local models by using multiple threads
in parallel.
This strategy is viable when the trained models are deployed on a
powerful server, and the prediction is performed server-side.
However, some enterprises (as it is the case for our industrial
partner) might prefer to integrate a form filling approach like \approach
\emph{locally}, on the machines used by data entry operators, to avoid network
latency during data entry sessions. In such a deployment configuration,
using multiple threads could significantly slow down the data entry
session, especially on machines with a limited computing power. 
For these reasons, we designed \approach with a lightweight model selection
strategy, instead of using alternatives like ensemble learning.

Integrating \approach into a data entry form requires providing a mapping between 
the field names and the column names in the dataset.
This mapping needs to be identified only once and therefore does not
have much impact on the practical application of \approach.
We remark that, in modern applications built using the MVC pattern for
the UI and an Object-Relational Mapping (ORM) framework for data persistence~\cite{chen2016empirical}, 
the mapping can be retrieved from the ORM framework configuration. 
Other sources for the mapping can be software design documentation, 
such as the database schema and the description of the UI widgets in
the data entry forms.

\subsubsection{End Users}
\label{sec:implications:end-users}
Form filling is time-consuming and error-prone for end users, which
can cause more than half of data errors~\cite{qian2020trend}.  These
errors seriously affect data-reliant software systems and even cause
loss of human
life~\cite{american2005data,qian2020trend,khajouei2010impact,bis2003general}.
In this study, we have proposed \approach to improve the accuracy and
efficiency of the data entry process executed by end users, when
filling categorical fields.  First, \approach uses an endorser module
to alleviate the cognitive load on users caused by wrong suggestions;
this module significantly improves the accuracy of suggestions.
Second, \approach helps users focus on the most-likely candidate
values in a list of values.  It decreases the number of candidate
values users need to browse.  \emph{We follow the
  practice of designing an effective recommender system} (e.g., a form
filling system) in software engineering, which is to avoid
``helpless'' suggestions to be ignored by users, but provide
them with a large number of ``helpful''
suggestions~\cite{robillard2009recommendation}.

Furthermore, in the experiments, \approach can provide suggestions
within at most {\SI{\LAFFPredictMax}{\milli\s}} for each target
field, 
which enables end users to conduct seamless interaction with the data entry form.

\subsubsection{System Administrators}\label{sec:implications:sys-admin}
The deployment of \approach in production requires system
administrators to configure its parameters. This can be achieved
with the help of domain experts, based on their domain knowledge.
A group of configurable parameters is set in the pre-processing step.
We implement this step based on best practices for predictive data mining~\cite{Alexandropoulos2019DataPI}.
A threshold that can affect auto-filling effectiveness is
$\theta$, for the endorser module.  This threshold
reflects how much uncertainty domain experts are willing to
accept regarding the suggestions provided by \approach.  Such 
configuration allows domain experts
to use \approach according to their requirements and application scenarios.

Overall, the configuration parameters represent the only
domain-specific aspect of \approach. Everything else about it is
domain-agnostic. Deploying \approach for a new
form only requires to collect the input instances of the form for a
certain period of time and use these instances to train the form filling
models.

\subsubsection{Researchers}
In this paper, we use a local modeling module to effectively learn the fine-grained dependencies on historical input instances.
Since existing approaches in software engineering perform local modeling on numerical data instances (e.g., software metrics~\cite{menzies2011local}),
we propose a novel solution to solve the problem of using local
modeling on categorical data instances.
We speculate that our proposed solution can inspire the adoption of
local modeling for different data types in many software engineering tasks.
In addition, we use an endorser module to decide if suggestions are
accurate enough to be provided to end users.
Such a module is important for algorithms where 100\% accuracy cannot
be achieved in practice.
For example, during form filling, 
predicting all the other fields based on only one filled field may
lead to inaccurate suggestions.
In this case, it is more practical to automatically remove
inaccurate suggestions using an endorser module.
We believe the endorser-based architecture discussed in
section~\ref{sec:filling-suggestion} can be adopted by other
recommender systems.

Furthermore, the error analysis for RQ1 described in
section~\ref{sec:rq1} suggests possible research directions, such as 
learning from users' corrections 
and using multi-objective optimization for form refactoring (with a new
order of fields) for improving form filling suggestions.

 \subsection{Limitations}
\label{sec:limitation}

\subsubsection{Type of Fields}
In this work, we have focused  on predicting the values of categorical
fields, based on the historical information available from the same software system;
\approach does not work with other types of fields.
Due to the unique characteristics of each type of field,
distinct solutions have been proposed in the literature 
(e.g., text auto-completion for textual fields, data testing for numerical fields).
However, these solutions cannot help users  fill categorical fields,
which is a critical task during form filling.
We acknowledge that other types of fields may contain critical data
that could cause major problems. However, as discussed in
section~\ref{sec:introduction}, 
empirical studies show that selection errors lead to more than half (54.5\%) of the data errors in a software system~\cite{qian2020trend}.
For example, the selection of wrong drugs~\cite{khajouei2010impact} or
the wrong modality of care~\cite{qian2020trend} in medical record systems 
can even cause loss of human life.
In addition, as shown in the NCBI and PROP datasets (in section~\ref{sec:dataset_and_setting}), 
the majority of categorical fields we evaluated (with the number of candidate values ranging from \numrange{13}{208}) are related to certain domains or business processes.
These fields are more difficult to fill than fields such as ``sex''
and ``age'', since users need to understand the meaning of candidate values.
Errors in these fields can cause significant  business problems.
For example, the selection of a wrong ``field of activity'' when opening a bank account 
may cause business loss between the company and the bank.
By knowing the actual
``field of activity'' of the company, the bank could have offered targeted products to its
customer since the beginning of the business relation.
Hence, there is a need for a semi-automated method that supports and guides users when filling categorical fields.

\subsubsection{Cognitive Load on Users}
The suggestions made by automated form filling tools like \approach
can increase the cognitive load on users, when the tool provides a
long list of suggested values/options for users to check or when the
suggestions do not include the correct value.
To reduce the cognitive load in the first case,
we configured \approach to suggest the top 5\% most likely candidate values  for each target field,
rather than reordering all the candidate values based on their probability.
Although \approach does not directly reduce the number of candidate values in a field,
it highlights the top 5\% values, on which users can focus.
As for the second case of cognitive load (suggestions not including the correct value), \approach includes an endorser module to
decide whether the suggestions are accurate enough to be returned to users.
As shown in \tablename~\ref{tab:component} and as part of the answer to RQ3, the endorser module significantly reduces the number of possibly inaccurate suggestions while increasing the $\mathit{MRR}$ values.
As a result, our experiments show that \SIrange{79.0}{90.1}{\percent} of the \approach  suggestions allow users to find the correct value among the top 5\% suggested values.

\subsubsection{Fields with Semantic Overlap}\label{sec:limitations:overlap}
Candidate values in categorical fields could have semantic overlaps.
For example, let us consider the case of field ``field of activity'' with two
possible values, ``banking service'' and ``financial services''. From
a semantic point of view, the former is a specific case of the
latter.  This form of semantic overlap affects \approach as follows.
During the model building phase, \approach
could be trained with inconsistent historical input instances (e.g.,
two historical instances that have same values in all fields but
``field of activity'', with one instance having ``banking service'' and the
other ``financial services'').  In this case, \approach cannot build,
with enough confidence, dependencies between the values in the field affected by a
semantic overlap (e.g., ``field of activity'') and the other fields.
During the form filling suggestion phase, when \approach identifies
both values as candidate values to suggest (e.g., with similar
probability), due to the endorser module, \approach may not provide
any suggestions or suggest both values to users (depending on the
number of suggested values $n_r$ and the threshold $\theta$).  
In both
cases, 
users need to decide which value to select by themselves.  
However, defining candidate values with semantic overlaps is
not a good practice for form design, since it increases the mental
load on users to decide which value is more appropriate (e.g., the
more specialized ``banking service'' or the more generic
``financial service'').  We suggest to address this issue by refining
the candidate values during the design phase of the system.

\subsubsection{Cases of Limited Accuracy}
As discussed in the ``Error analysis'' part of the answer to RQ1
(page~\pageref{sec:rq1:error}), there are \SIrange{9.9}{21}{\%} of
suggestions made by \approach for which the correct value is not in
the top 5\% suggested values.
We have identified two main reasons leading to low accuracy:
\approach tends to provide incorrect suggestions, when
\begin{inparaenum}[(1)]
	\item the number of filled fields used for prediction is small, and 
	\item the size of the training input instances for a target field is small.
\end{inparaenum}
We plan to address these limitations as part of future work, along the
lines mentioned on page~\pageref{sec:rq1:error}.

\subsubsection{Learning from Corrections}
Automated form filling is an interactive process between users and the
automated tool.  In the current version, \approach does not offer the
possibility, during a data entry session, for a user to correct
\approach's suggestions by selecting a candidate value that is not
presented in the top 5\% suggested values.  If a user makes many
corrections to suggestions during a data entry session, it means
that the learned probability table (BN) may not reflect the relations
of values in the current input instance.  The ability to learn from
these users' corrections could further improve the accuracy of
\approach.  One intuitive solution is to assign a higher weight on
such input instances when re-training \approach for learning new
relations.  For example, input instances with many inaccurate
suggestions can be oversampled to increase their proportion in the entire historical of input instances.

\subsubsection{Cold Start}\label{sec:limitations:cold}
\approach depends on the dataset associated with an input form; 
it always has to be trained for a new system. 
The size of the pool of historical input instances can affect the
accuracy of \approach, a common issue among learning 
algorithms. 
When there are no or only a few historical input instances (i.e., the cold start problem~\cite{lee2019melu}),
the accuracy of \approach is limited. 
As presented in RQ5 (in section~\ref{sec:rq5}),
$\mathit{MRR}$ values of \approach increase when the size of the training set increases; 
\approach achieves accurate suggestions with \num{30000} and \num{56000} historical input instances 
on the NCBI and PROP datasets, respectively.
Nevertheless, we do not expect the size of the training set to be a strong limitation on the feasibility of applying \approach.
First, \approach targets data entry functionalities in 
enterprise software, in which one can expect thousands of input
instances per day.
For example, nowadays the NCBI platform gets about \num{9600} input instances 
per month for a single species (i.e., ``Homo sapiens''). 
It is also the case for the system used by our industrial partner.  
Second, \approach does not require additional effort to label training data;
it uses the actual values filled by users as the ground truth to train the model.
With an adequate training set size,
\approach can directly provide accurate suggestions for users.

\section{Related Work}
\label{sec:related_work}

The approach proposed in this paper is mainly related to works
on automated form filling based on the information from \emph{the same software system},
which focuses on filling free-text fields and categorical fields.
Regarding the former, the main proposals use language models (e.g.,
n-gram and sequence-to-sequence learning) to learn relationships
between characters or words from historical textual
inputs~\cite{van2008efficient, salama2018text}; these relationships
are then used to provide word auto-completion based on the letters typed in a field~\cite{zhang2019text}.
As
for dealing with categorical fields, most of the approaches suggest
possible values from a list of candidate values.
\citet{martinez2019using}
use association rule mining (ARM) to uncover the hidden associations of
fields for real-time form filling; however, as shown in our experiments, 
ARM does not provide accurate suggestions when compared with \approach.  \citet{hermens1994machine} applied decision tree and hierarchical clustering for filling
forms in an electronic leave report system: the study reports that these  algorithms performed worse than the simple most-frequent method when the forms were filled out in a random order.
\citet{troiano2017modeling} trained a Bayesian network model on the historical inputs from a user for an online payment system. The trained model could auto-fill the payment form for the same user by reusing his historical payment records.
  The plain Bayesian network model is also trained by
  \citet{ali2009predictive} for auto-filling the bug submission form
  of a bug tracking system.
Compared to existing approaches, \approach performs automated form filling by mining field dependencies on input instances from different users. We proposed local modeling and a heuristic-based endorser to further improve the accuracy of form filling suggestions.

Automated form filling has also been investigated in the context of 
developing ``smart'' personal information management systems, to
support information exchange \emph{across software
systems}~\cite{firmenich2012supporting,
  winckler2011approach,araujo2010carbon, chusho2002automatic}. 
In this
context, the main challenge is the semantic mapping of fields across
software systems, e.g., how to map the ``postal code'' and ``zip code'' fields
(from two different software systems) to the same concept.  To address this
challenge, \citet{chusho2002automatic} manually construct rules to
merge similar concepts of commonly used field names.  Other works use
string-based matching~\cite{he2004discovering},
WordNet~\cite{araujo2010carbon}, and
Wikipedia~\cite{hartmann2009context} as additional resources to
calculate the similarity among fields.  \citet{wang2014automatic}
calculate field similarity based on the field names, form
topics, and names of neighbor fields.  They recently propose to use learning-to-rank algorithms to further improve form filling
effectiveness~\cite{wang2017context}. However, to learn mapping rules,
these algorithms
require access to the users' personal records from different software
systems. In contrast, we only use input instances from the targeted
software system; thus, our approach can be used when there are
legal or security constraints on sharing records across
systems~\cite{winckler2011approach}.
Moreover, the above approaches perform well
only on common fields like ``age'' and ``address'', and cannot cope
with fields that are domain-specific and used only in few
software systems.

\begin{table}[tb]
	\caption{Overview of related work}
	\centering
	\scriptsize
	
\begin{tabular}{p{1.5cm}lp{3.9cm }p{3.8cm}}
	\toprule
	Scenario&Reference&Task/Example&Difference with \approach\\
	\midrule
	Form filling under the same software system
	&\makecell[tl]{
		\citet{salama2018text}\\
		\citet{van2008efficient}\\
		\citet{zhang2019text}
	}
    &\textbf{[Task]} Predict or auto-complete the next character or word for textual inputs.
    \textbf{[Example]} They suggest \emph{Angeles} when users type the word \emph{Los}.
    & They build language models on the values in \emph{textual fields} instead of categorical fields; the user's initial input is required. \\
	\cmidrule{2-4}

	&\makecell[tl]{\citet{martinez2019using}\\
		\citet{hermens1994machine} \\
		\citet{troiano2017modeling} \\
		\citet{ali2009predictive}}
	&\textbf{[Task]} Suggest the correct value from a list of candidate values in \emph{categorical fields}.
	\textbf{[Example]} They suggest \emph{Leasing Service} from a list of options in the field ``primary field of activity''.
	&We compared \approach with algorithms
   MFM~\cite{hermens1994machine}, ARM~\cite{martinez2019using}, and Na{\"i}veDT ~\cite{hermens1994machine}  as
   part of RQ1; we compared to a plain BN (equivalent to the one
   proposed in~\cite{troiano2017modeling} and ~\cite{ali2009predictive}) as part of RQ3. \\ 
	\midrule

	Form filling across software systems
	&\makecell[tl]{
		~\citet{araujo2010carbon}\\
		~\citet{chusho2002automatic}\\
		~\citet{firmenich2012supporting}\\
		~\citet{winckler2011approach}\\
		~\citet{he2004discovering}\\
		~\citet{hartmann2009context}\\ 
		~\citet{wang2014automatic}\\
		~\citet{wang2017context}
	}
	&\textbf{[Task]} Prefill form fields by reusing the values filled in other web forms; they build the mapping or ontology of field names across web forms.
	\textbf{[Example]} They use the value filled in ``postal code'' in one web form to help the same user fill ``zip code'' in another web form.
	&These approaches cannot be applied to fields that are domain-specific and used only in few software systems (e.g., medical systems);
	enterprise information system may have constraints on visiting or sharing records across systems. 
	\approach is designed to work in these scenarios.\\
	\midrule

	Form refactoring
	&\makecell[tl]{
		\citet{chen2011usher} \\
		\citet{troiano2017modeling} \\
		\citet{akiki2016engineering} 
	}
	&\textbf{[Task]} Re-order fields to support effective form filling. 
	\textbf{[Example]} They refactor the field ``role'' as the first field, 
	if they find ``role'' is  informative to predict values of other fields.
	&It is a preliminary step, complementary to automated form filling\\
	\midrule

	Software artifacts information auto-completion
	&\makecell[tl]{
		\citet{umer2019cnn} \\
		\citet{burgueno2020nlp}
	}
	&\textbf{[Task]} Predict values of certain fields in software systems. 
	\textbf{[Example]} They predict ``bug priority'' based on the description of bug reports.
	&It relies on textual data (e.g., bug reports and requirement documents) for suggestions. \approach does not rely on such information.\\
	\midrule
	
	Data crawling
	&\makecell[tl]{
		\citet{hernandez2019deep}\\
		\citet{kantorski2015automatic}
	}
	&\textbf{[Task]} Crawl data from databases by form filling.
	\textbf{[Example]} They generate value combinations for fields in a job search site to crawl all the job information hidden in the database.
	&They aim to generate valid input values for data crawling, instead of helping users find the correct candidate value.\\
	\midrule
	
	Data mining
	&\makecell[tl]{
		\citet{diaz2013user} \\
		\citet{ kristjansson2004interactive}\\
		\citet{toda2010probabilistic}\\
	}
	&\textbf{[Task]} Extract data values from data-rich text to fill forms. \textbf{[Example]} They use a resume text file to fill several fields of forms in job search sites.
	&They rely on the extraction of information in data-rich files (e.g., text files and spreadsheets). \approach does not rely on such information.\\
	
	\bottomrule
\end{tabular}
 	\label{tab:related}
\end{table}

The filling order of fields influences the ability of form filling
algorithms~\cite{hermens1994machine}. Several works refactor data entry forms  to
provide effective supports for form filling, for example using fields
dependencies~\cite{chen2011usher,troiano2017modeling} or user
roles~\cite{akiki2016engineering}. 
All these approaches  identify the fields that a target field directly
depends on and change their (field) order so that they can be filled
before the target field, to increase the accuracy of predicting the latter.
Form refactoring can be regarded as a preliminary step to
our proposed approach.

Form filling can be considered as a task to auto-complete data entry for software systems, which in the literature is a popular research topic in many domains;
however, most of the existing approaches propose auto-completion for
different purposes or with different inputs than what we considered in
this paper.
In software engineering, several techniques have been proposed to
provide suggestions for certain fields of software systems (e.g.,
priority of bug reports in bug tracking systems~\cite{umer2019cnn} and
elements of a domain model when using domain-model designing
systems~\cite{burgueno2020nlp}). Suggestions are provided  by
analyzing the textual data (e.g., bug reports, requirement documents)
with natural language processing techniques; instead, in our work we analyze the dependencies of categorical fields.
In the field of information
retrieval, crawlers automatically fill and submit web forms to
crawl the data returned from databases~\cite{hernandez2019deep,
kantorski2015automatic}; in the context of data crawling, automated form filling aims to automatically generate input values that can pass the field validation and
retrieve more data, instead of helping users find the correct value they intend to fill. 
In the field of data mining, several
approaches~\cite{toda2010probabilistic, kristjansson2004interactive,
	diaz2013user} fill data entry forms with the metadata extracted from
data-rich text files. For example, these approaches can automatically use the metadata taken from a resume text file to fill several fields of forms in different job search sites~\cite{toda2010probabilistic}.
However, in this study we do not rely on data-rich text files to infer field values.

\tablename~\ref{tab:related} presents an overview of related work and their differences with \approach.
 
\section{Conclusion}
\label{sec:conclusion}

In this paper, we proposed \approach, an approach to automatically
suggest possible values of categorical fields in data entry
forms, which are common user interface features in many software systems. 
Our approach utilizes Bayesian Networks to learn field
dependencies from historical input instances. Moreover, \approach
relies on a clustering-based local modeling strategy to mine local
field dependencies from partitions of historical input instances, to
improve its learning ability. Furthermore, \approach uses a
heuristic-based endorser to ensure minimal accuracy for suggested
values.

We evaluated LAFF by assessing its effectiveness and efficiency in
form filling on two datasets, one of them proprietary from the banking domain. 
Evaluation results show that \approach can provide a large number of accurate
form filling suggestions, significantly outperforming state-of-the-art
approaches in terms of Mean Reciprocal Rank (MRR). Further, \approach
takes at most {\SI{\LAFFPredictMax}{\milli\s}} to provide a
suggestion and is therefore applicable in practical data-entry scenarios.

As part of future work, we plan to conduct a study from the point of view of both users and developers, to analyze the effect of
\approach on reducing form filling time, input errors, and the cost of developing data entry forms. We also plan
to assess the performance of \approach under different deployment
configurations and the accuracy of \approach for different datasets and data entry forms.

Finally, we are also going to investigate methods to reduce
the number of incorrect suggestions provided by \approach when the
number of filled fields used for prediction or the
size of the training set is small.

\begin{acks}
The authors would like to thank the anonymous referees for
their valuable comments and helpful suggestions. Financial support for this work was provided by the \grantsponsor{awf}{Alphonse Weicker Foundation}{https://www.bnpparibas.lu/en/bnp-paribas/corporate-philantropy/scientific-research/}.
\end{acks}

\bibliographystyle{ACM-Reference-Format}

\begin{thebibliography}{83}



\ifx \showCODEN    \undefined \def \showCODEN     #1{\unskip}     \fi
\ifx \showDOI      \undefined \def \showDOI       #1{#1}\fi
\ifx \showISBNx    \undefined \def \showISBNx     #1{\unskip}     \fi
\ifx \showISBNxiii \undefined \def \showISBNxiii  #1{\unskip}     \fi
\ifx \showISSN     \undefined \def \showISSN      #1{\unskip}     \fi
\ifx \showLCCN     \undefined \def \showLCCN      #1{\unskip}     \fi
\ifx \shownote     \undefined \def \shownote      #1{#1}          \fi
\ifx \showarticletitle \undefined \def \showarticletitle #1{#1}   \fi
\ifx \showURL      \undefined \def \showURL       {\relax}        \fi
\providecommand\bibfield[2]{#2}
\providecommand\bibinfo[2]{#2}
\providecommand\natexlab[1]{#1}
\providecommand\showeprint[2][]{arXiv:#2}

\bibitem[\protect\citeauthoryear{Aggarwal, Dhawan, and Kumar}{Aggarwal
  et~al\mbox{.}}{2007}]{aggarwal2007database}
\bibfield{author}{\bibinfo{person}{RB Aggarwal}, \bibinfo{person}{Amit Dhawan},
  {and} \bibinfo{person}{Jay~Shankar Kumar}.} \bibinfo{year}{2007}\natexlab{}.
\newblock \showarticletitle{Database-centric development of menus and graphic
  user interfaces}.
\newblock \bibinfo{journal}{\emph{Defence Science Journal}}
  \bibinfo{volume}{57}, \bibinfo{number}{1} (\bibinfo{year}{2007}),
  \bibinfo{pages}{133}.
\newblock


\bibitem[\protect\citeauthoryear{Akiki, Bandara, and Yu}{Akiki
  et~al\mbox{.}}{2016}]{akiki2016engineering}
\bibfield{author}{\bibinfo{person}{Pierre~A Akiki}, \bibinfo{person}{Arosha~K
  Bandara}, {and} \bibinfo{person}{Yijun Yu}.} \bibinfo{year}{2016}\natexlab{}.
\newblock \showarticletitle{Engineering adaptive model-driven user interfaces}.
\newblock \bibinfo{journal}{\emph{IEEE Transaction on Software Engineering}}
  \bibinfo{volume}{42}, \bibinfo{number}{12} (\bibinfo{year}{2016}),
  \bibinfo{pages}{1118--1147}.
\newblock


\bibitem[\protect\citeauthoryear{Alexandropoulos, Kotsiantis, and
  Vrahatis}{Alexandropoulos et~al\mbox{.}}{2019}]{Alexandropoulos2019DataPI}
\bibfield{author}{\bibinfo{person}{Stamatios-Aggelos~N. Alexandropoulos},
  \bibinfo{person}{Sotiris~B. Kotsiantis}, {and} \bibinfo{person}{Michael~N.
  Vrahatis}.} \bibinfo{year}{2019}\natexlab{}.
\newblock \showarticletitle{Data preprocessing in predictive data mining}.
\newblock \bibinfo{journal}{\emph{Knowledge Engineering Review}}
  \bibinfo{volume}{34} (\bibinfo{year}{2019}), \bibinfo{pages}{e1}.
\newblock


\bibitem[\protect\citeauthoryear{Ali and Meek}{Ali and Meek}{2009}]{ali2009predictive}
\bibfield{author}{\bibinfo{person}{Alnur Ali} {and} \bibinfo{person}{Chris
  Meek}.} \bibinfo{year}{2009}\natexlab{}.
\newblock \bibinfo{booktitle}{\emph{Predictive Models of Form Filling}}.
\newblock \bibinfo{type}{{T}echnical {R}eport} MSR-TR-2009-1.
  \bibinfo{institution}{Microsoft Research}.
\newblock
\urldef\tempurl \url{https://www.microsoft.com/en-us/research/publication/predictive-models-of-form-filling/}
\showURL{\tempurl}


\bibitem[\protect\citeauthoryear{{American Medical News}}{{American Medical
  News}}{2005}]{american2005data}
\bibfield{author}{\bibinfo{person}{{American Medical News}}.}
  \bibinfo{year}{2005}\natexlab{}.
\newblock \bibinfo{title}{Data entry is a top cause of medication errors}.
\newblock
  \bibinfo{howpublished}{\url{https://amednews.com/article/20050124/profession/301249959/4/}}.
\newblock


\bibitem[\protect\citeauthoryear{An, Hu, and Song}{An et~al\mbox{.}}{2012}]{an2012learning}
\bibfield{author}{\bibinfo{person}{Yuan An}, \bibinfo{person}{Xiaohua Hu},
  {and} \bibinfo{person}{Il-Yeol Song}.} \bibinfo{year}{2012}\natexlab{}.
\newblock \showarticletitle{Learning to discover complex mappings from web
  forms to ontologies}. In \bibinfo{booktitle}{\emph{Proceedings of the 21st
  ACM international conference on Information and knowledge management}}.
  \bibinfo{publisher}{ACM}, \bibinfo{address}{New York, NY, USA},
  \bibinfo{pages}{1253--1262}.
\newblock


\bibitem[\protect\citeauthoryear{{Android API Reference}}{{Android API
  Reference}}{2021}]{android2021android}
\bibfield{author}{\bibinfo{person}{{Android API Reference}}.}
  \bibinfo{year}{2021}\natexlab{}.
\newblock \bibinfo{title}{Android View Autofill}.
\newblock
  \bibinfo{howpublished}{\url{https://developer.android.com/reference/kotlin/android/view/autofill/package-summary}}.
\newblock


\bibitem[\protect\citeauthoryear{Ankan and Panda}{Ankan and Panda}{2015}]{ankan2015pgmpy}
\bibfield{author}{\bibinfo{person}{Ankur Ankan} {and} \bibinfo{person}{Abinash
  Panda}.} \bibinfo{year}{2015}\natexlab{}.
\newblock \showarticletitle{Pgmpy: probabilistic graphical models using
  python}. In \bibinfo{booktitle}{\emph{Proc. SCIPY'15}}.
  \bibinfo{publisher}{SCIPY}, \bibinfo{address}{Austin, Texas, USA},
  \bibinfo{pages}{6--11}.
\newblock


\bibitem[\protect\citeauthoryear{Araujo, Gao, Leonardi, and Houben}{Araujo
  et~al\mbox{.}}{2010}]{araujo2010carbon}
\bibfield{author}{\bibinfo{person}{Samur Araujo}, \bibinfo{person}{Qi Gao},
  \bibinfo{person}{Erwin Leonardi}, {and} \bibinfo{person}{Geert-Jan Houben}.}
  \bibinfo{year}{2010}\natexlab{}.
\newblock \showarticletitle{Carbon: domain-independent automatic web form
  filling}. In \bibinfo{booktitle}{\emph{Proc. ICWE'10}}
  \emph{(\bibinfo{series}{LNCS}, Vol.~\bibinfo{volume}{6189})}. Springer,
  \bibinfo{publisher}{Springer Berlin Heidelberg}, \bibinfo{address}{Berlin,
  Heidelberg, Germany}, \bibinfo{pages}{292--306}.
\newblock


\bibitem[\protect\citeauthoryear{Avazpour, Pitakrat, Grunske, and
  Grundy}{Avazpour et~al\mbox{.}}{2014}]{avazpour2014dimensions}
\bibfield{author}{\bibinfo{person}{Iman Avazpour}, \bibinfo{person}{Teerat
  Pitakrat}, \bibinfo{person}{Lars Grunske}, {and} \bibinfo{person}{John
  Grundy}.} \bibinfo{year}{2014}\natexlab{}.
\newblock \showarticletitle{Dimensions and metrics for evaluating
  recommendation systems}.
\newblock In \bibinfo{booktitle}{\emph{Recommendation Systems in Software
  Engineering}}. \bibinfo{publisher}{Springer}, \bibinfo{address}{Berlin,
  Heidelberg, Germany}, \bibinfo{pages}{245--273}.
\newblock


\bibitem[\protect\citeauthoryear{{Bank for International Settlements}}{{Bank
  for International Settlements}}{2003}]{bis2003general}
\bibfield{author}{\bibinfo{person}{{Bank for International Settlements}}.}
  \bibinfo{year}{2003}\natexlab{}.
\newblock \bibinfo{title}{General Guide to Account Opening and Customer
  Identification}.
\newblock
  \bibinfo{howpublished}{\url{https://www.bis.org/publ/bcbs85annex.htm}}.
\newblock


\bibitem[\protect\citeauthoryear{Barrett, Clark, Gevorgyan, Gorelenkov, Gribov,
  Karsch-Mizrachi, Kimelman, Pruitt, Resenchuk, Tatusova,
  et~al\mbox{.}}{Barrett et~al\mbox{.}}{2012}]{barrett2012bioproject}
\bibfield{author}{\bibinfo{person}{Tanya Barrett}, \bibinfo{person}{Karen
  Clark}, \bibinfo{person}{Robert Gevorgyan}, \bibinfo{person}{Vyacheslav
  Gorelenkov}, \bibinfo{person}{Eugene Gribov}, \bibinfo{person}{Ilene
  Karsch-Mizrachi}, \bibinfo{person}{Michael Kimelman}, \bibinfo{person}{Kim~D
  Pruitt}, \bibinfo{person}{Sergei Resenchuk}, \bibinfo{person}{Tatiana
  Tatusova}, {et~al\mbox{.}}} \bibinfo{year}{2012}\natexlab{}.
\newblock \showarticletitle{BioProject and BioSample databases at {NCBI}:
  facilitating capture and organization of metadata}.
\newblock \bibinfo{journal}{\emph{Nucleic acids research}}
  \bibinfo{volume}{40}, \bibinfo{number}{D1} (\bibinfo{year}{2012}),
  \bibinfo{pages}{D57--D63}.
\newblock


\bibitem[\protect\citeauthoryear{Breiman, Friedman, Stone, and Olshen}{Breiman
  et~al\mbox{.}}{1984}]{breiman1984classification}
\bibfield{author}{\bibinfo{person}{Leo Breiman}, \bibinfo{person}{Jerome
  Friedman}, \bibinfo{person}{Charles~J Stone}, {and}
  \bibinfo{person}{Richard~A Olshen}.} \bibinfo{year}{1984}\natexlab{}.
\newblock \bibinfo{booktitle}{\emph{Classification and regression trees}}.
\newblock \bibinfo{publisher}{CRC press}, \bibinfo{address}{Boca Raton,
  Florida, USA}.
\newblock


\bibitem[\protect\citeauthoryear{Burgue{\~n}o, Claris{\'o}, Li, G{\'e}rard, and
  Cabot}{Burgue{\~n}o et~al\mbox{.}}{2021}]{burgueno2020nlp}
\bibfield{author}{\bibinfo{person}{Loli Burgue{\~n}o}, \bibinfo{person}{Robert
  Claris{\'o}}, \bibinfo{person}{Shuai Li}, \bibinfo{person}{S{\'e}bastien
  G{\'e}rard}, {and} \bibinfo{person}{Jordi Cabot}.}
  \bibinfo{year}{2021}\natexlab{}.
\newblock \showarticletitle{A NLP-based architecture for the autocompletion of
  partial domain models}. In \bibinfo{booktitle}{\emph{Proc. CAiSE’21}}
  \emph{(\bibinfo{series}{LNCS})}. \bibinfo{publisher}{Springer},
  \bibinfo{address}{Berlin, Heidelberg, Germany}, \bibinfo{numpages}{15}~pages.
\newblock


\bibitem[\protect\citeauthoryear{Cardellini, Colajanni, and Yu}{Cardellini
  et~al\mbox{.}}{2000}]{cardellini2000geographic}
\bibfield{author}{\bibinfo{person}{Valeria Cardellini},
  \bibinfo{person}{Michele Colajanni}, {and} \bibinfo{person}{Philip~S Yu}.}
  \bibinfo{year}{2000}\natexlab{}.
\newblock \showarticletitle{Geographic load balancing for scalable distributed
  web systems}. In \bibinfo{booktitle}{\emph{Proc. MASCOTS'00}}.
  \bibinfo{publisher}{IEEE}, \bibinfo{address}{San Francisco, CA, USA},
  \bibinfo{pages}{20--27}.
\newblock


\bibitem[\protect\citeauthoryear{Castells, Vargas, and Wang}{Castells
  et~al\mbox{.}}{2011}]{castells2011novelty}
\bibfield{author}{\bibinfo{person}{Pablo Castells}, \bibinfo{person}{Sa{\'u}l
  Vargas}, {and} \bibinfo{person}{Jun Wang}.} \bibinfo{year}{2011}\natexlab{}.
\newblock \showarticletitle{Novelty and diversity metrics for recommender
  systems: choice, discovery and relevance}. In \bibinfo{booktitle}{\emph{Proc.
  DDR'11 - International Workshop on Diversity in Document Retrieval}}.
  \bibinfo{publisher}{self-published},
  \bibinfo{address}{\url{http://www.dcs.gla.ac.uk/workshops/ddr2011/ddr2011.proceedings.pdf}},
  \bibinfo{pages}{29--36}.
\newblock


\bibitem[\protect\citeauthoryear{Chen, Chen, Conway, Hellerstein, and
  Parikh}{Chen et~al\mbox{.}}{2011}]{chen2011usher}
\bibfield{author}{\bibinfo{person}{Kuang Chen}, \bibinfo{person}{Harr Chen},
  \bibinfo{person}{Neil Conway}, \bibinfo{person}{Joseph~M Hellerstein}, {and}
  \bibinfo{person}{Tapan~S Parikh}.} \bibinfo{year}{2011}\natexlab{}.
\newblock \showarticletitle{Usher: Improving data quality with dynamic forms}.
\newblock \bibinfo{journal}{\emph{IEEE Transaction on Knowledge and Data
  Engineering}} \bibinfo{volume}{23}, \bibinfo{number}{8}
  (\bibinfo{year}{2011}), \bibinfo{pages}{1138--1153}.
\newblock


\bibitem[\protect\citeauthoryear{Chen, Shang, Yang, Hassan, Godfrey, Nasser,
  and Flora}{Chen et~al\mbox{.}}{2016}]{chen2016empirical}
\bibfield{author}{\bibinfo{person}{Tse-Hsun Chen}, \bibinfo{person}{Weiyi
  Shang}, \bibinfo{person}{Jinqiu Yang}, \bibinfo{person}{Ahmed~E Hassan},
  \bibinfo{person}{Michael~W Godfrey}, \bibinfo{person}{Mohamed Nasser}, {and}
  \bibinfo{person}{Parminder Flora}.} \bibinfo{year}{2016}\natexlab{}.
\newblock \showarticletitle{An empirical study on the practice of maintaining
  object-relational mapping code in {Java} systems}. In
  \bibinfo{booktitle}{\emph{Proc. MSR'16}}. \bibinfo{publisher}{ACM},
  \bibinfo{address}{New York, NY, USA}, \bibinfo{pages}{165--176}.
\newblock


\bibitem[\protect\citeauthoryear{Chusho, Fujiwara, and Minamitani}{Chusho
  et~al\mbox{.}}{2002}]{chusho2002automatic}
\bibfield{author}{\bibinfo{person}{Takeshi Chusho}, \bibinfo{person}{Katsuya
  Fujiwara}, {and} \bibinfo{person}{Keiji Minamitani}.}
  \bibinfo{year}{2002}\natexlab{}.
\newblock \showarticletitle{Automatic filling in a form by an agent for web
  applications}. In \bibinfo{booktitle}{\emph{Proc. APSEC'02}}.
  \bibinfo{publisher}{IEEE}, \bibinfo{address}{Berlin, Heidelberg, Germany},
  \bibinfo{pages}{239--247}.
\newblock


\bibitem[\protect\citeauthoryear{Cockburn and Gutwin}{Cockburn and
  Gutwin}{2009}]{cockburn2009predictive}
\bibfield{author}{\bibinfo{person}{Andy Cockburn} {and} \bibinfo{person}{Carl
  Gutwin}.} \bibinfo{year}{2009}\natexlab{}.
\newblock \showarticletitle{A predictive model of human performance with
  scrolling and hierarchical lists}.
\newblock \bibinfo{journal}{\emph{Human--Computer Interaction}}
  \bibinfo{volume}{24}, \bibinfo{number}{3} (\bibinfo{year}{2009}),
  \bibinfo{pages}{273--314}.
\newblock


\bibitem[\protect\citeauthoryear{Dekel, Shamir, and Xiao}{Dekel
  et~al\mbox{.}}{2010}]{dekel2010learning}
\bibfield{author}{\bibinfo{person}{Ofer Dekel}, \bibinfo{person}{Ohad Shamir},
  {and} \bibinfo{person}{Lin Xiao}.} \bibinfo{year}{2010}\natexlab{}.
\newblock \showarticletitle{Learning to classify with missing and corrupted
  features}.
\newblock \bibinfo{journal}{\emph{Machine learning}} \bibinfo{volume}{81},
  \bibinfo{number}{2} (\bibinfo{year}{2010}), \bibinfo{pages}{149--178}.
\newblock


\bibitem[\protect\citeauthoryear{Diaz, Otaduy, and Puente}{Diaz
  et~al\mbox{.}}{2013}]{diaz2013user}
\bibfield{author}{\bibinfo{person}{Oscar Diaz}, \bibinfo{person}{Itziar
  Otaduy}, {and} \bibinfo{person}{Gorka Puente}.}
  \bibinfo{year}{2013}\natexlab{}.
\newblock \showarticletitle{User-driven automation of web form filling}. In
  \bibinfo{booktitle}{\emph{Proc. ICWE'13}} \emph{(\bibinfo{series}{LNCS},
  Vol.~\bibinfo{volume}{7977})}. \bibinfo{publisher}{Springer},
  \bibinfo{address}{Berlin, Heidelberg, Germany}, \bibinfo{pages}{171--185}.
\newblock


\bibitem[\protect\citeauthoryear{Emam}{Emam}{1999}]{emam1999benchmarking}
\bibfield{author}{\bibinfo{person}{Khaled~El Emam}.}
  \bibinfo{year}{1999}\natexlab{}.
\newblock \showarticletitle{Benchmarking Kappa: Interrater agreement in
  software process assessments}.
\newblock \bibinfo{journal}{\emph{Empirical Software Engineering}}
  \bibinfo{volume}{4}, \bibinfo{number}{2} (\bibinfo{year}{1999}),
  \bibinfo{pages}{113--133}.
\newblock


\bibitem[\protect\citeauthoryear{Fan, Geerts, and Jia}{Fan
  et~al\mbox{.}}{2008}]{fan2008revival}
\bibfield{author}{\bibinfo{person}{Wenfei Fan}, \bibinfo{person}{Floris
  Geerts}, {and} \bibinfo{person}{Xibei Jia}.} \bibinfo{year}{2008}\natexlab{}.
\newblock \showarticletitle{A revival of integrity constraints for data
  cleaning}.
\newblock \bibinfo{journal}{\emph{Proc. VLDB Endowment'08}}
  \bibinfo{volume}{1}, \bibinfo{number}{2} (\bibinfo{year}{2008}),
  \bibinfo{pages}{1522--1523}.
\newblock


\bibitem[\protect\citeauthoryear{Firmenich, Gaits, Gordillo, Rossi, and
  Winckler}{Firmenich et~al\mbox{.}}{2012}]{firmenich2012supporting}
\bibfield{author}{\bibinfo{person}{Sergio Firmenich}, \bibinfo{person}{Vincent
  Gaits}, \bibinfo{person}{Silvia Gordillo}, \bibinfo{person}{Gustavo Rossi},
  {and} \bibinfo{person}{Marco Winckler}.} \bibinfo{year}{2012}\natexlab{}.
\newblock \showarticletitle{Supporting users tasks with personal information
  management and web forms augmentation}. In \bibinfo{booktitle}{\emph{Proc.
  ICWE2012}} \emph{(\bibinfo{series}{LNCS}, Vol.~\bibinfo{volume}{7387})}.
  Springer, \bibinfo{publisher}{Oxford University Press},
  \bibinfo{address}{Berlin, Heidelberg, Germany}, \bibinfo{pages}{268--282}.
\newblock


\bibitem[\protect\citeauthoryear{Fowler and Stanwick}{Fowler and
  Stanwick}{2004}]{fowler2004web}
\bibfield{author}{\bibinfo{person}{Susan Fowler} {and} \bibinfo{person}{Victor
  Stanwick}.} \bibinfo{year}{2004}\natexlab{}.
\newblock \bibinfo{booktitle}{\emph{Web application design handbook: Best
  practices for web-based software}}.
\newblock \bibinfo{publisher}{Morgan Kaufmann}, \bibinfo{address}{Amsterdam,
  Boston, USA}.
\newblock


\bibitem[\protect\citeauthoryear{Friedman, Geiger, and Goldszmidt}{Friedman
  et~al\mbox{.}}{1997}]{friedman1997bayesian}
\bibfield{author}{\bibinfo{person}{Nir Friedman}, \bibinfo{person}{Dan Geiger},
  {and} \bibinfo{person}{Moises Goldszmidt}.} \bibinfo{year}{1997}\natexlab{}.
\newblock \showarticletitle{Bayesian network classifiers}.
\newblock \bibinfo{journal}{\emph{Machine learning}} \bibinfo{volume}{29},
  \bibinfo{number}{2-3} (\bibinfo{year}{1997}), \bibinfo{pages}{131--163}.
\newblock


\bibitem[\protect\citeauthoryear{Gafur}{Gafur}{2020}]{gafur2020updated}
\bibfield{author}{\bibinfo{person}{Abdul Gafur}.}
  \bibinfo{year}{2020}\natexlab{}.
\newblock \showarticletitle{Updated tabular key and improved browser-based
  interactive key to species of Pratylenchus Filipjev (Nematoda:
  Pratylenchidae)}.
\newblock \bibinfo{journal}{\emph{Biodiversitas Journal of Biological
  Diversity}} \bibinfo{volume}{21}, \bibinfo{number}{8} (\bibinfo{year}{2020}),
  \bibinfo{pages}{3780--3785}.
\newblock


\bibitem[\protect\citeauthoryear{G{\'a}mez, Mateo, and Puerta}{G{\'a}mez
  et~al\mbox{.}}{2011}]{gamez2011learning}
\bibfield{author}{\bibinfo{person}{Jos{\'e}~A G{\'a}mez},
  \bibinfo{person}{Juan~L Mateo}, {and} \bibinfo{person}{Jos{\'e}~M Puerta}.}
  \bibinfo{year}{2011}\natexlab{}.
\newblock \showarticletitle{Learning Bayesian networks by hill climbing:
  efficient methods based on progressive restriction of the neighborhood}.
\newblock \bibinfo{journal}{\emph{Data Mining and Knowledge Discovery}}
  \bibinfo{volume}{22}, \bibinfo{number}{1-2} (\bibinfo{year}{2011}),
  \bibinfo{pages}{106--148}.
\newblock


\bibitem[\protect\citeauthoryear{Ge, Delgado-Battenfeld, and Jannach}{Ge
  et~al\mbox{.}}{2010}]{ge2010beyond}
\bibfield{author}{\bibinfo{person}{Mouzhi Ge}, \bibinfo{person}{Carla
  Delgado-Battenfeld}, {and} \bibinfo{person}{Dietmar Jannach}.}
  \bibinfo{year}{2010}\natexlab{}.
\newblock \showarticletitle{Beyond accuracy: evaluating recommender systems by
  coverage and serendipity}. In \bibinfo{booktitle}{\emph{Proc. RecSys'10}}.
  \bibinfo{publisher}{ACM}, \bibinfo{address}{New York, NY, USA},
  \bibinfo{pages}{257--260}.
\newblock


\bibitem[\protect\citeauthoryear{Gon{\c{c}}alves, O'Connor,
  Mart{\'i}nez-Romero, Egyedi, Willrett, Graybeal, and Musen}{Gon{\c{c}}alves
  et~al\mbox{.}}{2017}]{goncalves17:_cedar_workb}
\bibfield{author}{\bibinfo{person}{Rafael~S. Gon{\c{c}}alves},
  \bibinfo{person}{Martin~J. O'Connor}, \bibinfo{person}{Marcos
  Mart{\'i}nez-Romero}, \bibinfo{person}{Attila~L. Egyedi},
  \bibinfo{person}{Debra Willrett}, \bibinfo{person}{John Graybeal}, {and}
  \bibinfo{person}{Mark~A. Musen}.} \bibinfo{year}{2017}\natexlab{}.
\newblock \showarticletitle{The {CEDAR} workbench: an ontology-assisted
  environment for authoring metadata that describe scientific experiments}. In
  \bibinfo{booktitle}{\emph{Proc. ISWC'17}} \emph{(\bibinfo{series}{LNCS},
  Vol.~\bibinfo{volume}{10588})}. \bibinfo{publisher}{Springer International
  Publishing}, \bibinfo{address}{Cham}, \bibinfo{pages}{103--110}.
\newblock


\bibitem[\protect\citeauthoryear{Google}{Google}{2008}]{chrome}
\bibfield{author}{\bibinfo{person}{Google}.} \bibinfo{year}{2008}\natexlab{}.
\newblock \bibinfo{title}{Chrome autofill forms}.
\newblock \bibinfo{howpublished}{\url{https://support.google.com/chrome}}.
\newblock
\newblock
\shownote{Accessed Feb 18, 2020.}


\bibitem[\protect\citeauthoryear{Gutwin and Cockburn}{Gutwin and
  Cockburn}{2006}]{gutwin2006improving}
\bibfield{author}{\bibinfo{person}{Carl Gutwin} {and} \bibinfo{person}{Andy
  Cockburn}.} \bibinfo{year}{2006}\natexlab{}.
\newblock \showarticletitle{Improving list revisitation with {L}ist{M}aps}. In
  \bibinfo{booktitle}{\emph{Proc. AVI'06}}. \bibinfo{publisher}{ACM},
  \bibinfo{address}{New York, NY, USA}, \bibinfo{pages}{396--403}.
\newblock


\bibitem[\protect\citeauthoryear{Hartmann and Muhlhauser}{Hartmann and
  Muhlhauser}{2009}]{hartmann2009context}
\bibfield{author}{\bibinfo{person}{Melanie Hartmann} {and} \bibinfo{person}{Max
  Muhlhauser}.} \bibinfo{year}{2009}\natexlab{}.
\newblock \showarticletitle{Context-aware form filling for web applications}.
  In \bibinfo{booktitle}{\emph{Proc. ICSC'09}}. \bibinfo{publisher}{IEEE},
  \bibinfo{address}{Berkeley, CA, USA}, \bibinfo{pages}{221--228}.
\newblock


\bibitem[\protect\citeauthoryear{He, Chang, and Han}{He et~al\mbox{.}}{2004}]{he2004discovering}
\bibfield{author}{\bibinfo{person}{Bin He}, \bibinfo{person}{Kevin Chen-Chuan
  Chang}, {and} \bibinfo{person}{Jiawei Han}.} \bibinfo{year}{2004}\natexlab{}.
\newblock \showarticletitle{Discovering complex matchings across web query
  interfaces: a correlation mining approach}. In
  \bibinfo{booktitle}{\emph{Proc. KDD'04}}. \bibinfo{publisher}{ACM},
  \bibinfo{address}{New York, NY, USA}, \bibinfo{pages}{148--157}.
\newblock


\bibitem[\protect\citeauthoryear{Heeter}{Heeter}{2000}]{heeter2000interactivity}
\bibfield{author}{\bibinfo{person}{Carrie Heeter}.}
  \bibinfo{year}{2000}\natexlab{}.
\newblock \showarticletitle{Interactivity in the context of designed
  experiences}.
\newblock \bibinfo{journal}{\emph{J. of Interactive Advertising}}
  \bibinfo{volume}{1}, \bibinfo{number}{1} (\bibinfo{year}{2000}),
  \bibinfo{pages}{3--14}.
\newblock


\bibitem[\protect\citeauthoryear{Herlocker, Konstan, Terveen, and
  Riedl}{Herlocker et~al\mbox{.}}{2004}]{herlocker2004evaluating}
\bibfield{author}{\bibinfo{person}{Jonathan~L Herlocker},
  \bibinfo{person}{Joseph~A Konstan}, \bibinfo{person}{Loren~G Terveen}, {and}
  \bibinfo{person}{John~T Riedl}.} \bibinfo{year}{2004}\natexlab{}.
\newblock \showarticletitle{Evaluating collaborative filtering recommender
  systems}.
\newblock \bibinfo{journal}{\emph{ACM Transaction on Information Systems}}
  \bibinfo{volume}{22}, \bibinfo{number}{1} (\bibinfo{year}{2004}),
  \bibinfo{pages}{5--53}.
\newblock


\bibitem[\protect\citeauthoryear{{Hermens} and {Shlimmer}}{{Hermens} and
  {Shlimmer}}{1994}]{hermens1994machine}
\bibfield{author}{\bibinfo{person}{L.~A. {Hermens}} {and}
  \bibinfo{person}{J.~C. {Shlimmer}}.} \bibinfo{year}{1994}\natexlab{}.
\newblock \showarticletitle{A machine-learning apprentice for the completion of
  repetitive forms}.
\newblock \bibinfo{journal}{\emph{IEEE Expert}} \bibinfo{volume}{9},
  \bibinfo{number}{1} (\bibinfo{year}{1994}), \bibinfo{pages}{28--33}.
\newblock


\bibitem[\protect\citeauthoryear{Hern{\'a}ndez, Rivero, and Ruiz}{Hern{\'a}ndez
  et~al\mbox{.}}{2019}]{hernandez2019deep}
\bibfield{author}{\bibinfo{person}{Inma Hern{\'a}ndez},
  \bibinfo{person}{Carlos~R Rivero}, {and} \bibinfo{person}{David Ruiz}.}
  \bibinfo{year}{2019}\natexlab{}.
\newblock \showarticletitle{Deep web crawling: a survey}.
\newblock \bibinfo{journal}{\emph{World Wide Web}} \bibinfo{volume}{22},
  \bibinfo{number}{4} (\bibinfo{year}{2019}), \bibinfo{pages}{1577--1610}.
\newblock


\bibitem[\protect\citeauthoryear{Horsky, Kaufman, Oppenheim, and Patel}{Horsky
  et~al\mbox{.}}{2003}]{horsky2003framework}
\bibfield{author}{\bibinfo{person}{Jan Horsky}, \bibinfo{person}{David~R
  Kaufman}, \bibinfo{person}{Michael~I Oppenheim}, {and}
  \bibinfo{person}{Vimla~L Patel}.} \bibinfo{year}{2003}\natexlab{}.
\newblock \showarticletitle{A framework for analyzing the cognitive complexity
  of computer-assisted clinical ordering}.
\newblock \bibinfo{journal}{\emph{Journal of Biomedical Informatics}}
  \bibinfo{volume}{36}, \bibinfo{number}{1-2} (\bibinfo{year}{2003}),
  \bibinfo{pages}{4--22}.
\newblock


\bibitem[\protect\citeauthoryear{Huang}{Huang}{1998}]{huang1998extensions}
\bibfield{author}{\bibinfo{person}{Zhexue Huang}.}
  \bibinfo{year}{1998}\natexlab{}.
\newblock \showarticletitle{Extensions to the k-means algorithm for clustering
  large data sets with categorical values}.
\newblock \bibinfo{journal}{\emph{Data Mining and Knowledge Discovery}}
  \bibinfo{volume}{2}, \bibinfo{number}{3} (\bibinfo{year}{1998}),
  \bibinfo{pages}{283--304}.
\newblock


\bibitem[\protect\citeauthoryear{Jarrett and Gaffney}{Jarrett and
  Gaffney}{2009}]{jarrett2009forms}
\bibfield{author}{\bibinfo{person}{Caroline Jarrett} {and}
  \bibinfo{person}{Gerry Gaffney}.} \bibinfo{year}{2009}\natexlab{}.
\newblock \bibinfo{booktitle}{\emph{Forms that work: Designing Web forms for
  usability}}.
\newblock \bibinfo{publisher}{Morgan Kaufmann}, \bibinfo{address}{Amsterdam,
  Boston, USA}.
\newblock


\bibitem[\protect\citeauthoryear{Jensen, Hansen, Eika, and Sandnes}{Jensen
  et~al\mbox{.}}{2020}]{jensen2020country}
\bibfield{author}{\bibinfo{person}{Emil~Thorstensen Jensen},
  \bibinfo{person}{Martin Hansen}, \bibinfo{person}{Evelyn Eika}, {and}
  \bibinfo{person}{Frode~Eika Sandnes}.} \bibinfo{year}{2020}\natexlab{}.
\newblock \showarticletitle{Country selection on web forms: a comparison of
  dropdown menus, radio buttons and text field with autocomplete}. In
  \bibinfo{booktitle}{\emph{Proc. IMCOM'20}}. IEEE, \bibinfo{publisher}{IEEE},
  \bibinfo{address}{Taiwan, China}, \bibinfo{pages}{1--4}.
\newblock


\bibitem[\protect\citeauthoryear{Jing, Qi, Wu, and Xu}{Jing
  et~al\mbox{.}}{2016}]{jing2016missing}
\bibfield{author}{\bibinfo{person}{Xiao-Yuan Jing}, \bibinfo{person}{Fumin Qi},
  \bibinfo{person}{Fei Wu}, {and} \bibinfo{person}{Baowen Xu}.}
  \bibinfo{year}{2016}\natexlab{}.
\newblock \showarticletitle{Missing data imputation based on low-rank recovery
  and semi-supervised regression for software effort estimation}. In
  \bibinfo{booktitle}{\emph{Proc. ICSE'16}}. \bibinfo{publisher}{IEEE},
  \bibinfo{address}{Austin, TX, USA}, \bibinfo{pages}{607--618}.
\newblock


\bibitem[\protect\citeauthoryear{Jou}{Jou}{2019}]{jou2019schema}
\bibfield{author}{\bibinfo{person}{Chichang Jou}.}
  \bibinfo{year}{2019}\natexlab{}.
\newblock \showarticletitle{Schema Extraction for Deep Web Query Interfaces
  Using Heuristics Rules}.
\newblock \bibinfo{journal}{\emph{Information Systems Frontiers}}
  \bibinfo{volume}{21}, \bibinfo{number}{1} (\bibinfo{year}{2019}),
  \bibinfo{pages}{163--174}.
\newblock


\bibitem[\protect\citeauthoryear{Kaminskas and Bridge}{Kaminskas and
  Bridge}{2016}]{kaminskas2016diversity}
\bibfield{author}{\bibinfo{person}{Marius Kaminskas} {and}
  \bibinfo{person}{Derek Bridge}.} \bibinfo{year}{2016}\natexlab{}.
\newblock \showarticletitle{Diversity, serendipity, novelty, and coverage: a
  survey and empirical analysis of beyond-accuracy objectives in recommender
  systems}.
\newblock \bibinfo{journal}{\emph{ACM Transactions on Interactive Intelligent
  Systems}} \bibinfo{volume}{7}, \bibinfo{number}{1} (\bibinfo{year}{2016}),
  \bibinfo{pages}{1--42}.
\newblock


\bibitem[\protect\citeauthoryear{Kantorski, Moreira, and Heuser}{Kantorski
  et~al\mbox{.}}{2015}]{kantorski2015automatic}
\bibfield{author}{\bibinfo{person}{Gustavo~Zanini Kantorski},
  \bibinfo{person}{Viviane~Pereira Moreira}, {and}
  \bibinfo{person}{Carlos~Alberto Heuser}.} \bibinfo{year}{2015}\natexlab{}.
\newblock \showarticletitle{Automatic filling of hidden web forms: a survey}.
\newblock \bibinfo{journal}{\emph{ACM SIGMOD Record}} \bibinfo{volume}{44},
  \bibinfo{number}{1} (\bibinfo{year}{2015}), \bibinfo{pages}{24--35}.
\newblock


\bibitem[\protect\citeauthoryear{Karimi, Jannach, and Jugovac}{Karimi
  et~al\mbox{.}}{2018}]{karimi2018news}
\bibfield{author}{\bibinfo{person}{Mozhgan Karimi}, \bibinfo{person}{Dietmar
  Jannach}, {and} \bibinfo{person}{Michael Jugovac}.}
  \bibinfo{year}{2018}\natexlab{}.
\newblock \showarticletitle{News recommender systems--Survey and roads ahead}.
\newblock \bibinfo{journal}{\emph{Information Processing \& Management}}
  \bibinfo{volume}{54}, \bibinfo{number}{6} (\bibinfo{year}{2018}),
  \bibinfo{pages}{1203--1227}.
\newblock


\bibitem[\protect\citeauthoryear{Khajouei and Jaspers}{Khajouei and
  Jaspers}{2010}]{khajouei2010impact}
\bibfield{author}{\bibinfo{person}{Reza Khajouei} {and} \bibinfo{person}{MWM
  Jaspers}.} \bibinfo{year}{2010}\natexlab{}.
\newblock \showarticletitle{The impact of CPOE medication systems’ design
  aspects on usability, workflow and medication orders}.
\newblock \bibinfo{journal}{\emph{Methods of Information in Medicine}}
  \bibinfo{volume}{49}, \bibinfo{number}{01} (\bibinfo{year}{2010}),
  \bibinfo{pages}{03--19}.
\newblock


\bibitem[\protect\citeauthoryear{Kristjansson, Culotta, Viola, and
  McCallum}{Kristjansson et~al\mbox{.}}{2004}]{kristjansson2004interactive}
\bibfield{author}{\bibinfo{person}{Trausti Kristjansson}, \bibinfo{person}{Aron
  Culotta}, \bibinfo{person}{Paul Viola}, {and} \bibinfo{person}{Andrew
  McCallum}.} \bibinfo{year}{2004}\natexlab{}.
\newblock \showarticletitle{Interactive information extraction with constrained
  conditional random fields}. In \bibinfo{booktitle}{\emph{Proc. AAAI'04}},
  Vol.~\bibinfo{volume}{4}. \bibinfo{publisher}{ACM}, \bibinfo{address}{New
  York, NY, USA}, \bibinfo{pages}{412--418}.
\newblock


\bibitem[\protect\citeauthoryear{Kunaver and Po{\v{z}}rl}{Kunaver and
  Po{\v{z}}rl}{2017}]{kunaver2017diversity}
\bibfield{author}{\bibinfo{person}{Matev{\v{z}} Kunaver} {and}
  \bibinfo{person}{Toma{\v{z}} Po{\v{z}}rl}.} \bibinfo{year}{2017}\natexlab{}.
\newblock \showarticletitle{Diversity in recommender systems--A survey}.
\newblock \bibinfo{journal}{\emph{Knowledge-based Systems}}
  \bibinfo{volume}{123} (\bibinfo{year}{2017}), \bibinfo{pages}{154--162}.
\newblock


\bibitem[\protect\citeauthoryear{Lee, Im, Jang, Cho, and Chung}{Lee
  et~al\mbox{.}}{2019}]{lee2019melu}
\bibfield{author}{\bibinfo{person}{Hoyeop Lee}, \bibinfo{person}{Jinbae Im},
  \bibinfo{person}{Seongwon Jang}, \bibinfo{person}{Hyunsouk Cho}, {and}
  \bibinfo{person}{Sehee Chung}.} \bibinfo{year}{2019}\natexlab{}.
\newblock \showarticletitle{Melu: Meta-learned user preference estimator for
  cold-start recommendation}. In \bibinfo{booktitle}{\emph{Proc. SIGKDD'19}}.
  \bibinfo{publisher}{ACM}, \bibinfo{address}{New York, NY, USA},
  \bibinfo{pages}{1073--1082}.
\newblock


\bibitem[\protect\citeauthoryear{Mart{\'\i}nez-Romero, O'Connor, Egyedi,
  Willrett, Hardi, Graybeal, and Musen}{Mart{\'\i}nez-Romero
  et~al\mbox{.}}{2019}]{martinez2019using}
\bibfield{author}{\bibinfo{person}{Marcos Mart{\'\i}nez-Romero},
  \bibinfo{person}{Martin~J O'Connor}, \bibinfo{person}{Attila~L Egyedi},
  \bibinfo{person}{Debra Willrett}, \bibinfo{person}{Josef Hardi},
  \bibinfo{person}{John Graybeal}, {and} \bibinfo{person}{Mark~A Musen}.}
  \bibinfo{year}{2019}\natexlab{}.
\newblock \showarticletitle{Using association rule mining and ontologies to
  generate metadata recommendations from multiple biomedical databases}.
\newblock \bibinfo{journal}{\emph{Database J. Biol. Databases Curation}}
  \bibinfo{volume}{2019} (\bibinfo{year}{2019}), \bibinfo{numpages}{25}~pages.
\newblock


\bibitem[\protect\citeauthoryear{{McIntosh} and {Kamei}}{{McIntosh} and
  {Kamei}}{2017}]{mcintosh2017fix}
\bibfield{author}{\bibinfo{person}{S. {McIntosh}} {and} \bibinfo{person}{Y.
  {Kamei}}.} \bibinfo{year}{2017}\natexlab{}.
\newblock \showarticletitle{Are fix-inducing changes a moving target? a
  longitudinal case study of just-in-time defect prediction}.
\newblock \bibinfo{journal}{\emph{IEEE Transaction on Software Engineering}}
  \bibinfo{volume}{44}, \bibinfo{number}{5} (\bibinfo{year}{2017}),
  \bibinfo{pages}{412--428}.
\newblock


\bibitem[\protect\citeauthoryear{Menzies, Butcher, Marcus, Zimmermann, and
  Cok}{Menzies et~al\mbox{.}}{2011}]{menzies2011local}
\bibfield{author}{\bibinfo{person}{Tim Menzies}, \bibinfo{person}{Andrew
  Butcher}, \bibinfo{person}{Andrian Marcus}, \bibinfo{person}{Thomas
  Zimmermann}, {and} \bibinfo{person}{David Cok}.}
  \bibinfo{year}{2011}\natexlab{}.
\newblock \showarticletitle{Local vs. global models for effort estimation and
  defect prediction}. In \bibinfo{booktitle}{\emph{Proc. ASE'11}}.
  \bibinfo{publisher}{IEEE}, \bibinfo{address}{Lawrence, KS, USA},
  \bibinfo{pages}{343--351}.
\newblock


\bibitem[\protect\citeauthoryear{{Microsoft}}{{Microsoft}}{2013}]{change2013microsoft}
\bibfield{author}{\bibinfo{person}{{Microsoft}}.}
  \bibinfo{year}{2013}\natexlab{}.
\newblock \bibinfo{title}{Change the default tab order for controls on a form}.
\newblock
  \bibinfo{howpublished}{\url{https://support.microsoft.com/en-us/office/change-the-default-tab-order-for-controls-on-a-form-03d1599a-debf-4b66-a95b-e3e744210afe}}.
\newblock


\bibitem[\protect\citeauthoryear{Mu{\c{s}}lu, Brun, and Meliou}{Mu{\c{s}}lu
  et~al\mbox{.}}{2015}]{mucslu2015preventing}
\bibfield{author}{\bibinfo{person}{K{\i}van{\c{c}} Mu{\c{s}}lu},
  \bibinfo{person}{Yuriy Brun}, {and} \bibinfo{person}{Alexandra Meliou}.}
  \bibinfo{year}{2015}\natexlab{}.
\newblock \showarticletitle{Preventing data errors with continuous testing}. In
  \bibinfo{booktitle}{\emph{Proc. ISSTA'15}}. \bibinfo{publisher}{ACM},
  \bibinfo{address}{New York, NY, USA}, \bibinfo{pages}{373--384}.
\newblock


\bibitem[\protect\citeauthoryear{Ore, Elbaum, Detweiler, and Karkazis}{Ore
  et~al\mbox{.}}{2018}]{ore2018assessing}
\bibfield{author}{\bibinfo{person}{John-Paul Ore}, \bibinfo{person}{Sebastian
  Elbaum}, \bibinfo{person}{Carrick Detweiler}, {and} \bibinfo{person}{Lambros
  Karkazis}.} \bibinfo{year}{2018}\natexlab{}.
\newblock \showarticletitle{Assessing the type annotation burden}. In
  \bibinfo{booktitle}{\emph{Proc. ASE'18}}. \bibinfo{publisher}{ACM},
  \bibinfo{address}{New York, NY, USA}, \bibinfo{pages}{190--201}.
\newblock


\bibitem[\protect\citeauthoryear{Pearson, Campos, Just, Fraser, Abreu, Ernst,
  Pang, and Keller}{Pearson et~al\mbox{.}}{2017}]{pearson2017evaluating}
\bibfield{author}{\bibinfo{person}{Spencer Pearson}, \bibinfo{person}{Jos{\'e}
  Campos}, \bibinfo{person}{Ren{\'e} Just}, \bibinfo{person}{Gordon Fraser},
  \bibinfo{person}{Rui Abreu}, \bibinfo{person}{Michael~D Ernst},
  \bibinfo{person}{Deric Pang}, {and} \bibinfo{person}{Benjamin Keller}.}
  \bibinfo{year}{2017}\natexlab{}.
\newblock \showarticletitle{Evaluating and improving fault localization}. In
  \bibinfo{booktitle}{\emph{Proc. ICSE'17}}. \bibinfo{publisher}{IEEE},
  \bibinfo{address}{Buenos Aires, Argentina}, \bibinfo{pages}{609--620}.
\newblock


\bibitem[\protect\citeauthoryear{Pedregosa, Varoquaux, Gramfort, Michel,
  Thirion, Grisel, Blondel, Prettenhofer, Weiss, Dubourg, Vanderplas, Passos,
  Cournapeau, Brucher, Perrot, and Duchesnay}{Pedregosa et~al\mbox{.}}{2011}]{scikit-learn}
\bibfield{author}{\bibinfo{person}{F. Pedregosa}, \bibinfo{person}{G.
  Varoquaux}, \bibinfo{person}{A. Gramfort}, \bibinfo{person}{V. Michel},
  \bibinfo{person}{B. Thirion}, \bibinfo{person}{O. Grisel},
  \bibinfo{person}{M. Blondel}, \bibinfo{person}{P. Prettenhofer},
  \bibinfo{person}{R. Weiss}, \bibinfo{person}{V. Dubourg}, \bibinfo{person}{J.
  Vanderplas}, \bibinfo{person}{A. Passos}, \bibinfo{person}{D. Cournapeau},
  \bibinfo{person}{M. Brucher}, \bibinfo{person}{M. Perrot}, {and}
  \bibinfo{person}{E. Duchesnay}.} \bibinfo{year}{2011}\natexlab{}.
\newblock \showarticletitle{Scikit-learn: Machine Learning in {P}ython}.
\newblock \bibinfo{journal}{\emph{Journal of Machine Learning Research}}
  \bibinfo{volume}{12} (\bibinfo{year}{2011}), \bibinfo{pages}{2825--2830}.
\newblock


\bibitem[\protect\citeauthoryear{Qian, Munyisia, Reid, Hailey, Pados, and
  Yu}{Qian et~al\mbox{.}}{2020}]{qian2020trend}
\bibfield{author}{\bibinfo{person}{Siyu Qian}, \bibinfo{person}{Esther
  Munyisia}, \bibinfo{person}{David Reid}, \bibinfo{person}{David Hailey},
  \bibinfo{person}{Jade Pados}, {and} \bibinfo{person}{Ping Yu}.}
  \bibinfo{year}{2020}\natexlab{}.
\newblock \showarticletitle{Trend in data errors after the implementation of an
  electronic medical record system: A longitudinal study in an Australian
  regional Drug and Alcohol Service}.
\newblock \bibinfo{journal}{\emph{International Journal of Medical
  Informatics}}  \bibinfo{volume}{144} (\bibinfo{year}{2020}),
  \bibinfo{pages}{104292}.
\newblock


\bibitem[\protect\citeauthoryear{Raftery}{Raftery}{1995}]{raftery1995bayesian}
\bibfield{author}{\bibinfo{person}{Adrian~E Raftery}.}
  \bibinfo{year}{1995}\natexlab{}.
\newblock \showarticletitle{Bayesian model selection in social research}.
\newblock \bibinfo{journal}{\emph{Sociological Methodology}}
  \bibinfo{volume}{25} (\bibinfo{year}{1995}), \bibinfo{pages}{111--163}.
\newblock


\bibitem[\protect\citeauthoryear{Robillard, Walker, and Zimmermann}{Robillard
  et~al\mbox{.}}{2009}]{robillard2009recommendation}
\bibfield{author}{\bibinfo{person}{Martin Robillard}, \bibinfo{person}{Robert
  Walker}, {and} \bibinfo{person}{Thomas Zimmermann}.}
  \bibinfo{year}{2009}\natexlab{}.
\newblock \showarticletitle{Recommendation systems for software engineering}.
\newblock \bibinfo{journal}{\emph{IEEE Software}} \bibinfo{volume}{27},
  \bibinfo{number}{4} (\bibinfo{year}{2009}), \bibinfo{pages}{80--86}.
\newblock


\bibitem[\protect\citeauthoryear{Rukzio, Noda, De~Luca, Hamard, and
  Coskun}{Rukzio et~al\mbox{.}}{2008}]{rukzio2008automatic}
\bibfield{author}{\bibinfo{person}{Enrico Rukzio}, \bibinfo{person}{Chie Noda},
  \bibinfo{person}{Alexander De~Luca}, \bibinfo{person}{John Hamard}, {and}
  \bibinfo{person}{Fatih Coskun}.} \bibinfo{year}{2008}\natexlab{}.
\newblock \showarticletitle{Automatic form filling on mobile devices}.
\newblock \bibinfo{journal}{\emph{Pervasive and Mobile Computing}}
  \bibinfo{volume}{4}, \bibinfo{number}{2} (\bibinfo{year}{2008}),
  \bibinfo{pages}{161--181}.
\newblock


\bibitem[\protect\citeauthoryear{Sakal and Rakovic}{Sakal and Rakovic}{2012}]{sakal2012errors}
\bibfield{author}{\bibinfo{person}{Marton Sakal} {and} \bibinfo{person}{Lazar
  Rakovic}.} \bibinfo{year}{2012}\natexlab{}.
\newblock \showarticletitle{Errors in building and using electronic tables:
  Financial consequences and minimisation techniques}.
\newblock \bibinfo{journal}{\emph{Strategic Management}} \bibinfo{volume}{17},
  \bibinfo{number}{3} (\bibinfo{year}{2012}), \bibinfo{pages}{29--35}.
\newblock


\bibitem[\protect\citeauthoryear{Salama, Ala{\c{c}}am, and Menzel}{Salama
  et~al\mbox{.}}{2018}]{salama2018text}
\bibfield{author}{\bibinfo{person}{Amr~Rekaby Salama}, \bibinfo{person}{Ozge
  Ala{\c{c}}am}, {and} \bibinfo{person}{Wolfgang Menzel}.}
  \bibinfo{year}{2018}\natexlab{}.
\newblock \showarticletitle{Text completion using a context-integrating
  dependency parser}. In \bibinfo{booktitle}{\emph{Proc. RepL4NLP'18}}.
  \bibinfo{publisher}{ACL}, \bibinfo{address}{Melbourne, Australia},
  \bibinfo{pages}{41--49}.
\newblock


\bibitem[\protect\citeauthoryear{Schr{\"o}der, Thiele, and Lehner}{Schr{\"o}der
  et~al\mbox{.}}{2011}]{schroder2011setting}
\bibfield{author}{\bibinfo{person}{Gunnar Schr{\"o}der}, \bibinfo{person}{Maik
  Thiele}, {and} \bibinfo{person}{Wolfgang Lehner}.}
  \bibinfo{year}{2011}\natexlab{}.
\newblock \showarticletitle{Setting goals and choosing metrics for recommender
  system evaluations}. In \bibinfo{booktitle}{\emph{UCERSTI2 workshop at the
  5th ACM conference on recommender systems}}, Vol.~\bibinfo{volume}{23}.
  \bibinfo{publisher}{ACM}, \bibinfo{address}{New York, NY, USA},
  \bibinfo{pages}{53}.
\newblock


\bibitem[\protect\citeauthoryear{Sears and Zha}{Sears and Zha}{2003}]{sears2003data}
\bibfield{author}{\bibinfo{person}{Andrew Sears} {and} \bibinfo{person}{Ying
  Zha}.} \bibinfo{year}{2003}\natexlab{}.
\newblock \showarticletitle{Data entry for mobile devices using soft keyboards:
  Understanding the effects of keyboard size and user tasks}.
\newblock \bibinfo{journal}{\emph{J. of Human-Computer Interaction}}
  \bibinfo{volume}{16}, \bibinfo{number}{2} (\bibinfo{year}{2003}),
  \bibinfo{pages}{163--184}.
\newblock


\bibitem[\protect\citeauthoryear{Toda, Cortez, da~Silva, and de~Moura}{Toda
  et~al\mbox{.}}{2010}]{toda2010probabilistic}
\bibfield{author}{\bibinfo{person}{Guilherme~A Toda}, \bibinfo{person}{Eli
  Cortez}, \bibinfo{person}{Altigran~S da Silva}, {and} \bibinfo{person}{Edleno
  de Moura}.} \bibinfo{year}{2010}\natexlab{}.
\newblock \showarticletitle{A probabilistic approach for automatically filling
  form-based web interfaces}.
\newblock \bibinfo{journal}{\emph{Proc. of the VLDB Endowment}}
  \bibinfo{volume}{4}, \bibinfo{number}{3} (\bibinfo{year}{2010}),
  \bibinfo{pages}{151--160}.
\newblock


\bibitem[\protect\citeauthoryear{Tolley, Forde, Coffey, Sittig, Ash, Husband,
  Bates, and Slight}{Tolley et~al\mbox{.}}{2018}]{tolley2018factors}
\bibfield{author}{\bibinfo{person}{Clare~L Tolley}, \bibinfo{person}{Niamh~E
  Forde}, \bibinfo{person}{Katherine~L Coffey}, \bibinfo{person}{Dean~F
  Sittig}, \bibinfo{person}{Joan~S Ash}, \bibinfo{person}{Andrew~K Husband},
  \bibinfo{person}{David~W Bates}, {and} \bibinfo{person}{Sarah~P Slight}.}
  \bibinfo{year}{2018}\natexlab{}.
\newblock \showarticletitle{Factors contributing to medication errors made when
  using computerized order entry in pediatrics: a systematic review}.
\newblock \bibinfo{journal}{\emph{Journal of the American Medical Informatics
  Association}} \bibinfo{volume}{25}, \bibinfo{number}{5}
  (\bibinfo{year}{2018}), \bibinfo{pages}{575--584}.
\newblock


\bibitem[\protect\citeauthoryear{Troiano, Birtolo, and Armenise}{Troiano
  et~al\mbox{.}}{2017}]{troiano2017modeling}
\bibfield{author}{\bibinfo{person}{Luigi Troiano}, \bibinfo{person}{Cosimo
  Birtolo}, {and} \bibinfo{person}{Roberto Armenise}.}
  \bibinfo{year}{2017}\natexlab{}.
\newblock \showarticletitle{Modeling and predicting the user next input by
  Bayesian reasoning}.
\newblock \bibinfo{journal}{\emph{Soft Computing}} \bibinfo{volume}{21},
  \bibinfo{number}{6} (\bibinfo{year}{2017}), \bibinfo{pages}{1583--1600}.
\newblock


\bibitem[\protect\citeauthoryear{Umer, Liu, and Illahi}{Umer
  et~al\mbox{.}}{2019}]{umer2019cnn}
\bibfield{author}{\bibinfo{person}{Qasim Umer}, \bibinfo{person}{Hui Liu},
  {and} \bibinfo{person}{Inam Illahi}.} \bibinfo{year}{2019}\natexlab{}.
\newblock \showarticletitle{CNN-based automatic prioritization of bug reports}.
\newblock \bibinfo{journal}{\emph{IEEE Transaction on Reliability}}
  \bibinfo{volume}{69}, \bibinfo{number}{4} (\bibinfo{year}{2019}),
  \bibinfo{pages}{1341--1354}.
\newblock


\bibitem[\protect\citeauthoryear{Van Den~Bosch and Bogers}{Van Den~Bosch and
  Bogers}{2008}]{van2008efficient}
\bibfield{author}{\bibinfo{person}{Antal Van Den~Bosch} {and}
  \bibinfo{person}{Toine Bogers}.} \bibinfo{year}{2008}\natexlab{}.
\newblock \showarticletitle{Efficient context-sensitive word completion for
  mobile devices}. In \bibinfo{booktitle}{\emph{Proc. MobileHCI'08}}.
  \bibinfo{publisher}{ACM}, \bibinfo{address}{New York, NY, USA},
  \bibinfo{pages}{465--470}.
\newblock


\bibitem[\protect\citeauthoryear{{W3C School}}{{W3C School}}{2021}]{w3chtml2021html}
\bibfield{author}{\bibinfo{person}{{W3C School}}.}
  \bibinfo{year}{2021}\natexlab{}.
\newblock \bibinfo{title}{HTML <input> autocomplete Attribute}.
\newblock
  \bibinfo{howpublished}{\url{https://www.w3schools.com/tags/att_input_autocomplete.asp}}.
\newblock


\bibitem[\protect\citeauthoryear{W3CSchools}{W3CSchools}{2017}]{w3cschools2017html}
\bibfield{author}{\bibinfo{person}{W3CSchools}.}
  \bibinfo{year}{2017}\natexlab{}.
\newblock \bibinfo{title}{{HTML} DOM input text object}.
\newblock
  \bibinfo{howpublished}{\url{https://www.w3schools.com/jsref/dom\_obj\_text.asp}}.
\newblock


\bibitem[\protect\citeauthoryear{Wang, Zou, Keivanloo, Upadhyaya, Ng, and
  Ng}{Wang et~al\mbox{.}}{2014}]{wang2014automatic}
\bibfield{author}{\bibinfo{person}{Shaohua Wang}, \bibinfo{person}{Ying Zou},
  \bibinfo{person}{Iman Keivanloo}, \bibinfo{person}{Bipin Upadhyaya},
  \bibinfo{person}{Joanna Ng}, {and} \bibinfo{person}{Tinny Ng}.}
  \bibinfo{year}{2014}\natexlab{}.
\newblock \showarticletitle{Automatic reuse of user inputs to services among
  end-users in service composition}.
\newblock \bibinfo{journal}{\emph{IEEE Transaction on Services Computing}}
  \bibinfo{volume}{8}, \bibinfo{number}{3} (\bibinfo{year}{2014}),
  \bibinfo{pages}{343--355}.
\newblock


\bibitem[\protect\citeauthoryear{Wang, Zou, Ng, and Ng}{Wang
  et~al\mbox{.}}{2017}]{wang2017context}
\bibfield{author}{\bibinfo{person}{Shaohua Wang}, \bibinfo{person}{Ying Zou},
  \bibinfo{person}{Joanna Ng}, {and} \bibinfo{person}{Tinny Ng}.}
  \bibinfo{year}{2017}\natexlab{}.
\newblock \showarticletitle{Context-aware service input ranking by learning
  from historical information}.
\newblock \bibinfo{journal}{\emph{IEEE Transaction on Services Computing}}
  \bibinfo{volume}{14}, \bibinfo{number}{1} (\bibinfo{year}{2017}),
  \bibinfo{pages}{97--110}.
\newblock


\bibitem[\protect\citeauthoryear{Westbrook, Baysari, Li, Burke, Richardson, and
  Day}{Westbrook et~al\mbox{.}}{2013}]{westbrook2013safety}
\bibfield{author}{\bibinfo{person}{Johanna~I Westbrook},
  \bibinfo{person}{Melissa~T Baysari}, \bibinfo{person}{Ling Li},
  \bibinfo{person}{Rosemary Burke}, \bibinfo{person}{Katrina~L Richardson},
  {and} \bibinfo{person}{Richard~O Day}.} \bibinfo{year}{2013}\natexlab{}.
\newblock \showarticletitle{The safety of electronic prescribing:
  manifestations, mechanisms, and rates of system-related errors associated
  with two commercial systems in hospitals}.
\newblock \bibinfo{journal}{\emph{Journal of the American Medical Informatics
  Association}} \bibinfo{volume}{20}, \bibinfo{number}{6}
  (\bibinfo{year}{2013}), \bibinfo{pages}{1159--1167}.
\newblock


\bibitem[\protect\citeauthoryear{Winckler, Gaits, Vo, Sergio, and
  Rossi}{Winckler et~al\mbox{.}}{2011}]{winckler2011approach}
\bibfield{author}{\bibinfo{person}{Marco Winckler}, \bibinfo{person}{Vicent
  Gaits}, \bibinfo{person}{Dong-Bach Vo}, \bibinfo{person}{Firmenich Sergio},
  {and} \bibinfo{person}{Gustavo Rossi}.} \bibinfo{year}{2011}\natexlab{}.
\newblock \showarticletitle{An approach and tool support for assisting users to
  fill-in web forms with personal information}. In
  \bibinfo{booktitle}{\emph{Proc. SIGDOC'11}}. \bibinfo{publisher}{ACM},
  \bibinfo{address}{New York, NY, USA}, \bibinfo{pages}{195--202}.
\newblock


\bibitem[\protect\citeauthoryear{Yang, Lo, Xia, Bao, and Sun}{Yang
  et~al\mbox{.}}{2016}]{yang2016combining}
\bibfield{author}{\bibinfo{person}{Xinli Yang}, \bibinfo{person}{David Lo},
  \bibinfo{person}{Xin Xia}, \bibinfo{person}{Lingfeng Bao}, {and}
  \bibinfo{person}{Jianling Sun}.} \bibinfo{year}{2016}\natexlab{}.
\newblock \showarticletitle{Combining word embedding with information retrieval
  to recommend similar bug reports}. In \bibinfo{booktitle}{\emph{Proc.
  ISSRE'16}}. \bibinfo{publisher}{IEEE}, \bibinfo{address}{Ottawa, ON, Canada},
  \bibinfo{pages}{127--137}.
\newblock


\bibitem[\protect\citeauthoryear{Ye, Bunescu, and Liu}{Ye
  et~al\mbox{.}}{2014}]{ye2014learning}
\bibfield{author}{\bibinfo{person}{Xin Ye}, \bibinfo{person}{Razvan Bunescu},
  {and} \bibinfo{person}{Chang Liu}.} \bibinfo{year}{2014}\natexlab{}.
\newblock \showarticletitle{Learning to rank relevant files for bug reports
  using domain knowledge}. In \bibinfo{booktitle}{\emph{Proc. FSE'14}}.
  \bibinfo{publisher}{ACM}, \bibinfo{address}{New York, NY, USA},
  \bibinfo{pages}{689--699}.
\newblock


\bibitem[\protect\citeauthoryear{Zhang, Zhai, and Wobbrock}{Zhang
  et~al\mbox{.}}{2019}]{zhang2019text}
\bibfield{author}{\bibinfo{person}{Mingrui~Ray Zhang}, \bibinfo{person}{Shumin
  Zhai}, {and} \bibinfo{person}{Jacob~O. Wobbrock}.}
  \bibinfo{year}{2019}\natexlab{}.
\newblock \showarticletitle{Text entry throughput: towards unifying speed and
  accuracy in a single performance metric}. In \bibinfo{booktitle}{\emph{Proc.
  CHI'19}}. \bibinfo{publisher}{ACM}, \bibinfo{address}{New York, NY, USA},
  \bibinfo{pages}{1--13}.
\newblock


\bibitem[\protect\citeauthoryear{Zhou}{Zhou}{2021}]{zhou2021ensemble}
\bibfield{author}{\bibinfo{person}{Zhi-Hua Zhou}.}
  \bibinfo{year}{2021}\natexlab{}.
\newblock \showarticletitle{Ensemble learning}.
\newblock In \bibinfo{booktitle}{\emph{Machine Learning}}.
  \bibinfo{publisher}{Springer}, \bibinfo{address}{Singapore},
  \bibinfo{pages}{181--210}.
\newblock


\end{thebibliography}

\end{document}